\documentclass{aa}
\usepackage{txfonts}

\usepackage{amsmath}
\usepackage{natbib}
\usepackage{graphicx}
\usepackage{color}
\usepackage{appendix}
\usepackage{fancyhdr}
\usepackage{multirow}

\defcitealias{Schrabback}{S10}
\defcitealias{Schrabback_2018}{S18a}
\defcitealias{Schrabback_2018b}{S18b}
\defcitealias{Schrabback_2019}{S20}
\defcitealias{Diemer_2019}{D19}

\begin{document}

\title{Constraining the masses of high-redshift clusters with weak lensing: Revised shape calibration testing for the impact of stronger shears and increased blending}

\author{B. Hern\'andez-Mart\'in, 
          \inst{1}
          T. Schrabback\inst{1},
          H. Hoekstra\inst{2}
        \and
        N. Martinet\inst{3}
        \and
        J. Hlavacek-Larrondo\inst{4}
        \and
        L. E. Bleem\inst{5,6}
        \and
        M. D. Gladders\inst{6,7}
        \and
        B. Stalder\inst{8,9}
        \and
        A. A. Stark\inst{9}
        \and
        M. Bayliss\inst{10}}
        
   \institute{Argelander Institut f\"ur Astronomie, Universit\"at Bonn,
              Auf dem H\"ugel 71, D-53121 Bonn \\
              \email{bhernandez@astro.uni-bonn.de}
              \and 
              Leiden Observatory, Leiden University,
              P.O. Box 9513, 2300 RA Leiden, The Netherlands
              \and
              Aix-Marseille Univ., CNRS, CNES, LAM, Marseille, France
              \and 
              D\'epartement de Physique, Universit\'e de Montr\'eal,
Montr\'eal, QC, Canada
\and
High Energy Physics Division, Argonne National Laboratory, 9700 South Cass Avenue, Lemont, IL 60439, USA
\and
Kavli Institute for Cosmological Physics, University of Chicago, 5640 South Ellis Avenue, Chicago, IL 60637, USA
\and
Department of Astronomy and Astrophysics, University of Chicago, 5640 South Ellis Avenue, Chicago, IL 60637, USA
\and
Rubin Observatory Project Office, 950 N. Cherry Ave, Tucson, AZ 85719, USA
\and
Center for Astrophysics $|$ Harvard \& Smithsonian, 60 Garden Street, Cambridge, MA 02138, USA
\and
Department of Physics, University of Cincinnati, Cincinnati, OH 45221, USA}

   \date{Received 28 February 2020; Accepted 27 May 2020}

\titlerunning{Constraining the masses of high-redshift clusters with weak lensing}
\authorrunning{Hernandez-Martin et al.}

\abstract
{ Weak lensing (WL) measurements suffer from well-known shear estimation biases, which can be partially corrected for with the use of image simulations.
In this work we present an analysis of simulated images that mimic Hubble Space telescope/Advance Camera for Surveys (HST/ACS) observations of high-redshift galaxy clusters, including cluster specific issues such as non-weak shear and increased blending.
Our synthetic galaxies have been generated to have similar observed properties as the background-selected source samples studied in the real images.
First, we used simulations with galaxies placed on a grid to determine a revised signal-to-noise-dependent ($S/N_\mathrm{KSB}$) correction for multiplicative shear measurement bias,
and to quantify the sensitivity of our KSB+ bias calibration to mismatches of galaxy or PSF properties between the real data and the simulations.
Next, we studied the impact of increased blending and light contamination from cluster and foreground galaxies, finding it to be negligible for high-redshift (\mbox{$z>0.7$}) clusters, whereas shear measurements can be affected at the \mbox{$\sim 1\%$} level for lower redshift clusters given their brighter member galaxies.
Finally, we studied the impact of fainter neighbours and selection bias using a set of simulated images that mimic the positions and magnitudes of galaxies in Cosmic Assembly Near-IR Deep Extragalactic Legacy Survey (CANDELS) data, thereby including realistic clustering.
While the initial \texttt{SExtractor} object detection causes a multiplicative shear selection bias of \mbox{$ -0.028 \pm 0.002$}, this is reduced to \mbox{$ -0.016 \pm 0.002$} by further cuts applied in our pipeline.
Given the limited depth of the CANDELS data, we compared our CANDELS-based estimate for the impact of faint neighbours
on the multiplicative shear measurement bias
to a grid-based analysis, to which we added clustered galaxies
to even fainter magnitudes based on Hubble Ultra Deep Field (HUDF) data, yielding a refined estimate of \mbox{$\sim -0.013$}.
Our sensitivity analysis suggests that our pipeline is calibrated to an accuracy of \mbox{$\sim 0.015$} once all corrections are applied, which is fully sufficient for current and near-future weak lensing studies of high-redshift clusters.
As an application, we used it for a refined analysis of three highly relaxed clusters from the South Pole Telescope Sunyaev-Zeldovich (SPT-SZ) survey, where we now included measurements down to the cluster core (\mbox{$r>200$} kpc) as enabled by our work. Compared to previously employed scales (\mbox{$r>500$} kpc), this tightens the cluster mass constraints by a factor 1.38 on average.
}

\maketitle

\section{Introduction}

Galaxy clusters are among the most massive structures in the Universe and are key to our cosmological understanding. One of the techniques used to find them is the Sunyaev-Zeldovich (SZ) effect \citep{Sunyaev}. In particular, the South Pole Telescope (SPT) with its 2500 deg$^2$ SPT-SZ survey \citep{Bleem} has been very efficient in finding massive clusters out to the highest redshifts where they exist. This is useful to obtain robust constraints on the number of clusters as a function of redshift and mass, which provide a sensitive route to constrain cosmological parameters (e.g. \citealp{Bocquet}). 

Weak lensing (WL, \citealp{Bartelmann}) provides a powerful tool to obtain accurate cluster mass measurements, which are required to constrain the cosmology.
The images of background galaxies are distorted due to the gravitational potential of the matter along their light path. Constraining this distortion, in particular the anisotropic shear, allows us to constrain the total mass distribution inside the structures, including dark matter. In particular, weak lensing is used to directly constrain the masses of clusters and calibrate mass-observable scaling relations (e.g. \citealp{Zhang}, \citealp{Dietrich}), which can then be used to estimate the masses of other clusters from their X-ray, SZ, or optical properties. The changes in the shapes of galaxies due to WL are typically small compared to the intrinsic ellipticity of the galaxies, which leads to the so-called 'shape noise', but we can azimuthally average the shear estimates obtained from many background galaxies in order to tightly constrain the cluster shear profile in different radial bins around the cluster centre. This measurement can then be used to estimate the total mass and the mass profile of the cluster. 

In this work, we study high-redshift clusters, which require deep high-resolution imaging to measure the shapes of the typically small and faint galaxies in the distant background. We follow the setup from \citet[][\citetalias{Schrabback_2018}  henceforth]{Schrabback_2018}, who study the WL signatures of 13 distant ($z \gtrsim 0.6$) SPT-SZ clusters in mosaic Hubble Space Telescope (HST) observations obtained using the Advance Camera for Surveys (ACS), employing shape measurements based on the KSB+ formalism (\citealp{Kaiser}, \citealp{Luppino}, \citealp{Hoekstra}). 

Our main aim is to carry out the calibration of the shear estimation for weak lensing mass modelling of galaxy clusters. The difference between the real and measured shear is known as the shear bias and simulations are needed in order to calibrate it. This allows us to control the inputs and systematics, which helps us to understand how the methods behave. However, the main drawback is the need to mimic the real data as closely as possible in order to serve as proper calibration and not introduce additional biases \citep[see][]{Hoekstra_2015, Hoekstra_2017}. Any assumptions that simplify the simulated images should be justified and their impact studied. In particular, the presence of selection bias is often ignored, but it can have a large influence. Here, we aim to build on previous work and further understand the impact that the choices made to create our simulations have on the final results. 

\citet[][\citetalias{Schrabback}  hereafter]{Schrabback} employed the
STEP2 simulations \citep{Massey} to obtain a correction
for noise-related multiplicative shear measurement bias which was
also applied in the cluster studies presented by \citetalias{Schrabback_2018} and \citet[][\citetalias{Schrabback_2018b}  henceforth]{Schrabback_2018b}.
These simulations mimicked ground-based observations with a galaxy selection that did not accurately match the background-selected source samples from \citetalias{Schrabback_2018}. Because of this, \citetalias{Schrabback_2018} assumed a larger shear uncertainty to account for possible discrepancies. In this work, we create more realistic HST-like simulations in order to update the noise bias correction and focus on calibrating our shape measurement algorithm for the use in galaxy cluster studies. Most previous work on shear calibration with simulations focused on the measurement of cosmic shear (e.g. \citealp{Hoekstra_2017}, \citealp{Pujol}, \citealp{Kannawadi}). While cosmic shear typically has tighter requirements on systematic error control, these simulations usually do not include the cluster regime of stronger shears and increased blending. In this paper we explicitly investigate these two effects.

This paper is organised as follows: in Sect.\thinspace\ref{sec:KSB} we summarise the formalism of the moment-based shape measurement method employed in our analysis. In Sect.\thinspace\ref{sec:grid} we introduce our simulations and investigate how the different systematics affect the shear recovery for simulations with source galaxies placed on a grid. In Sect.\thinspace\ref{sec:clustering} we study the impact of bright cluster members on our bias estimations, also as a function of their redshift. In Sect.\thinspace\ref{sec:neighbours} we investigate the influence of blending and selection bias by mimicking CANDELS data, for which we also estimate the signal-to-noise ratio-dependent correction used for the real data. In Sect.\thinspace\ref{sec:final} we summarise the different bias contributions and estimate the final residual bias. Our analysis enables a robust shear recovery also in the inner parts of massive clusters, where stronger shears and increased blending occur. As a demonstration, we present an HST WL analysis of three of the most relaxed SPT-SZ clusters in Sect.\thinspace\ref{sec:mass}, showing how much the mass constraints tighten by including the inner cluster regions, which have been excluded in the analysis of large samples presented by \citetalias{Schrabback_2018} and \citet[][\citetalias{Schrabback_2019} below]{Schrabback_2019}.

In this paper all magnitudes are in the AB system. When fitting cluster shear profiles and computing cluster mass constraints we assume a flat $\Lambda$CDM cosmology characterised by $\Omega_{\mathrm{m}}=0.3$, $\Omega_{\Lambda}=0.7$, and $H_0= 70\, \mathrm{km} \,\mathrm{s}^{-1}\, \mathrm{Mpc}^{-1}$.

\section{Shape measurement method} \label{sec:KSB}
There are different approaches to estimate the shear induced by the lensing effect. Some results are based on model-fitting (e.g. im2shape in \citealt{Bridle}, lensfit in \citealt{Miller}), whereas others employ moment-based methods, such as the Kaiser, Squires, and Broadhurst (KSB+) formalism (\citealp{Kaiser}, \citealp{Luppino}, \citealp{Hoekstra}) used in this work. KSB+ is based on the computation of higher-order moments of the light distribution using a Gaussian weight function. Both kinds of methods can suffer from a bias in the recovered shear due to noise bias \citep{Refregier}. Extensive work has been done in the past to calibrate WL methods and correct for this and other biases. Some pioneering work which compares different methods includes the Shear TEsting Program (STEP, \citealt{Heymans}) and the GRavitational lEnsing Accuracy Testing (\citealp{Bridle_2010}; \citealp{Mandelbaum_2015}) and more recent work was presented for example in \cite{Hoekstra_2017}, \cite{Fenech-Conti}, \cite{Mandelbaum_2018}, \cite{Pujol}, \cite{Martinet} and \citet{Kannawadi}.

We employ the KSB+ formalism. It is a moment-based algorithm that determines the shapes of galaxies by performing a correction for the PSF using the stars in the field. KSB+ has been widely used for cluster studies (e.g. \citealp{Hoekstra_2015}, \citealp{Von_der_linden}, \citetalias{Schrabback_2018}, \citetalias{Schrabback_2018b}, \citealp{Dietrich}, \citealp{Herbonnet}). The cosmic shear community is actively developing newer methods that aim to reach the $\sim 10^{-3}$ accuracy requirements of future cosmic shear experiments (e.g. \citealp{Bernstein}, \citealp{Tewes}). Such accuracy is however not required for current or near-future cluster studies, where an accuracy at the 1--2\% level is sufficient.

The KSB+ method measures the shape of each background galaxy through the determination of the weighted quadrupole moments $Q_{ij}$ of the light distribution in the form of
\begin{equation}
 Q_{ij} =\frac{\int \mathrm{d}^2 \boldsymbol{\theta}\, W(\boldsymbol{\theta})\, I(\boldsymbol{\theta}) \,\theta_i\, \theta_j}{\int \mathrm{d}^2 \boldsymbol{\theta}\, W(\boldsymbol{\theta}) \,I(\boldsymbol{\theta})} \, , \label{eq:moments}
\end{equation}
where $W(\boldsymbol{\theta})$ is a Gaussian weight function with scale length $r_g$, which we choose as the \texttt{SExtractor} \citep{Bertin} FLUX\_RADIUS, and $I(\boldsymbol{\theta})$ is the surface brightness.
The two polarisation components $e_{\alpha}$ are related to the moments as
\begin{eqnarray}
 e_1 &=&\frac{Q_{11}-Q_{22}}{Q_{11}+Q_{22}} \, , \nonumber \\
 e_2 &=&\frac{2Q_{12}}{Q_{11}+Q_{22}}  \, . \label{eq:axis}
\end{eqnarray}

The PSF impact on the ellipticity of a galaxy can be approximated as a convolution of a isotropic smearing PSF and an anisotropic kernel. We can define the PSF anisotropy kernel $p_{\mu}$, which was computed using the polarisation of stars $e_{\alpha}^{* \, \rm{obs}}$ (hence the overscript $^*$) as
\begin{equation}
p_{\mu} = (P^{\rm{sm} \,*})^{-1}_{\mu \alpha} e_{\alpha}^{* \, \rm{obs}} \, .
\end{equation}
$P^{\rm{sm} \,*}$ is the stellar smear polarisability tensor.

In the case of stars, following \cite{Hoekstra}, the weight function for the moment computation in Eq.\thinspace(\ref{eq:moments}) is adjusted to match the $r_g$ of the corresponding galaxy.

The PSF anisotropy-corrected polarisation can be defined as
\begin{equation}
 e_{\alpha}^{\rm{cor}} = e_{\alpha}^{\rm{obs}} - P_{\alpha \beta}^{\rm{sm}} p_{\beta} \, ,
\end{equation}
where $P_{\alpha \beta}^{\rm{sm}}$ is the smear polarisability tensor of the galaxy which describes the sensitivity to the circular smearing caused by the PSF.

We define the pre-seeing shear polarisability tensor $P^{g}$ as:
\begin{equation}
P^{g}_{\alpha \beta} = P_{\alpha \beta}^{\rm{sh}} - P_{\alpha \mu}^{\rm{sm}} (P^{\rm{sm *}})^{-1}_{\mu \delta} P_{\delta \beta}^{\rm{sh*}}\, ,
\end{equation}
where $P^{\rm{sh}}$ is the shear polarisability tensor from \cite{Hoekstra}, which measures the response of the galaxy ellipticity to shear in the absence of PSF effects and $P^{\rm{sh*}}$ is the stellar shear polarisability tensor.

With all these ingredients we can now define the KSB+ shear estimator as
\begin{equation}
 \epsilon_{\alpha}^{\rm{iso}} = (P^{g})_{\alpha \beta}^{-1} [e_{\beta}^{\rm{obs}}  - P_{\beta \mu}^{\rm{sm}} p_{\mu}] \, .
\end{equation}
This provides an estimate for the reduced gravitational shear $g$. In the absence of shape measurement biases $\langle \epsilon_{\alpha}^{\rm{iso}} \rangle = g$.
In our implementation, we make the approximation $(P^{g})^{-1}= 2/\rm{Tr}[P^{g}]$ to reduce noise, following \cite{Erben}.

To identify the galaxies in the images we used \texttt{SExtractor}, and for the moment measurement the code analyseldac \citep{Erben}. The object detection in \texttt{SExtractor} introduces a selection bias which is studied in Sect.\thinspace\ref{sec:selection_bias}.

Following \citet{Erben} we defined the KSB signal-to-noise ratio as
\begin{equation}
S/N_{\mathrm{KSB}} = \frac{\sum_i  \, W_i \, I_i}{\sigma \sqrt{\sum_i  \, W^2_i}} \, , \label{eq:sn_ksb} 
\end{equation}
where we sum over the pixels\footnote{In fact, the \citet{Erben} KSB+ implementation employed in our analysis uses sub-pixel interpolation to reduce the impact of pixel sampling when computing brightness moments and other relevant quantities such as $S/N_\mathrm{KSB}$.} $i$ and $W_i$ is the weight function evaluated in each pixel, $I_i$ is the intensity measured in this pixel, and $\sigma$ is the single-pixel dispersion of the sky background.

We applied similar cuts in our analysis as \citetalias{Schrabback}. We also required $r_{\mathrm{h}} >1.2 \, r_{\mathrm{h}}^{*}$, where $r_{\mathrm{h}}^{*}$ is the measured half-light radius of the stars in the field. We only considered galaxies with $\text{Tr} \, P^{g}/2 >0.1$ (to exclude nearly unresolved galaxies with very large PSF corrections) and $r_{\mathrm{g}}<10$ pixels (to ensure a sufficient coverage by the postage stamps\footnote{\citetalias{Schrabback_2018} use an even more restrictive cut $r_{\mathrm{g}}<7$ pixels. We find that switching between their and our cut changes the resulting multiplicative bias by less than $0.1\%$ for our colour-selected sample of mostly high-redshift distant background galaxies.}). We employed weights to down-weight the contributions of galaxies with noisy ellipticity estimates. Following \citetalias{Schrabback_2018} we defined weights via the magnitude-dependent RMS ellipticity, thereby avoiding biases that can occur for ellipticity-dependent weights.

Knowing the input shear ($g^{\text{true}}$) in our simulations we can run our pipeline and compare it to the recovered values ($g^{\text{obs}}$) assuming the following relation:
\begin{equation}
g_{\alpha}^{\text{obs}}= (1+m_{\alpha}) \, g_{\alpha}^{\text{true}} + c_{\alpha} \, , \label{eq:bias}
\end{equation}
where $\alpha$ refers to each component. In the case of cosmic shear the additive bias ($c$) is important, but for cluster analyses it typically cancels out for the azimuthally averaged tangential shear profiles\footnote{Additive biases can have a net impact if cluster centres are always placed at similar detector coordinates and similar PSF patterns occur with a significant net positive or negative average tangential PSF ellipticity (see e.g. \citealp{Dietrich}). For HST/ACS thermal breathing can cause substantial PSF ellipticity in individual frames. However, averaged over many exposures the mean tangential PSF ellipticity with respect to either the camera centre or the corners (which are typically used to place cluster centres, see S18a and S20) is very small (compare Fig.\thinspace8 in S10). Given the low level of additive biases for our method (see Table 2), typical cluster shears, and the accuracy requirements of our cluster studies,
we therefore conclude that we can safely ignore the impact of additive biases.}. In the following sections, we therefore mostly concentrate the analysis on the multiplicative bias ($m$). A consistency check for a possible quadratic dependence showed that this is negligible (see Sect.\thinspace\ref{sec:PSF}). 

The shear measurement suffers from noise bias \citep{Viola}. It is caused by the presence of noise in the data and the lack of knowledge on the exact parameters of the background galaxies (e.g. centroid, size and ellipticity). This shear measurement method is particularly biased at low signal-to-noise ratios. This was partly corrected for in \citetalias{Schrabback}, as
\begin{equation}
  g^{\textrm{cor}}=\frac{g^{\mathrm{biased}}}{1+m^{\mathrm{cor}}}   \, .
\end{equation}
Their correction ($m_{\mathrm{cor}}$) was computed using the STEP2 simulations \citep{Massey} of ground-based images and also tested for ACS-like simulations, which match the ACS resolution and detector characteristics, but do not accurately represent the background-selected source population of high-redshift cluster studies. Approximating the correction as a power law, \citetalias{Schrabback} obtained:
\begin{equation}
  m^{\mathrm{cor}}=-0.078 \left( \frac{S/N_{\mathrm{KSB}}}{2}\right) ^{-0.38} \, . \label{eq:or_sn_corr}
\end{equation}
This correction depends on the KSB+ signal-to-noise ratio, which is defined in Eq.\thinspace(\ref{eq:sn_ksb}). The effect of blends and neighbours complicates this simple picture (e.g. \citealp{Hoekstra_2015}). Our aim in the next section is to estimate a correction which depends on the signal-to-noise ratio and has been computed using isolated galaxies. This does not fully capture all bias effects but it removes the main contribution of noise bias.

An alternative signal-to-noise ratio to $S/N_{\mathrm{KSB}}$ can be defined using the \texttt{SExtractor} parameters. This definition is more widely used since it does not depend of the particulars of the shape measurement implementation:
\begin{equation}
    S/N_{\mathrm{flux}}= \frac{\mathrm{FLUX\_AUTO}}{\mathrm{FLUXERR\_AUTO}} \, . \label{eq:s/n_flux}
\end{equation}
We use this definition primarily to apply signal-to-noise cuts to the galaxies.

\section{Galaxies on a grid} \label{sec:grid}

Any shear measurement method needs to be calibrated through simulations to test for differences that might arise between the input and the recovered shear. For an accurate WL measurement, we need to understand and correct for such biases. We created customised simulations, which allowed us to test cluster-specific issues such as stronger shears, but also the impact of choices regarding the input parameters. An alternative option would be to use metacalibration \citep{Sheldon}, which does not use simulations, but it requires large quantities of data to avoid having very noisy results. We varied different aspects of the simulations, in order to investigate the sensitivities of our analysis to modelling details. The aim of this section is to understand the main simulation parameters that change the bias estimation, aiming to calibrate it to a tolerance level of around $1 \%$. Here, we use a grid placement for our galaxies in order to speed up the computation, to test how the input choices alone  (without the impact of neighbours) affect the bias, and for easier comparison to previous work. A more realistic approach, where we look at the effect of neighbours and selection bias is used in Sect. \ref{sec:clustering} and \ref{sec:neighbours}.

\subsection{Details on the creation of the simulations}
We created the simulations with the python package \texttt{GALSIM}\footnote{\url{https://github.com/GalSim-developers/GalSim}} \citep{Rowe} and used S\'ersic profiles for the galaxy light distribution:
\begin{equation}
    I(R)=I_{\mathrm{e}} \exp{(-b_n [(R/R_{\mathrm{e}})^{1/n}-1])}\, , \label{eq:sersic}
\end{equation}
with the model half-light radius $R_{\mathrm{e}}$, the intensity at that radius $I_e$ and the parameter $b_n\approx 2n-1/3$, with $n$ being the so-called S\'ersic index. The light profile of each galaxy was obtained from parametric fits to COSMOS galaxy images which are included in \texttt{GALSIM}. These provide input values for the S\'ersic index and the intrinsic ellipticity. This allowed us to employ a realistic input shape distribution for the creation of our galaxies. A more in depth discussion of this choice is presented in Sect.\thinspace\ref{sec:light}.

We simulated galaxies as they would be observed in HST/ACS images, shearing them either by a constant shear or, for the simulations discussed in Sect.\thinspace\ref{sec:clustering}, a theoretical cluster shear profile. We used an ACS-like PSF obtained from the software Tiny Tim \citep{Krist},  using re-fitted optical parameters \citep{Gillis} and including the effect of charge diffusion, which is discussed in more detail in Sect.\thinspace\ref{sec:PSF}.

We present here our 'grid simulation', which is used as a reference for the rest of this section when we vary the inputs and test how the results change. The simulation settings are summarised in Table \ref{tab:sims_param}, while the importance of these choices is investigated throughout the next subsections. In Sect.\thinspace\ref{sec:neighbours}, we discuss a more realistic scenario where galaxies suffer from neighbour contamination.

For their creation we extended the normally studied range of shear values to the non-weak regime of $|g|<0.4$, although the bias estimates reported in this work have been computed from the restricted range $|g|<0.2$ unless explicitly stated differently, since that is the regime used in cluster studies such as \citet{Schrabback_2018}. We also note that \texttt{GALSIM} does not allow for the inclusion of flexion effects (\citealp{Goldberg_2002}, \citealp{Goldberg}, \citealp{Bacon}). This is another reason why we limited our primary analysis to the $|g|<0.2$ regime, where flexion only plays a minor role .
 
To compute the bias, each image was assigned a different value of input shear, which is constant throughout that image. We used 50 different shear values in the range $-0.4<g<0.4$ with $10^4$ galaxies per value. We created a grid of $100 \times 100$ stamps of size $100 \times 100$ pixels and applied a random shift with a uniform distribution from $-0.5$ to $0.5$ at the sub-pixel level for the galaxy position to have a small displacement with respect to the pixel centre. In order to reduce shape noise we created a second set of galaxies, which are identical except for a 90 degree rotation of the input intrinsic ellipticity \citep{Massey}. This makes sure that the mean intrinsic ellipticity of the input population is 0, which reduces the number of galaxies needed, allowing us to constrain multiplicative biases to the few $\times 10^{-3}$ level. Multiple copies of the galaxies, rotated by $\pm 45$ degrees can also be created (e.g. \citealp{Fenech-Conti}), but in our case, due to the faintness of the galaxies, the noise dominates and the inclusion of further rotated simulations does not lead to a significant improvement, hence why we chose to only implement pairs.

In this part of the analysis we only consider galaxies that provide shape estimates in both the normal and the rotated frame. By doing this, we effectively cancel the effects of selection bias, which can be quite important \citep{Kannawadi}. In this section, however, we concern ourselves only with the changes in the residual bias estimates due to the input choices for the creation of the simulations. Using matched rotated pairs reduces the number of galaxies needed for these estimates. Our study of selection biases will be presented in Sect.\thinspace\ref{sec:selection_bias}. 
\begin{figure*}[t]
  \centering
 \begin{minipage}{0.45\linewidth}
  
  \includegraphics[width=\linewidth]{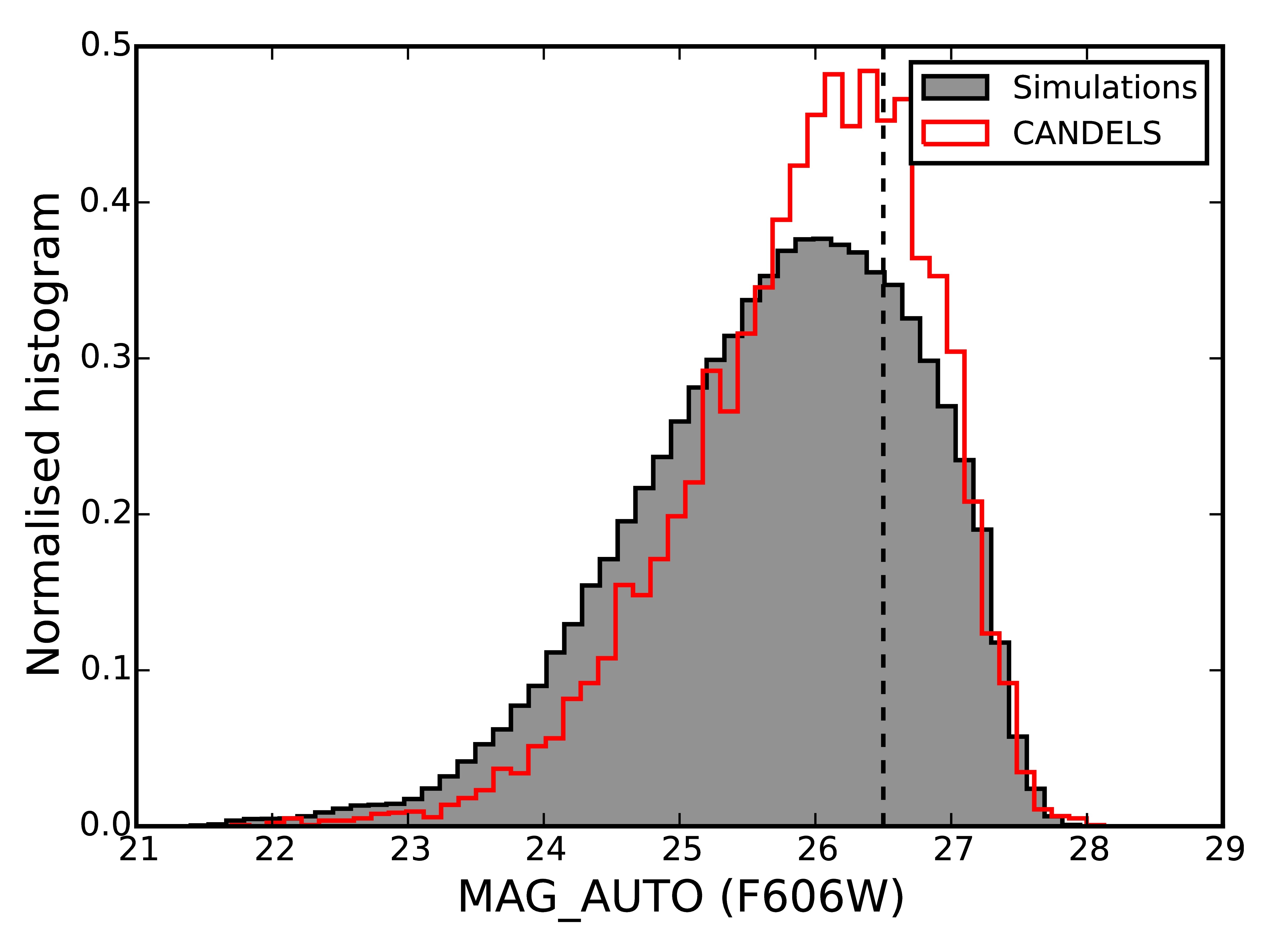}

  \end{minipage}\begin{minipage}{0.45\linewidth}
 
  \includegraphics[width=\linewidth]{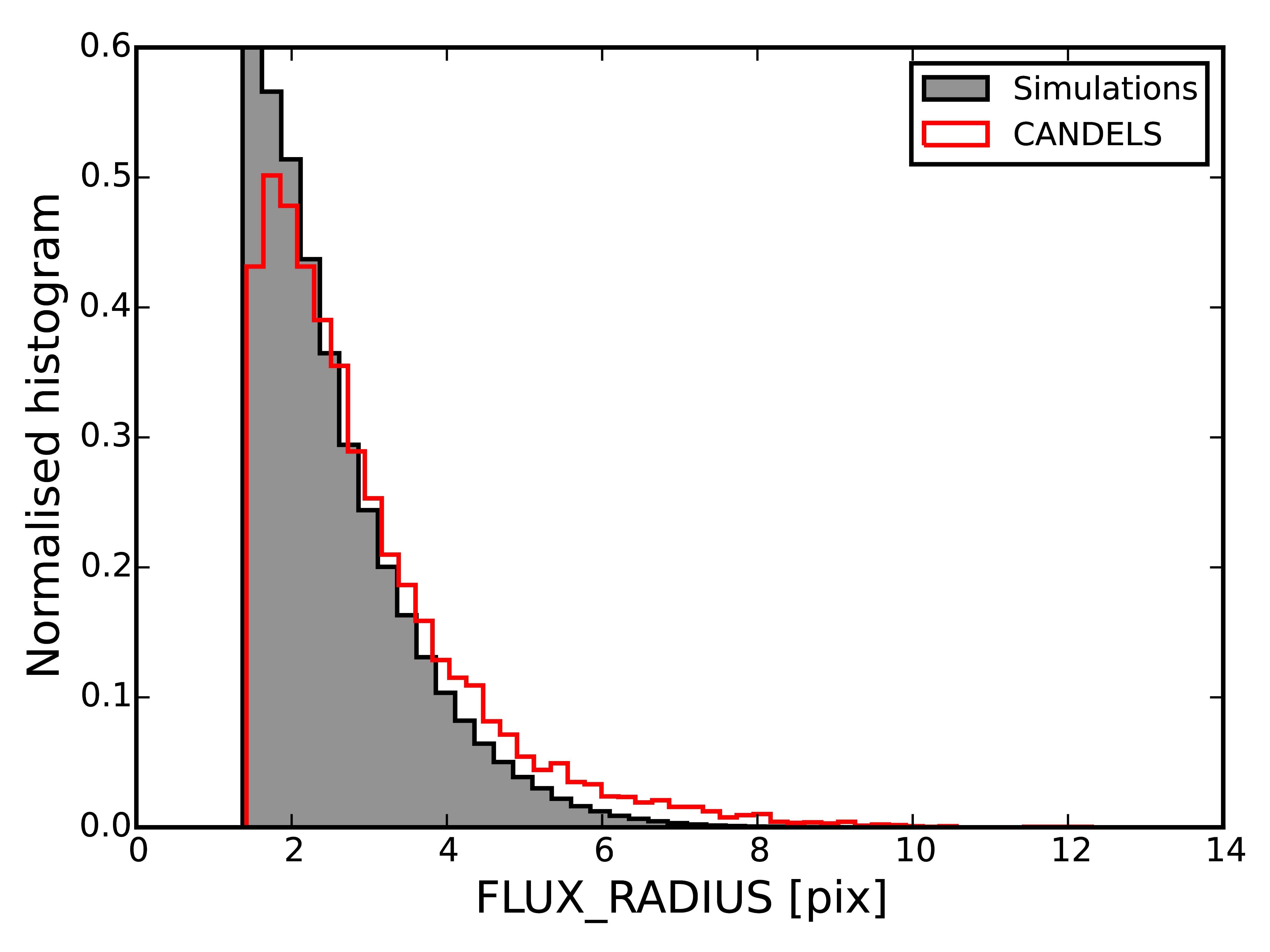}

  \end{minipage}
  \begin{minipage}{0.45\linewidth}
 
  \includegraphics[width=\linewidth]{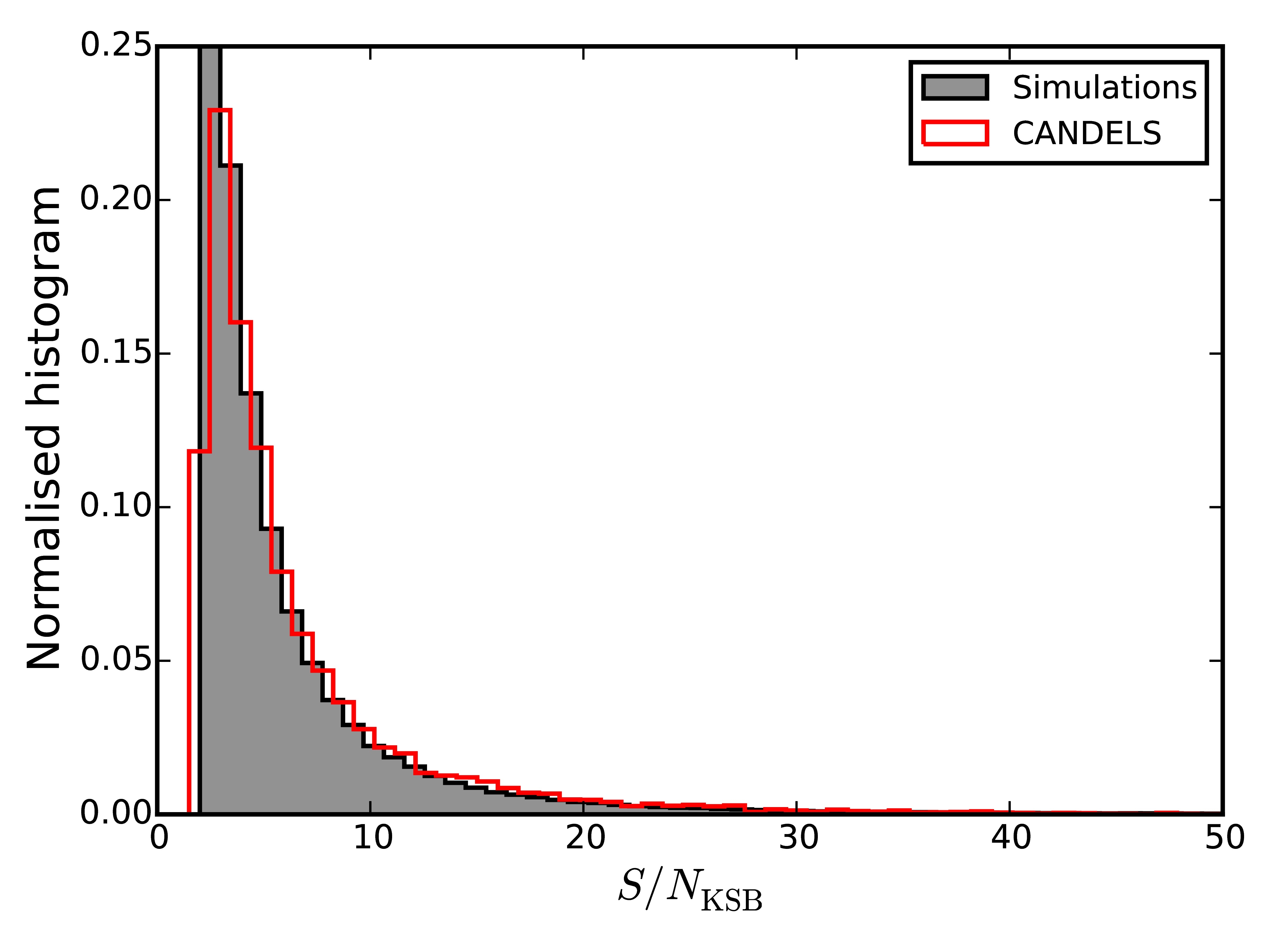}

  \end{minipage}\begin{minipage}{0.45\linewidth}
 
  \includegraphics[width=\linewidth]{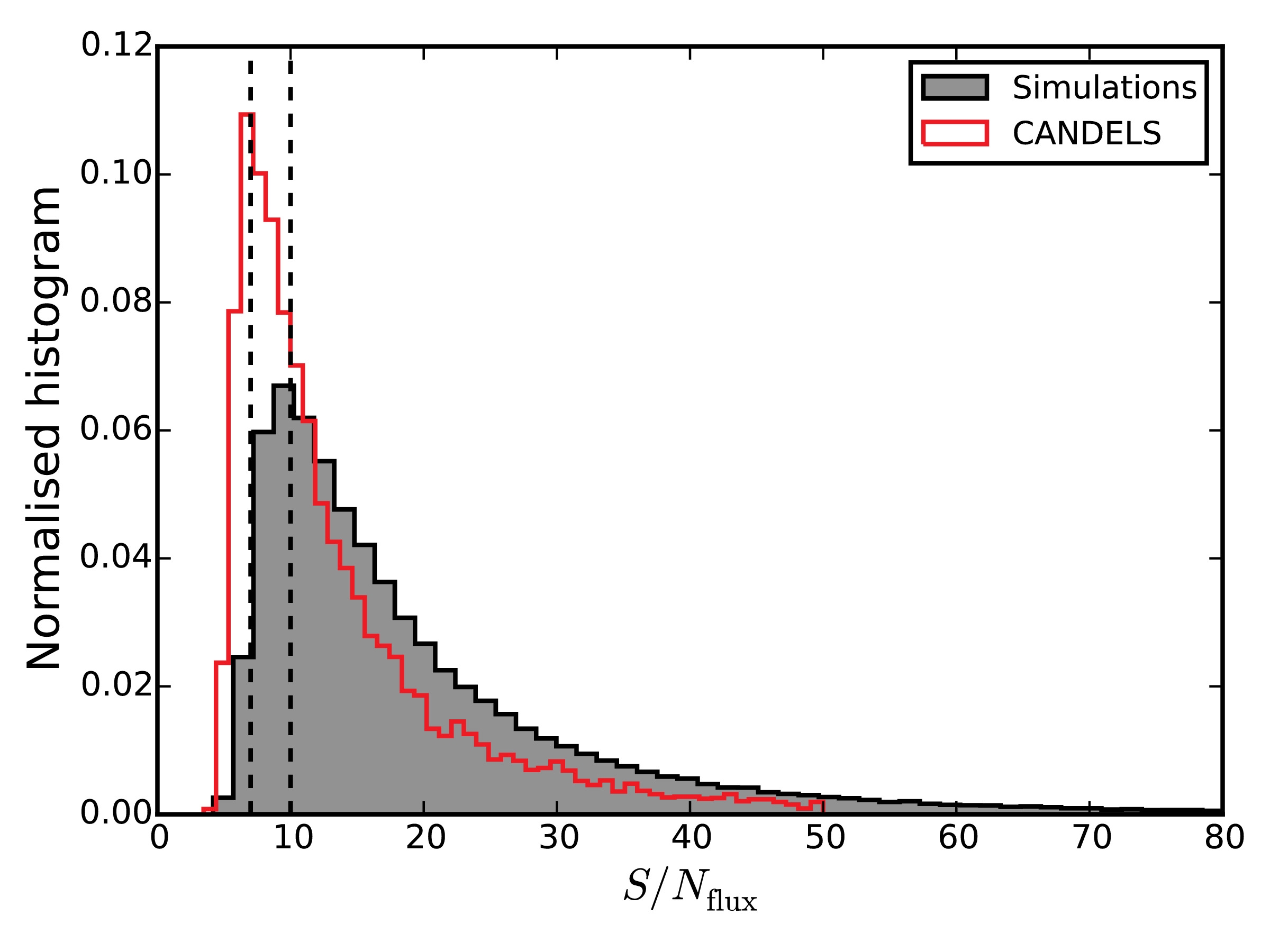}

  \end{minipage}
  \caption{Comparisons between the measured distributions in our simulations and the CANDELS data for the F606W magnitudes, the half-light radius, the KSB signal-to-noise ratio $S/N_{\mathrm{KSB}}$, and the \texttt{SExtractor} signal-to-noise ratio $S/N_{\mathrm{flux}}$. The vertical lines in the MAG\_AUTO distribution indicate the magnitude cuts we introduce for the bias estimation. The two vertical lines in the $S/N_{\mathrm{flux}}$ distribution show the two different signal-to-noise ratio cuts that we apply in this work.} \label{fig:candels}
 \end{figure*}

\begin{table*}[t]
 \caption{Summary of the input parameters used for the creation of the mock galaxies of the reference simulation.} \label{tab:sims_param}
\centering
\begin{tabular}{c c c}

\textbf{Parameter} & \textbf{Type} & \textbf{Values} \\
\hline
\hline
PSF model & HST/ACS-like & FWHM$=0\farcs1$\\
PSF ellipticity & HST/ACS-like & $|e|=0.072$\\
Noise & Gaussian & $\sigma=0.14 \, e^-/$s\\
Light distribution & Parametric fit to COSMOS$^*$ & Sersic index: 0.3 -- 6\\
Ellipticity distribution & Parametric fit to COSMOS$^*$ &  $|\epsilon|=$ 0 -- 1 \\
Half-light radius & Fit to CANDELS & 0\farcs05 -- 0\farcs3 \\
Magnitudes & Fit to CANDELS & 21.7 -- 28.1\\
Pixel scale & HST-like & $0\farcs05$/pixel \\
 \hline
\end{tabular}

$^*$ Drawn from parametric fits to individual COSMOS galaxies.
\end{table*}
 
The simulations should aim to resemble the real galaxies as closely as possible to make sure that we are not introducing any artificial bias due to our choice of input parameters. In order to achieve this, we compared our measured distributions of the signal-to-noise ratio, size and magnitude with the ones obtained using data from the Cosmic Assembly Near-IR Deep Extragalactic Legacy Survey (CANDELS) fields (\citealp{Grogin}, \citealp{Koekemoer}) as analysed by \citetalias{Schrabback_2018} (see Fig.\thinspace\ref{fig:candels}). In particular, we employed catalogues from \citetalias{Schrabback_2018}, which are based on ACS F606W stacks that approximately match the depth of our cluster field observations. \citetalias{Schrabback_2018} applied the same selection to these catalogues as to the cluster field observations in terms of galaxy shape parameters, magnitudes, signal-to-noise ratios, and colours. Importantly, their colour selection was tuned to provide a robust cluster member removal, selecting mostly background galaxies at $z\gtrsim1.4$. By matching the measured source properties of these catalogues with our simulations we therefore make sure to adequately resemble the source properties in the cluster field WL data. 

We compare the measured distributions for the galaxies in CANDELS and the simulations in Fig.\thinspace\ref{fig:candels}. The resulting distributions in $S/N_{\mathrm{KSB}}$ and the half-light radius are well matched. The simulation contains a slightly lower fraction of faint ($26\leq \mathrm{MAG\_AUTO} \leq 27$) galaxies with $S/N_{\mathrm{flux}}< 10$. We expect that this slight mismatch is caused by incompleteness: we used the CANDELS magnitude distribution (which itself is incomplete) as an input, but at faint magnitudes our simulation analysis recovered only an incomplete fraction of these, causing the discrepancy. Employing a reweighting scheme we verified that this minor discrepancy affects our results at a negligible ($\leq 0.3\%$) level only, so we can safely ignore it for our analysis. It is important to note that this issue can be reduced by using deeper input catalogues. In fact, we will do this in Sect.\thinspace\ref{sec:faint}, where the full depth CANDELS photometric catalogue will be used as an input, rather than the single-orbit depth shape catalogue from \citetalias{Schrabback_2018}. If we were to use our shear measurement pipeline with images from different telescopes, new sets of simulations matched to the desired setup should be created to check that the results are still applicable, as we do in Appendix \ref{sec:HAWKI} for a setup that approximately resembles the VLT/HAWK-I WL data from \citetalias{Schrabback_2018b}.

We computed the magnitude input distribution as well as the size distribution by measuring the properties of real CANDELS galaxies and then drawing randomly from it. The decision to obtain the size distributions from the KSB+ measured CANDELS data and not from the parametric fits to COSMOS is motivated by the difference in depth between the datasets, where the COSMOS catalogue is limited to F$_{814W}<25.2$.

We included uncorrelated Gaussian noise, where the level is tuned to provide a good match in the measured $S/N_{\mathrm{KSB}}$ distributions between CANDELS and the simulations. Our simple assumptions regarding the noise plus the slight underrepresentation of galaxies with faint measured magnitudes in the simulation (see the top panels of Fig.\thinspace\ref{fig:candels}) may be the reason for the lack of low $S/N_{\mathrm{flux}}$ galaxies seen in the lower right panel of Fig.\thinspace\ref{fig:candels}. 

We drew from independent magnitude and size distributions, which means we did not fully capture the correlation between galaxy parameters. The importance of these correlations was discussed in \citet{Kannawadi}, who put emphasis on simulating galaxies with joint distributions. This is not as important in our analysis as we do not perform tomographic cuts and always average over our full population. Also, \citet{Kannawadi} show that it is especially important to account for such correlations for lower redshift sources, but that the impact becomes small for their highest redshift sources bin, which is the closest to the galaxies we are simulating. We suspect that this redshift dependence is caused by the fact that a broad range of different morphological types contribute to lower redshift source samples, while the highest-redshift sources are largely dominated by star-forming late-type galaxies. 

As mentioned above, the input COSMOS catalogues used here are not as deep as our simulations. Thus, the input galaxies are brighter than the galaxies we aim to simulate. Since we modified the inputs slightly to match the CANDELS catalogues which have approximately the same depth as our real cluster catalogues this should not be a problem.

\subsection{Revised noise bias correction} \label{sec:noise_bias}

\begin{figure}[t]
 \includegraphics[width=\linewidth]{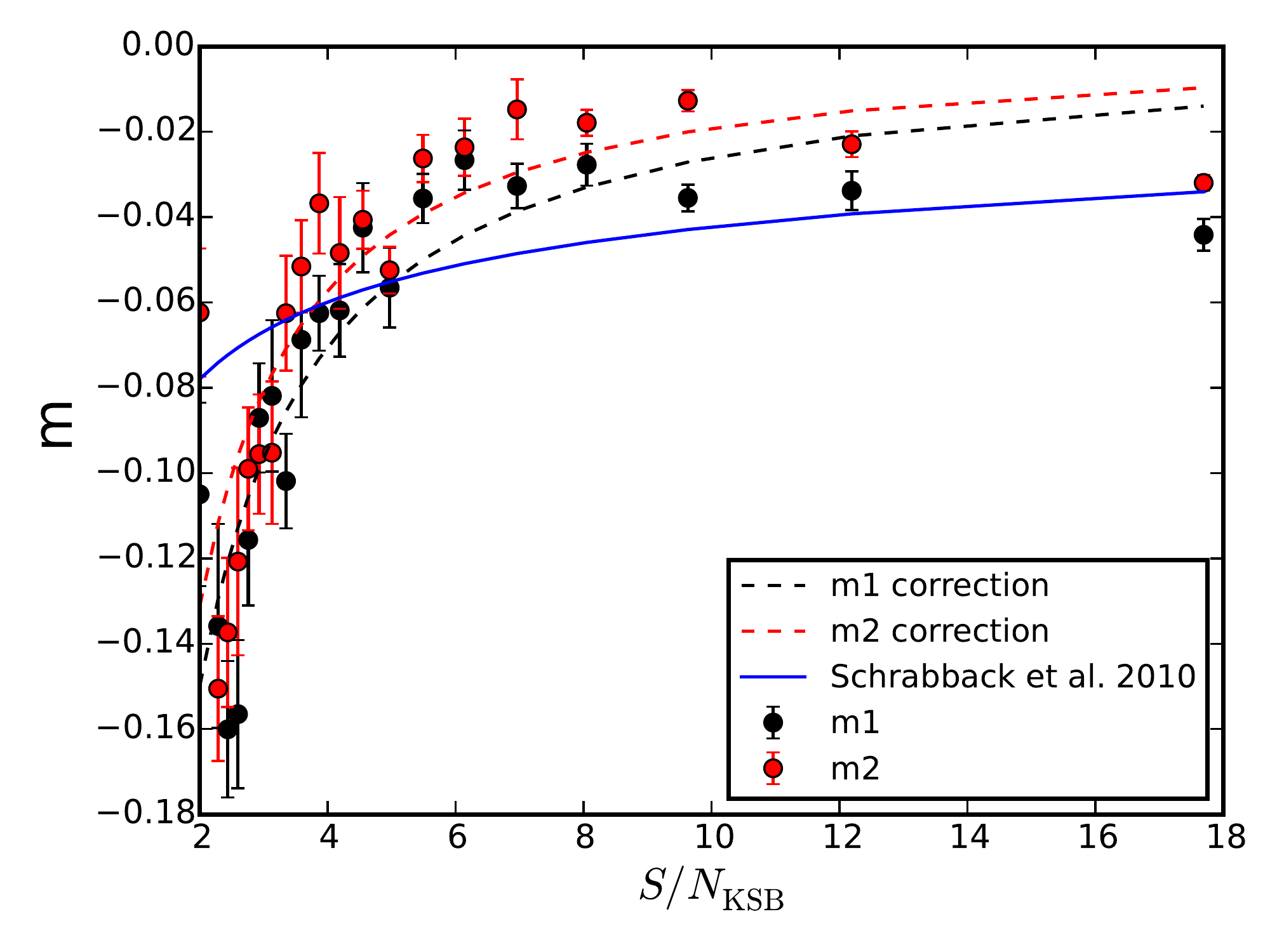}
 \caption{Dependence of the multiplicative bias on $S/N_{\mathrm{KSB}}$ and comparison between the $S/N_{\mathrm{KSB}}$ correction from \citetalias{Schrabback} (shown in blue) and the new correction presented in Eq.\thinspace(\ref{eq:sn_corr}). The red and black dashed lines show the correction for the two shear components. We binned the galaxies according to their $S/N_{\mathrm{KSB}}$ and computed the bias for each bin separately. We include no $S/N_{\mathrm{flux}}$ cuts here but note that the inclusion of a $S/N_{\mathrm{flux}}>10$ cut  changes the results only slightly.}\label{fig:sn_corr}
 \end{figure}

\citetalias{Schrabback} established that the multiplicative bias of our KSB+ implementation shows a strong dependence on \mbox{$S/N_\mathrm{KSB}$}.
This should be compensated for  using a
\mbox{$S/N_\mathrm{KSB}$}-dependent correction in order to weaken the  requirements on how well the simulations have to match the
real data.
However, \citetalias{Schrabback} obtained their correction from the STEP2 simulations of ground-based weak lensing data \citep{Massey}. These differ substantially from the HST cluster data we are interested in, likely affecting the required correction.
Indeed, Fig.\thinspace\ref{fig:sn_corr} shows that our analysis of HST-like simulations with background-selected isolated galaxies yields a steeper dependence on \mbox{$S/N_\mathrm{KSB}$}, which is not well described by the \citetalias{Schrabback} correction. In fact, when using the correction from \citetalias{Schrabback} we measure a multiplicative residual bias of$\sim -0.0221 \pm 0.0042$.
Assuming the same power-law function form as in Eq. (\ref{eq:or_sn_corr}) we computed a revised correction

\begin{eqnarray}
\begin{aligned}
m_{1\text{corr}}=-0.358 \left( S/N_{\mathrm{KSB}}\right) ^{-1.145}\, , \\
m_{2\text{corr}}=-0.357 \left( S/N_{\mathrm{KSB}}\right) ^{-1.298}\, , \label{eq:sn_corr}
\end{aligned}
\end{eqnarray}
which describes the dependence reasonably well and differs slightly for the two components.
We applied this revised noise bias correction in the remaining analysis,
where we always applied a correction on a galaxy by galaxy basis, scaling the KSB+ shear estimates by a factor \mbox{$1/(1+m_{\alpha\mathrm{corr}})$} depending on the \mbox{$S/N_\mathrm{KSB}$} of the individual galaxy.

We note that not yet all relevant effects
(e.g.\thinspace selection bias)
have been included
in this analysis, which is used to derive the correction.
We therefore have to add additional residual bias corrections below. We find, however, that these additional effects do not change the \mbox{$S/N_\mathrm{KSB}$} dependence significantly (see e.g.~Sect.\thinspace\ref{sec:faint}), which is why we
can account for them using constant (\mbox{$S/N_\mathrm{KSB}$}-independent) bias offsets.

In Table \ref{tab:bias} we present the bias estimates obtained from the grid reference simulation, computed for two different shear ranges. We used $|g|<0.4$ in order to study the behaviour for strong shears, but for typical datasets the $|g|<0.2$ regime is more relevant, so we included this estimation, which will then be used later. Nevertheless, we obtained consistent bias results for both limits. 
The residual multiplicative bias is smaller than $1\%$ in both cases after applying the $S/N_{\mathrm{KSB}}$-dependent corrections from Eq.\thinspace(\ref{eq:sn_corr}) and a cut on \mbox{$S/N_{\mathrm{flux}}>10$} (which is the default \mbox{$S/N_{\mathrm{flux}}$} cut from \citetalias{Schrabback_2018}). Errors were estimated by bootstrapping the galaxies in each image. In the following sections all bias estimates by default correspond to the $|g|<0.2$ setup unless explicitly stated differently. The estimates from this grid reference simulation will be used to study throughout the rest of Sect.\thinspace\ref{sec:grid} how the bias shifts as we change the input parameters.

\begin{center}
\begin{table}[ht]
\caption{Residual multiplicative and additive bias for the simulation with galaxies placed on a grid after a \mbox{$S/N_{\mathrm{flux}}>10$} cut and applying the correction from Eq.\thinspace(\ref{eq:sn_corr}).} \label{tab:bias}
\centering
\resizebox{9cm}{!}{
\begin{tabular}{c|c c c c}
Limits &\textbf{$\mathrm{m}_{1,\mathrm{grid}}$} & \textbf{$\mathrm{c}_{1,\mathrm{grid}}$} &\textbf{$\mathrm{m}_{2,\mathrm{grid}}$} & \textbf{$\mathrm{c}_{2,\mathrm{grid}}$}\\
 & \multicolumn{4}{c}{[$\times 10^{3}$]} \\
\hline
\hline
$|g|<0.4$ & $-7.3 \pm 2.0$ & $-1.2 \pm 0.4$  & $-6.7\pm 1.3$  & $-4.3 \pm 0.3$ \\
\hline
$|g|<0.2$ & $-8.1 \pm 3.8$ & $-1.0 \pm 0.4$ & $-5.1\pm 3.3$  & $-3.5 \pm 0.3$ \\
\hline
\end{tabular}}
\end{table}
\end{center}

\subsection{Residual dependence on $S/N_{\text{flux}}$ and magnitude} \label{sec:sn_flux}

\begin{figure}[t]
\includegraphics[width=\linewidth]{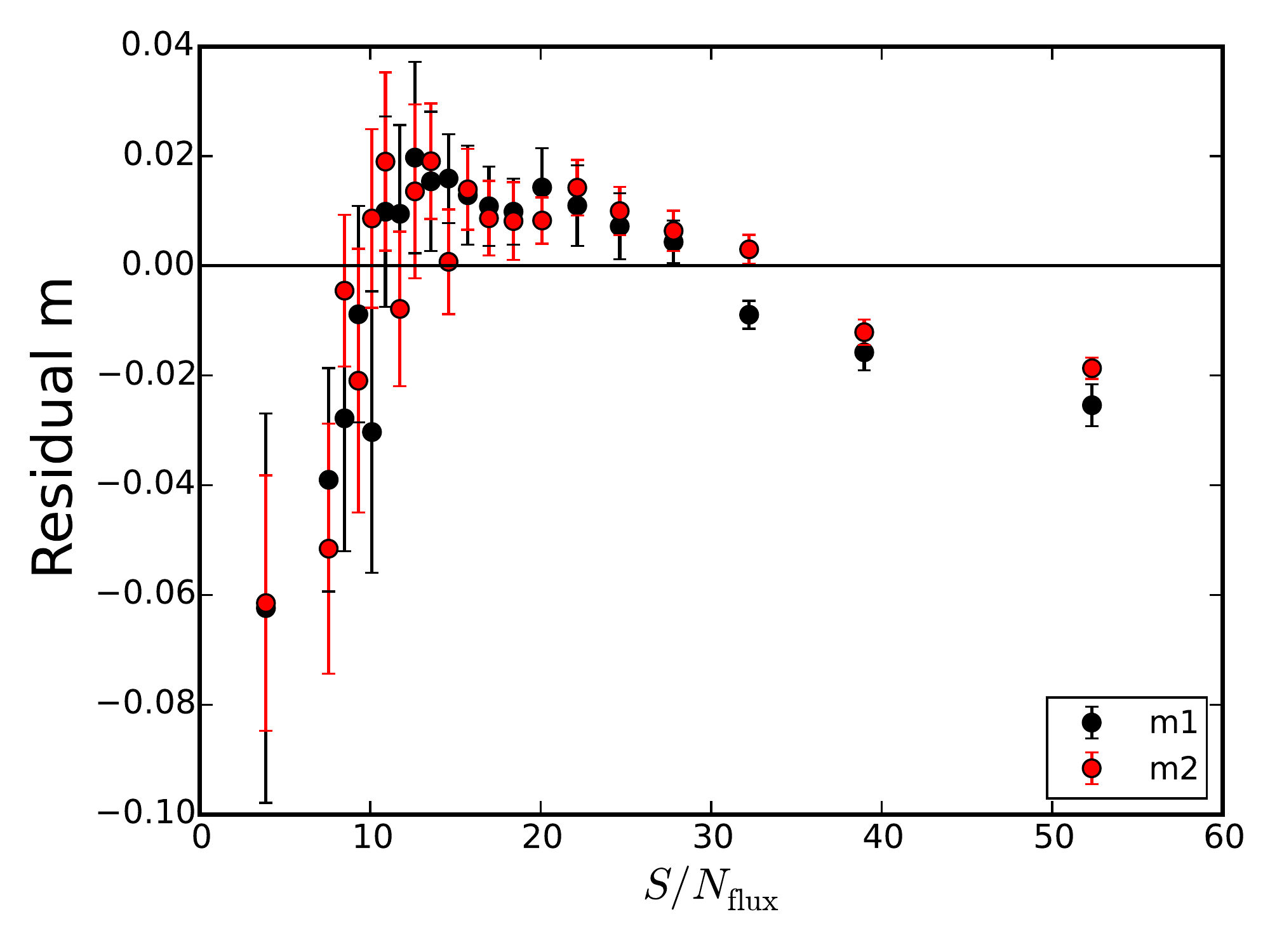}
\caption{Residual dependence of bias on the $S/N_{\mathrm{flux}}$, shown here after the $S/N_{\mathrm{KSB}}$-dependent correction is applied. The residual dependence is caused by a slight under- or over-corrections in different $S/N_{\mathrm{KSB}}$ regimes in Fig.\thinspace\ref{fig:sn_corr}, as well as differences in the definitions of the signal-to-noise ratios. }\label{fig:sn_sex}
\end{figure}

The use of the $S/N_{\mathrm{KSB}}$ correction defined in Eq.\thinspace(\ref{eq:sn_corr}), should compensate for the strongest bias dependence. Here we investigate residual dependences on other parameters. In Fig.\thinspace\ref{fig:sn_sex} we show the dependence of the residual multiplicative bias on $S/N_{\mathrm{flux}}$ by measuring the bias in 20 $S/N_{\mathrm{flux}}$ bins. The residual dependence comes from the fact that the $S/N_{\mathrm{KSB}}$-dependent correction over-corrects at $5\leq S/N_{\mathrm{KSB}}\leq 8$, which corresponds with the positive bias at $10 \leq S/N_{\mathrm{flux}} \leq 20$. Similarly, the bias is slightly under-corrected at both very low and very high $S/N_{\mathrm{KSB}}$, which can also be seen in the $S/N_{\mathrm{flux}}$ dependence.

In \citetalias{Schrabback_2018}, a conservative approach was used, restricting the galaxies to \mbox{$S/N_{\mathrm{flux}}>10$}. We tested the influence of different cuts by selecting the galaxies surviving each cut and performing an independent computation of the residual bias. A cut on  \mbox{$S/N_{\mathrm{flux}}>7$} increases the bias slightly from  $-0.0073 \pm 0.0020$ to $-0.0109 \pm 0.0018$ for $m_1$ and from $-0.0067 \pm 0.0013$ to $-0.0118 \pm 0.0013$ for $m_2$. Since we are still in the $\sim 1 \%$ bias regime this justifies the inclusion of noisier galaxies in the mass determination of the three relaxed clusters in Sect.\thinspace\ref{sec:mass}, tightening the constraints via an increased source density. By lowering our $S/N_{\mathrm{flux}}$ cut from 10 to 7 we increase the number of galaxies in our simulation analysis by $15\%$.

\begin{figure}[t]

  \includegraphics[width=\linewidth]{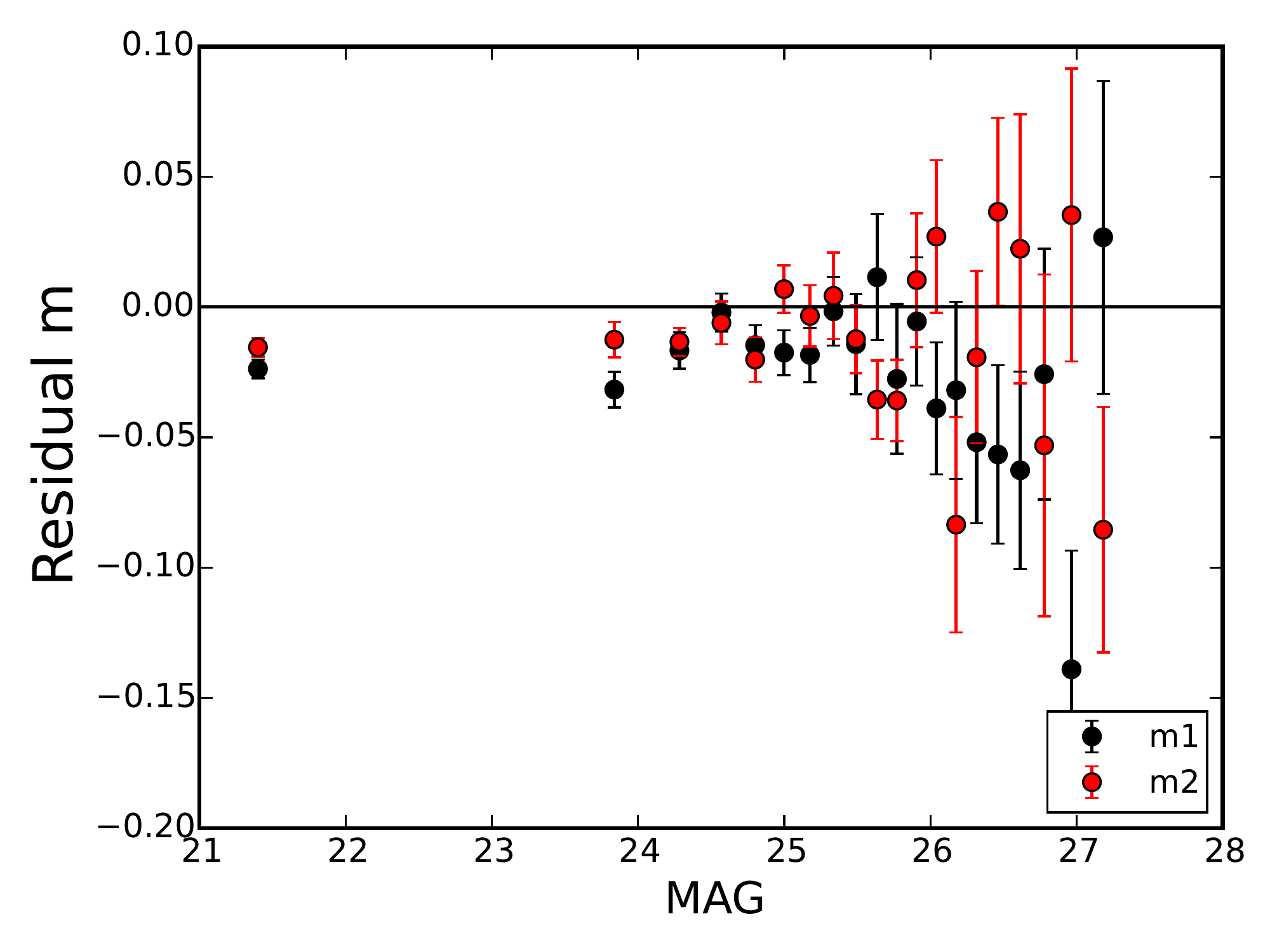}

  \caption{Dependence of the residual multiplicative bias on magnitude for the two ellipticity components. Here we have applied the $S/N_{\mathrm{KSB}}$-dependent correction and the cut \mbox{$S/N_{\mathrm{flux}}>10$}. The bins are created to have the same number of galaxies, but we have fewer galaxies for brighter magnitudes, which explains the different separation of points. } \label{fig:mag_dep}
 \end{figure}

We also investigated the dependence of the bias on the magnitude of the galaxies, finding no significant trend after applying the $S/N_{\mathrm{KSB}}$-dependent correction (see Fig.\thinspace\ref{fig:mag_dep}). This is not surprising since magnitude and signal-to-noise ratios are strongly correlated. We also investigated the residual bias as a function of the measured galaxy flux radius, finding only a weak dependence (see Sect. \ref{sec:realistic}), which can be safely ignored given our accuracy target, but see \citet{Hoekstra_2015} for a correction scheme that also depends on galaxy size.

\subsection{Light profile} \label{sec:light}
One of the critical parameters is the light distribution of the galaxies. Since we are measuring the changes in ellipticities caused by the lensing effect, the intrinsic shapes significantly affect our results. When using a S\'ersic profile to describe the light distribution (see Eq.\thinspace\ref{eq:sersic}), larger values of $n$ describe a more centrally-concentrated profile. These differences in concentration can affect biases in the moment measurement and the PSF correction even if galaxies with the same ellipticity are simulated.

\begin{figure}[t]

 \includegraphics[width=\linewidth]{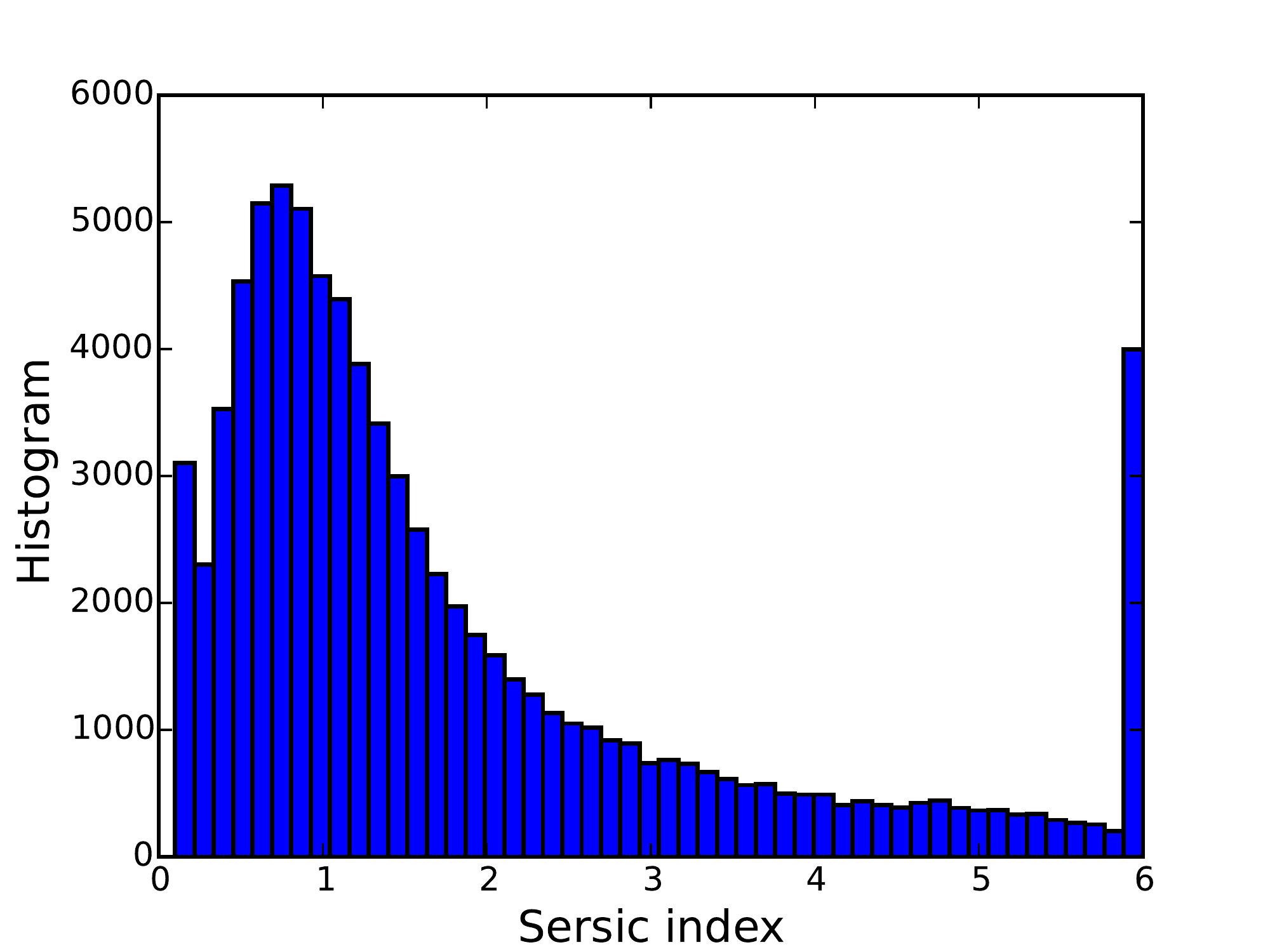}
 \caption{Distribution of S\'ersic indices in the parametric fit to real colour-selected COSMOS galaxies. The peak in the last bin is due to the limits in the index allowed by \texttt{GALSIM}. Everything larger than the \texttt{GALSIM} maximum S\'ersic index (6.2) is added to that bin.} \label{fig:sersic_index}
  
 \end{figure}

In order to investigate the dependencies on the light profile
we generated four sets of simulations with different S\'ersic index distributions. These included a set of pure exponentials (\mbox{$n=1$}), a set of pure De Vaucouleurs  (\mbox{$n=4$}), a set using a flat distribution (uniform S\'ersic index distribution between 0.3 and 6), and the more realistic setup employed in our reference simulation. The latter uses a S\'ersic index distribution (see Fig.\thinspace \ref{fig:sersic_index}) based on S\'ersic profile fits to COSMOS galaxies as provided by \texttt{GALSIM} (\citealp{Mandelbaum_2012}; \citealp{Rowe}).
Importantly, we did not use their full sample of \mbox{$I_{814}<25.2$} galaxies, but applied
a \mbox{$V_{606}-I_{814}<0.4$} colour selection to approximately match the selections applied in \citetalias{Schrabback_2018} and \citetalias{Schrabback_2019}\footnote{For most of their clusters and source galaxies these studies employ a \mbox{$V_{606}-I_{814}<0.3$} colour selection. The minor difference in the colour selection has a negligible impact on the S\'ersic index distribution.}.
The resulting distribution is skewed towards lower S\'ersic indices, with a median value of 1.24.

\begin{figure}[t]

 \includegraphics[width=\linewidth]{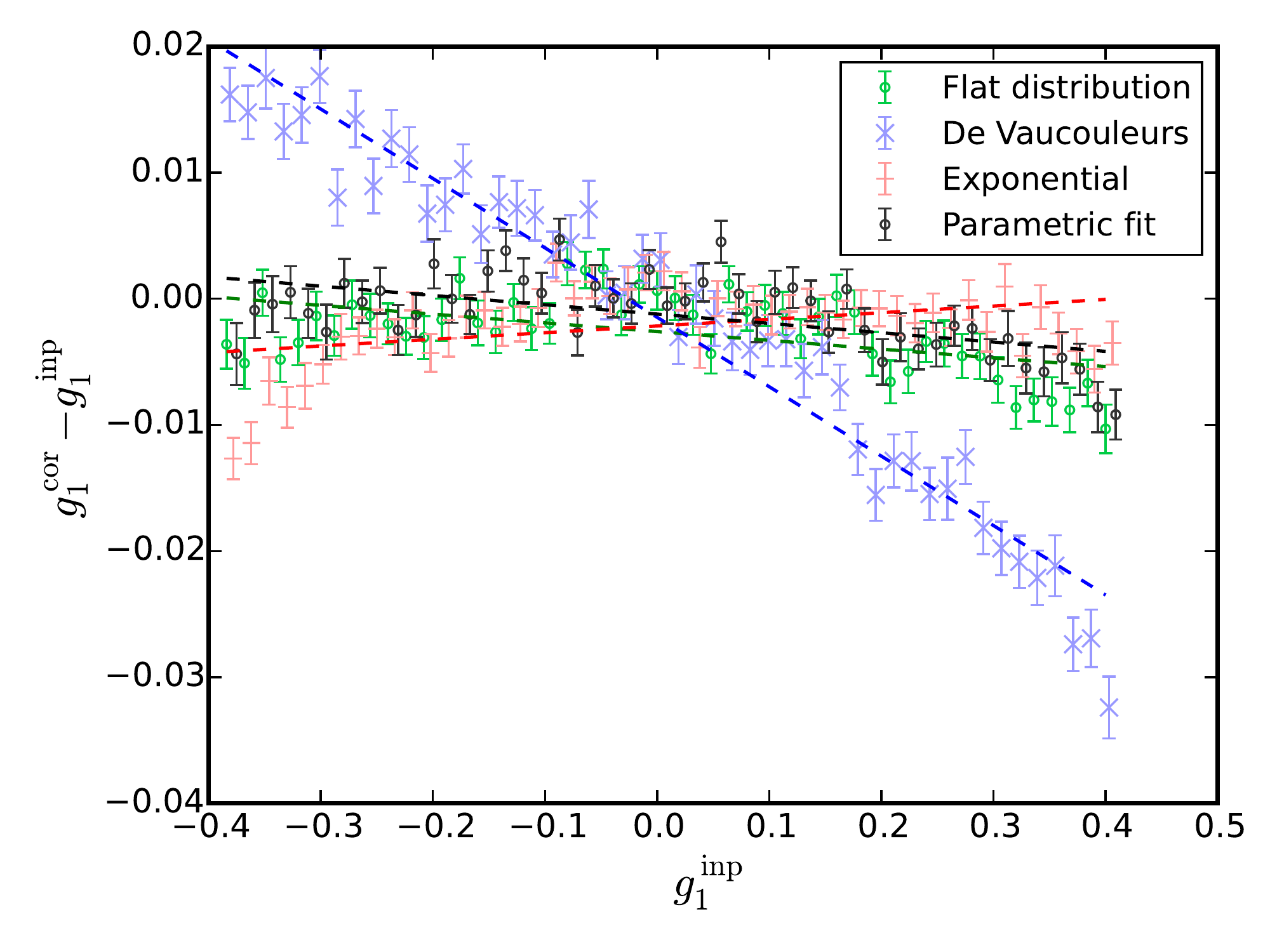}
 \caption{Residual bias estimation for simulations with different input light profile distributions. The different symbols correspond to the different light profiles used to create the mock galaxies. We show the cases of a flat S\'ersic index distribution in green, a purely De Vaucouleurs profile in blue, a purely exponential in red, and the more realistic case of the parametric fit to the COSMOS galaxies in grey. For the result shown here the $S/N_{\mathrm{KSB}}$-dependent correction from Eq.\thinspace(\ref{eq:sn_corr}) and the \mbox{$S/N_{\mathrm{flux}}>10$} cut have been applied. The causes of the deviation of some points at strong shear are discussed in Sect.\thinspace\ref{sec:PSF}. The indicated fits have been computed from the $|g|<0.2$ range, but we present the estimates for the full $|g|<0.4$ range.} \label{fig:light}
  
 \end{figure}

For all analyses we applied the noise bias correction calibrated on the reference simulation.
Fig.\thinspace\ref{fig:light} shows the resulting deviations between the recovered shear and the input shear as a function of the
input shear. The lines indicate the  linear fits according to Eq.\thinspace(\ref{eq:bias}), where the offsets and slopes correspond to the additive and multiplicative biases.
The multiplicative bias differences are relatively small except for the  simulation containing only De Vaucouleurs profiles, which exhibits a substantial residual multiplicative bias of \mbox{$\sim -5\%$}.
From this we conclude that the input galaxy light profiles can play a relevant role, but that minor differences do not have a major impact.

We note that the COSMOS S\'ersic index distribution was derived from a slightly brighter colour-selected galaxy sample. Accordingly, it might not exactly match the distribution of our fainter colour-selected source sample. Nevertheless we expect that the differences are small, with both populations being dominated by late-type galaxies.
We used the bias difference \mbox{$|\Delta m| \simeq 0.5\%$} between the fairly realistic simulation employing the parametric fits and the fairly unrealistic simulation using a flat distribution as a conservative estimate for the systematic uncertainty associated with the galaxy light profile assumptions (see Sect.\thinspace \ref{sec:final}).

When creating the mock galaxies with different light profiles we obtained slightly different distributions of the parameters shown in Fig.\thinspace\ref{fig:candels}, especially for the size distribution. Performing a reweighting of the size distribution to match it to our CANDELS reference did not change the bias significantly, so the differences shown in Fig.\thinspace\ref{fig:light} are not caused by these minor differences.
 
 \begin{figure*}[t!]
  
\centering
 \begin{minipage}{0.49\linewidth}
 
  \includegraphics[width=\linewidth]{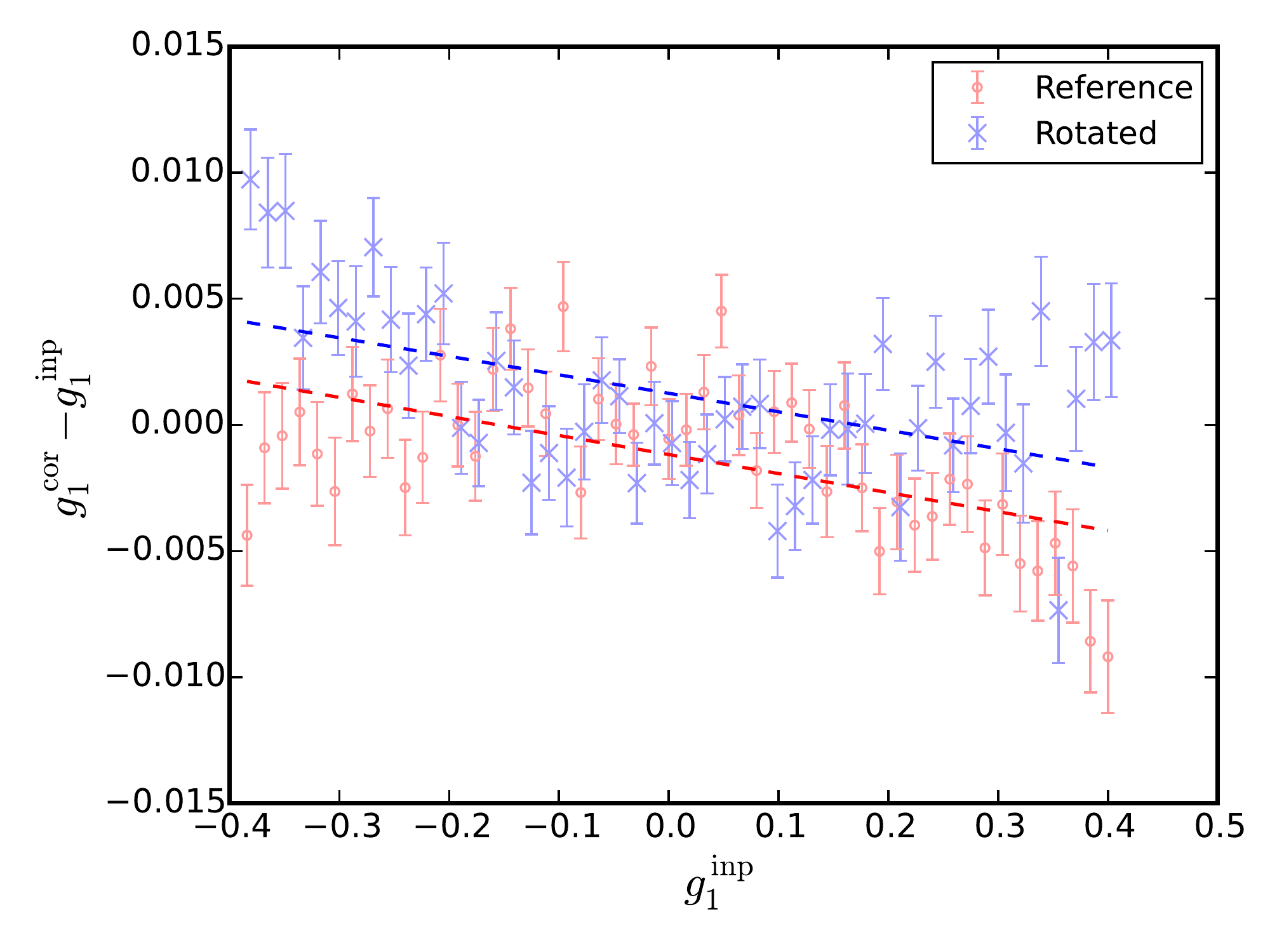}
  \end{minipage}
  \begin{minipage}{0.49\linewidth}
  
    \includegraphics[width=\linewidth]{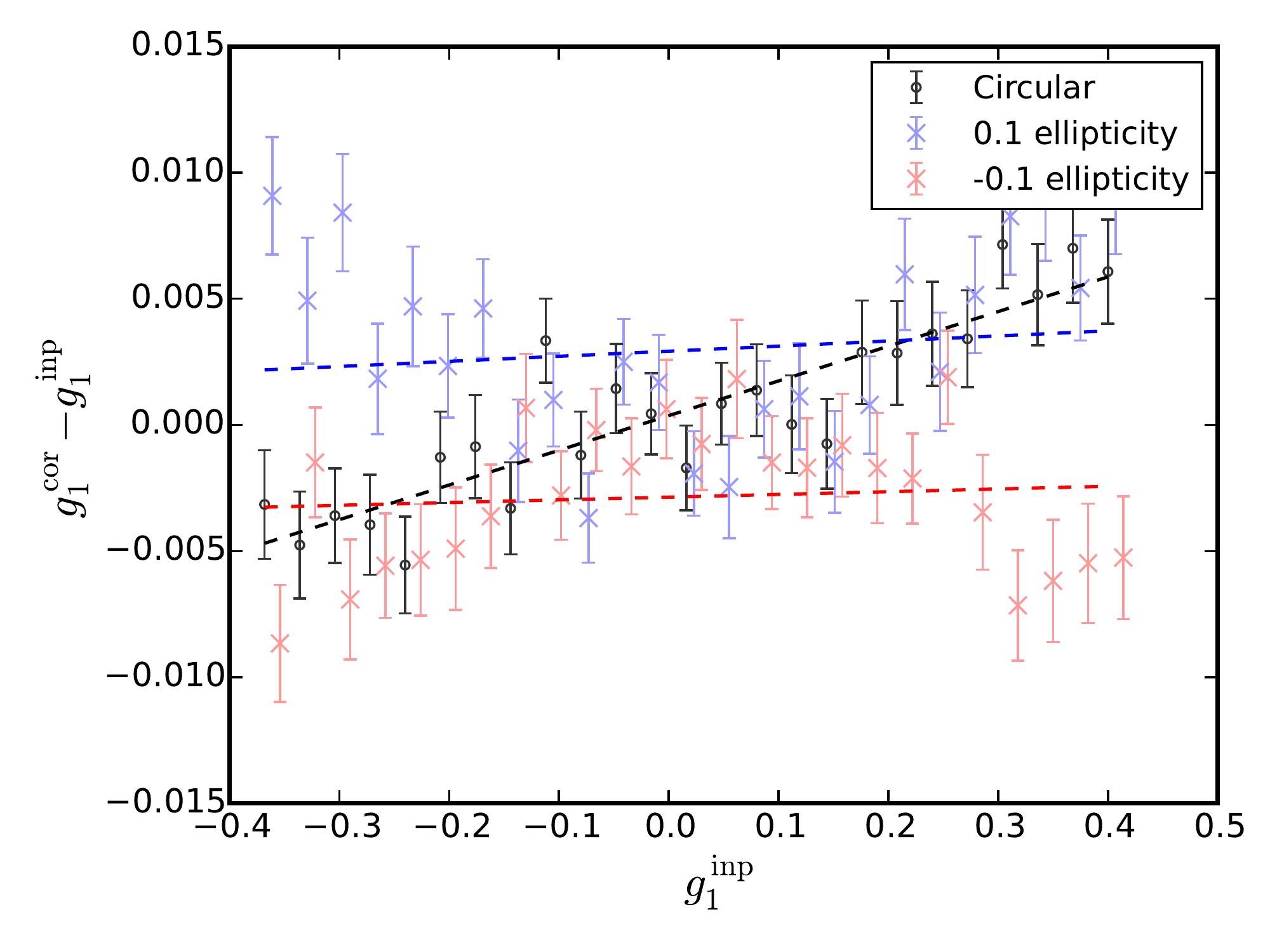}
      \end{minipage}

  \caption{\textit{Left}: Comparison of the residual bias from the reference simulations (red points) and a simulation in which the PSF has been rotated by 90 degrees (blue crosses). \textit{Right}: Comparison of the residual bias obtained for a circular Moffat PSF in grey, a modified Moffat with an $e_1=0.1$ ellipticity in blue, and a Moffat PSF with an $e_1=-0.1$ ellipticity in red. The indicated fits have been computed from the full $|g|<0.4$ range.} \label{fig:rotated}
 \end{figure*} 
 
\subsection{Intrinsic ellipticity dispersion} \label{sec:ellip}

All galaxies have some intrinsic ellipticity. For the reference simulations we used parametric fits to COSMOS galaxies to determine the galaxy input ellipticities, in order to have more realistic ellipticities. 
In this section, we test if the input intrinsic ellipticity distribution of the galaxies in the simulated images can play a significant role for the performance of the KSB+ algorithm (see e.g. \citealp{Viola}). To test this, we set up four sets of simulations with the same parameters, except for the intrinsic ellipticity. Each one had an input Gaussian ellipticity distribution for each component $\epsilon_{\alpha}$, with different RMS modulus $\sigma (|\epsilon|)$ values ranging from 0.2 to 0.35 (computed from both ellipticity components together). The biases between different sets show only minor ($<0.5 \%$) differences. These results are also consistent with the ones obtained for the reference simulation (which has a $\sigma (|\epsilon|)=0.28$), so we conclude this choice does not play a big role in the bias determination.

\subsection{PSF shape and deviations at stronger shears} \label{sec:PSF}

The KSB+ formalism makes the simplifying assumption that the PSF can be described as an isotropic function convolved with a small anisotropic kernel, which is not strictly met for many realistic PSFs. In fact, the Tiny Tim PSF we used has an ellipticity with $|e|=0.072$. Here we therefore investigate how sensitively our bias estimates depend on the details of the PSF model that is employed in the simulation. For our simulations we used a typical PSF model created with Tiny Tim \citep{Krist}. We selected the PSF model parameters according to the best fit to real HST/ACS starfield images obtained by \citet{Gillis}. In particular, we employed an ACS-like setup, with a subsampling by a factor 3 to avoid pixelation issues when convolving with the galaxy light profiles. The subsampled pixel scale for the PSF was $0\farcs0165$ /pixel.

The ACS PSF changes due to focus variations caused by the thermal fluctuations that happen in orbit (\citealp{Heymans_2005}; \citealp{Rhodes}; \citealp{Schrabback_2007}). We tested the influence of this in our analysis, finding that the bias varies less than $1\%$ within the typically expected focus ranges and spatial PSF variations across the field of view. For this reason, we did not model a position-dependent PSF and rather just took one particular model as input. As this default model we chose a typical average focus value of $-1\,  \mu m$, as well as a position of $x=1000$, $y=1000$ in chip 1, which leads to a PSF with ellipticity \mbox{$e_1=$0.018}, \mbox{$e_2=$0.063} (when measured with a weight function scale \mbox{$r_\mathrm{g}=2.0$} pixels). 

Another effect that influences the results is charge diffusion \citep{Krist_2003}. Electrons near the edges of pixels have a chance to travel to neighbouring pixels, effectively creating a blurring effect which should be included in our models. This is implemented by using a Gaussian kernel. While testing this, we found that excluding this effect for our realistic PSF increased the residual bias to $\sim - 4 \%$.
The more realistic charge diffusion-affected PSF was implemented into our simulations and compared to a Moffat profile to assess how incorrect PSF modelling affects our shape measurements. We scaled the Moffat PSF to a half-light radius of $0\farcs07$, which corresponds to the half-light radius of the ACS PSF used. The Moffat PSF changes the bias by $\sim + 0.018$. The employed ACS PSF model is fairly realistic (see \citealp{Gillis}), while the Moffat PSF clearly only constitutes a crude approximation. We therefore consider half of this bias difference as an estimate for the systematic uncertainty associated with the realism of the PSF shape in the simulation (see Sect.\thinspace\ref{sec:final}). 

For a number of consistency checks we also generated a set of simulations in which the PSF had
been rotated by 90 degrees compared to the default reference simulation. For both setups the shear recovery is compared in the left panel of Fig.\thinspace\ref{fig:rotated}. After the $S/N_{\mathrm{KSB}}$-dependent bias correction, we obtained similar parameters for the multiplicative bias ($m_1=-0.0076 \pm 0.0023$ and  $m_2= -0.0069 \pm 0.013$) for the simulations with the rotated PSF, and a change in the sign for the additive bias ($c_1=0.0012 \pm 0.0003$ and  $c_2= 0.0045 \pm 0.0003$), compared to the bias in Table \ref{tab:bias}. This comparison allows us to investigate the cause for the deviations from a linear trend that can be seen at large input shears in Figs. \ref{fig:light} and \ref{fig:rotated}. If this effect was caused by a non-linear response to shear, we would expect the deviation to be point-symmetric (with respect to the origin) for negative and positive shear values, due to the sign of the shear simply being the orientation (parallel or perpendicular) with respect to the component axis. For example, if the difference between the recovered and input shear is positive for positive shear values, then we should expect a negative difference for negative shear. However, in Fig.\thinspace\ref{fig:rotated} the difference for the reference simulation is negative both for positive and negative shear values. This suggests that it is not a quadratic response to shear but rather a dependency of the additive bias on the input shear. This hypothesis is supported by the fact that the sign of the difference switches for the simulations with rotated PSF. To further test the origin of this effect, we set up three sets of simulations using a Moffat PSF with ellipticities 0.1, 0 and $-0.1$. As visible in the right panel of Fig.\thinspace\ref{fig:rotated}, the deviation from the linear behaviour depends on the ellipticity of the PSF, being non-existent for the circular PSF and having a different sign for $-0.1$ and 0.1 PSF ellipticities. We conclude from these observations that our KSB+ implementation does not suffer from a non-linear shear response, but that the additive bias appears to be shear dependent in the non-weak shear regime.

\section{Cluster blending} \label{sec:clustering}

 \begin{figure*}[t!]
   \centering
   \includegraphics[width=\linewidth]{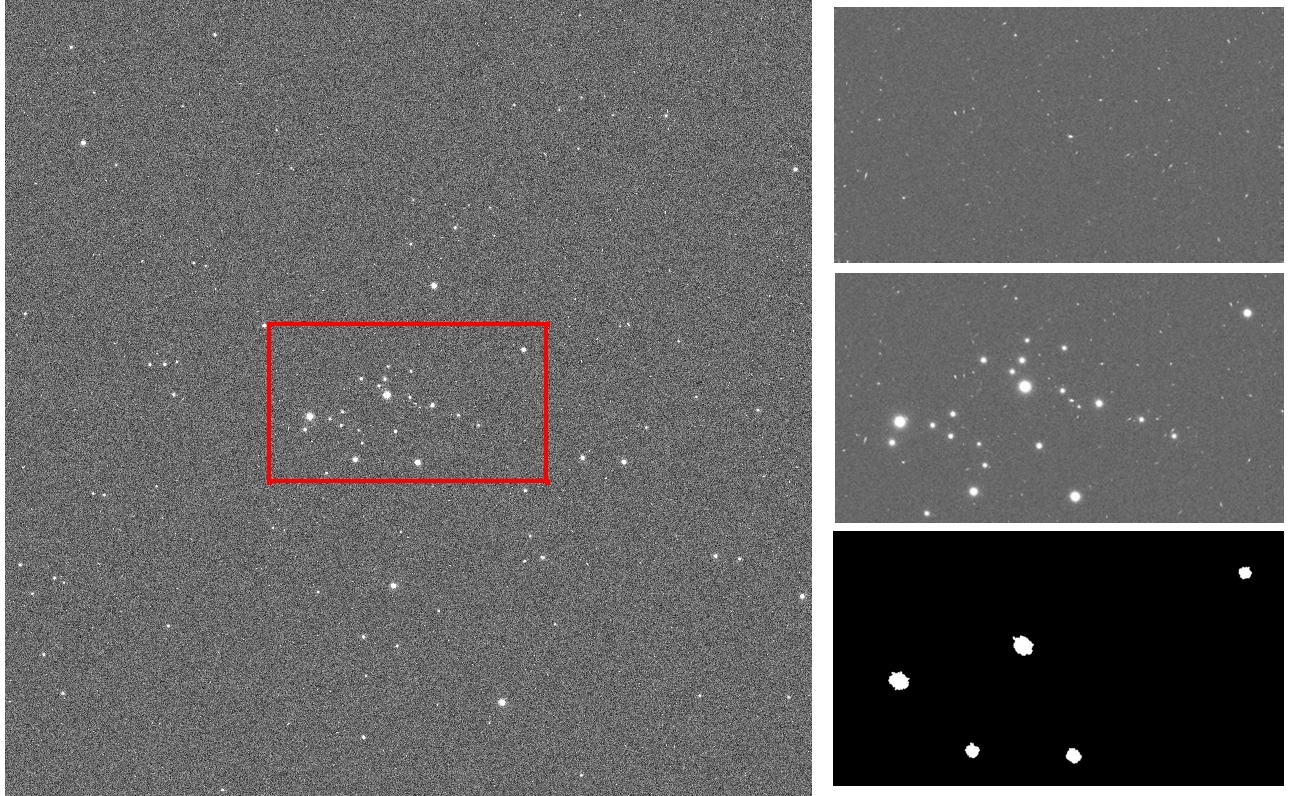}
   \caption{Example image of a simulated cluster at $z=0.72$. A cut-out of the full image, shown in red, can be seen in the right for the simulations with background galaxies only (top), with added cluster members (middle), and showing the mask used to remove bright objects (bottom). The full image and the cut outs span $300 '' \times 300''$ and $100'' \times 55''$, respectively. 
   A similar image for a simulated low-redshift cluster is provided in Fig.\thinspace\ref{fig:028_cluster} in the Appendix.} \label{fig:examples}
 \end{figure*}
 
 \begin{figure}[t!]
   \centering
  \includegraphics[width=\linewidth]{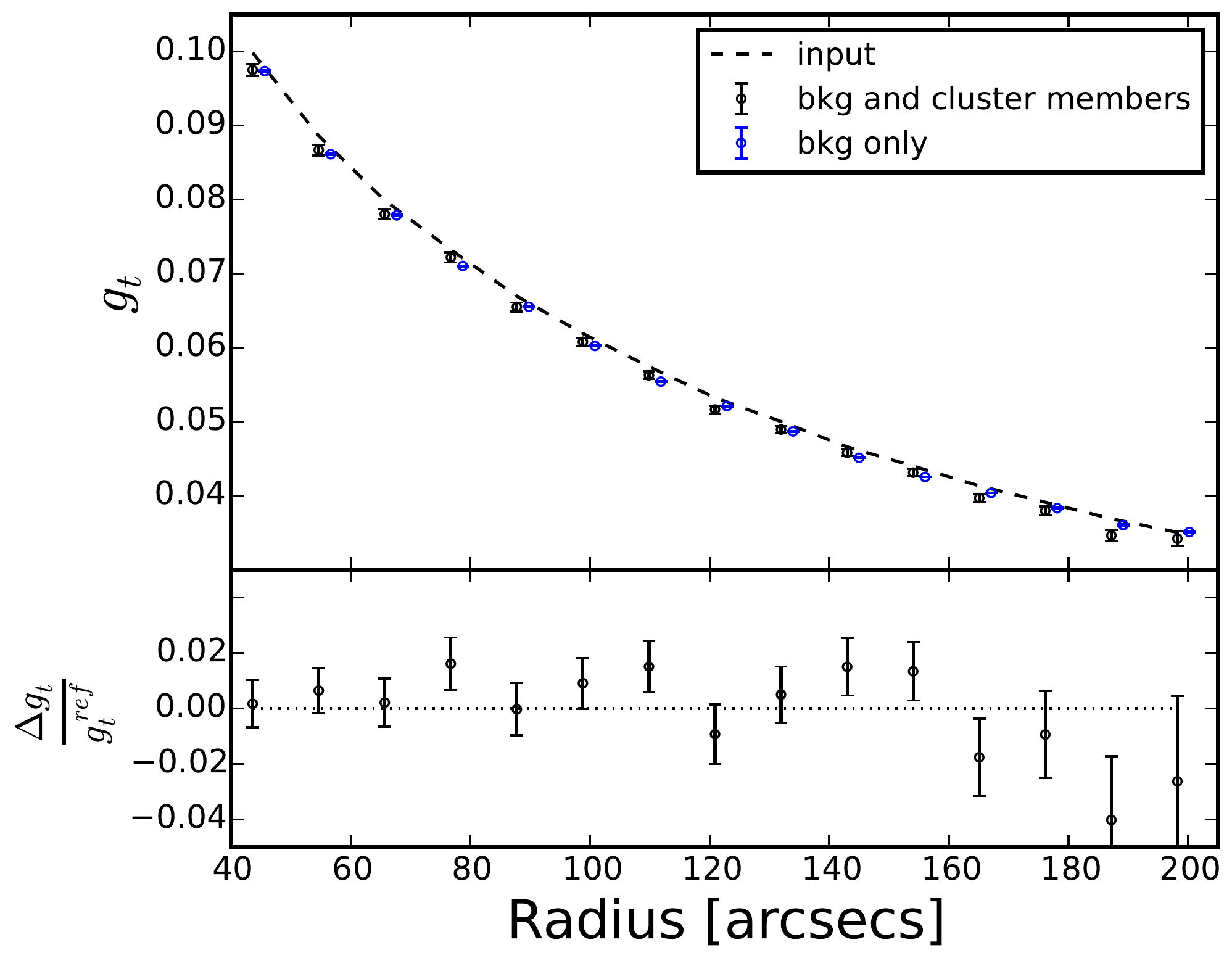}

  \caption{Averaged tangential shear profile of multiple simulations of background galaxies sheared following an NFW profile with (in black) and without (in blue) the presence of cluster member galaxies. Here we average the profiles of all simulated clusters at  $z>0.7$. The dashed line represents the average input profile. The bottom panel shows the relative difference between the tangential shears of the simulations with and without cluster member galaxies as a function of radius. } \label{fig:clustering_one}

  \end{figure}

  \begin{figure}[t!]
 
  \includegraphics[width=\linewidth]{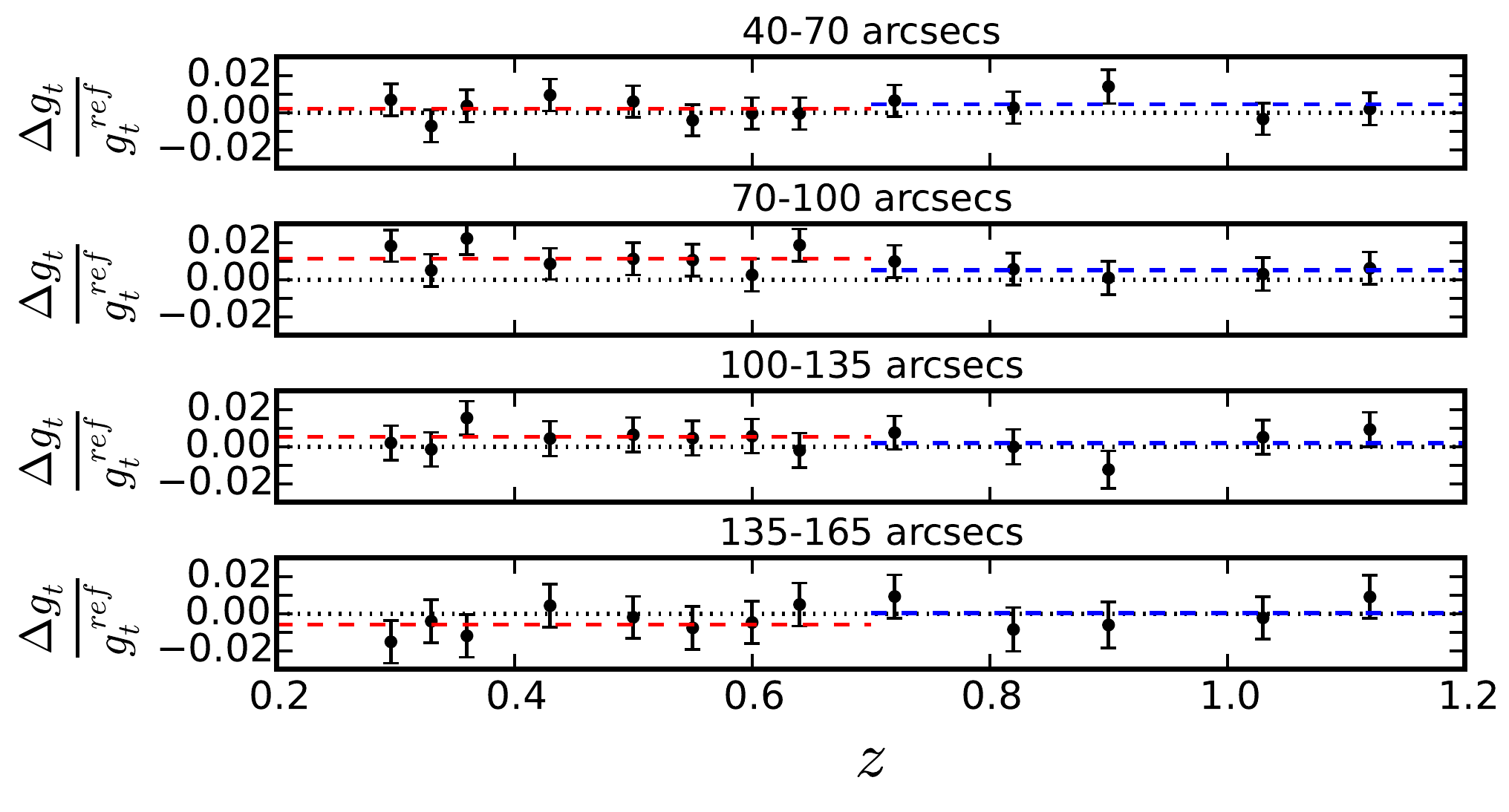}

   \caption{Relative difference of the measured tangential shears of the simulations with and without cluster member galaxies as a function of cluster redshift. The four panels show the values in different distance bins. The dotted black line represents the zero line for reference and the dashed line shows the mean of the points for clusters at $z<0.7$ (red) and $z>0.7$ (blue). } \label{fig:clustering_all}
 \end{figure}
 
In order to increase the realism of our simulations, we took a first step by studying the impact that the presence of bright cluster members has on the bias estimation. For example, the wings of their extended light profiles might contaminate the light distribution of nearby sources, potentially leading to biased shape estimates. This is a cluster-specific issue that is not generally studied in this kind of work. Different from bright foreground galaxies, which are randomly positioned, cluster galaxies have a higher number density in the cluster cores where shears are stronger.

For this analysis, we created two sets of simulations, one containing only lensed background galaxies, and one that also contains cluster member galaxies and other foreground galaxies. For the background galaxies we used the same galaxy properties as in Sect.\thinspace\ref{sec:grid}, but we placed them randomly in the image rather than on a grid. We note that this is not yet a fully realistic scenario, since the simulations are still missing the clustering of source galaxies, as well as very faint galaxies below the detection threshold. These effects are studied in Sect.\thinspace\ref{sec:neighbours}, but are not needed for the current step, where we are investigating the impact of bright cluster and foreground galaxies.
The $S/N_{\mathrm{KSB}}$-dependent correction as well as the \mbox{$S/N_{\mathrm{flux}}>10$} cut were implemented in the analysis, although we alternatively repeated the same analysis with \mbox{$S/N_{\mathrm{flux}}>7$}, which changed the results only marginally (the final bias estimation is presented later in Table \ref{tab:final}).
Instead of images with constant shear we used a more realistic approach where background galaxies were sheared according to their relative position to the centre of the cluster assuming a spherically-symmetric NFW profile \citep{Navarro} with the overdensity mass \mbox{$M_\mathrm{200c}=5\times 10^{14}\mathrm{M}_\odot$}\footnote{We note that \mbox{$M_{\Delta\mathrm{c}}$} corresponds to the mass contained within a sphere of radius \mbox{$r_{\Delta\mathrm{c}}$}, inside of which the mean density equals to $\Delta$ times the critical density of the Universe at the redshift of the lens.}, the concentration $c_{\mathrm{200c}}=4$, the redshift of the lens $z_{\mathrm{lens}}=0.3$, and of the source $z_{\mathrm{source}}=0.6$\footnote{This configuration corresponds to typical measurements of low-redshift clusters using ground-based weak lensing data, but the resulting level of shears is in fact comparable to typical HST measurements for high-redshift clusters (compare Fig.\thinspace\ref{fi:bea:shearprofiles}).}. We still created an identical image with a 90 degree rotation of the intrinsic ellipticity of each galaxy before the shear was applied, which is matched in the analysis to reduce shape noise, and therefore the number of simulations required to reach the desired precision on the bias estimate. For the properties of the cluster member galaxies we used catalogues from the MAGELLAN/PISCO \citep{Stalder} follow-up of SPT clusters of various redshifts ($z=0.3 - 1.1$, see \citealp{Bleem_2019}), using cluster redshifts from \cite{Bocquet}. The cluster galaxies have $r$-band magnitudes from $\sim 18$ to $\sim 24$, which are brighter than our background galaxies, and they have a slightly lower input ellipticity dispersion $\sigma(|\epsilon|)=0.2$. 
We created masks around very bright and extended objects, mostly cluster members, following the analysis of real clusters.\footnote{To create this mask, we perform a \texttt{SExtractor} object detection, requiring 2000 pixels with a flux that is higher than the background by 0.5$\sigma$. We then extend the mask in all directions by 3 pixels. This will exclude areas around the brightest objects from our shear estimates.}
A simulated image for one of the clusters (at $z=0.72$) and an example cut-out of the background-only, the background+cluster members image, and the corresponding mask are shown in Fig.\thinspace\ref{fig:examples} (see Fig.\thinspace\ref{fig:028_cluster} for a simulated cluster at $z=0.28$).
We created multiple images with different realisations of background galaxies and different cluster catalogues in order to stack the profiles and obtain a more significant result, independent of the particulars of each cluster. 

Fig.\thinspace\ref{fig:clustering_one} shows the mean input and recovered shear profiles for the two simulation setups (with and without bright galaxies) averaged over all clusters with \mbox{$z>0.7$} in the top panel.
The bottom panel shows the relative difference between the two recovered profiles, which corresponds to the change in the multiplicative bias caused by the presence of the bright foreground and cluster galaxies.
The same relative difference is shown as a function of cluster redshift in Fig.\thinspace\ref{fig:clustering_all}, split into four distance intervals.

\begin{table}[t!]
 \caption{Detailed estimate of the bias due to the presence of bright galaxies for different cluster-centric distance bins and two redshift bins. These estimates are obtained using galaxies with \mbox{$S/N_{\mathrm{flux}}>10$}.} \label{tab:clustering}
\centering
\begin{tabular}{c c c }
\textbf{Distance} & \textbf{Cluster redshift} & \textbf{Bias estimate}\\
\hline
40$''$-70$''$&$z<0.7$ & $0.0021 \pm 0.0033$\\
40$''$-70$''$&$z>0.7$ & $0.0044 \pm 0.0039$ \\
70$''$-100$''$&$z<0.7$ & $0.0113 \pm 0.0033$ \\
70$''$-100$''$&$z>0.7$ & $0.0052 \pm 0.0039$ \\
100$''$-135$''$&$z<0.7$ & $0.0054 \pm 0.0035$ \\
100$''$-135$''$&$z>0.7$ & $0.0020 \pm 0.0042$ \\
135$''$-165$''$&$z<0.7$ & $-0.0058 \pm 0.0043$ \\
135$''$-165$''$&$z>0.7$ & $0.0003 \pm 0.0053$ \\
 \hline
\end{tabular}
\end{table}

We generally find that adding the bright foreground and cluster members only has a minor impact on the shear recovery (see Table \ref{tab:clustering}). The biggest impact is detected for lower redshift clusters (\mbox{$z<0.7$}) at scales $70-100$ arcsec, amounting to a \mbox{$ 1.13\% \pm 0.33\%$} positive multiplicative bias. We expect that the impact decreases for higher redshift clusters given the stronger cosmological dimming of their cluster members. Indeed, computed over one broad bin between 70--165 arcsecs we find a very minor bias of \mbox{$0.48\% \pm 0.38\%$} for the simulated \mbox{$z>0.7$} clusters, which approximately corresponds to the scales and redshifts used in \citetalias{Schrabback_2018}.
When applying a \mbox{$S/N_{\mathrm{flux}}>7$} selection, this bias shifts to $0.25 \% \pm 0.40 \%$.

Given the redshift dependence of the bias caused by cluster galaxies, we decided to treat it separately, and not include the cluster galaxies in the simulations that are used in Sect.\thinspace\ref{sec:neighbours} to investigate the impact of nearby fainter galaxies and selection effects. We verified that the presence of bright galaxies does not lead to a significant shift in the estimates of selection bias for our method.

\section{Selection bias and the influence of blends and neighbours} \label{sec:neighbours}
\citet{Hoekstra_2017} and \citet{Martinet} demonstrate that faint sources below the selection threshold affect shape measurements. Using a Euclid-like setup selecting galaxies with $i<24.5$ and accounting for realistic galaxy and clustering properties calibrated using Hubble Ultra Deep Field data (HUDF, \citealp{Beckwith}), \citet{Martinet} show that these faint sources cause an additional shape measurement multiplicative bias for our KSB+ implementation of \mbox{$\Delta m = -0.0149 \pm 0.0002 $}. In Sect.\thinspace\ref{sec:grid} we placed the galaxies on a grid, which avoids any contamination by neighbours and any selection bias effect since we require matched pairs. In Sect.\thinspace\ref{sec:clustering} we improved the realism of our simulation by adding bright galaxies, but the background galaxies were placed randomly and without a realistic clustering. In real images, we have a larger number of partially blended galaxies and nearby galaxies due to the fact that faint galaxies are clustered together, which was not included in the previous setups. As an example, for deep optical data with ground-based resolution, \cite{Mandelbaum_2018} finds that the impact of nearby galaxies may affect the shape calibration at the $\sim 10\%$ level. Similarly, for the Dark Energy Survey \citet{Samuroff} finds that neighbours can affect multiplicative biases by $3-9\%$. In our pipeline, to be conservative, we used masks to remove bright and extended objects and additionally apply a neighbour rejection, which reduces the impact of neighbours. If two galaxies were detected in the catalogue with a separation $<0\farcs75$ we only kept the brighter one. We expect that this, together with the fact that we were simulating high resolution data, reduces the impact of neighbours compared to previous works, but we tested it here nevertheless.

In this section we created a more realistic scenario with blended galaxies and contamination from neighbours as well as clustering of galaxies. We placed the mock galaxies following the positions and magnitudes of the galaxies in the \citet{Skelton} 3D-HST/CANDELS catalogues. This was done to account for a realistic clustering of the galaxies. As mentioned before, we did not include here the effect of bright galaxies from Sect.\thinspace\ref{sec:clustering} given the dependence on cluster redshift. Instead, we compute the resulting net bias for the correction of the real data in Sect.\thinspace\ref{sec:final}, adding all the different contributions.

The 3D-HST/CANDELS catalogues used here are NIR detected, but the F606W number counts indicate that they do not suffer from significant incompleteness until \mbox{$V_{606} \sim 27.5$}, which is 1 magnitude fainter than the \mbox{$V_{606} =26.5$} limit of our source sample. We therefore expect that they allow us to capture most of the impact of faint galaxies, but we revisit this issue in Sect.\thinspace\ref{sec:faint}, using models that are based on much deeper HUDF data. 

We used the same galaxy property setup as in Sect.\thinspace\ref{sec:grid}. This means we obtained the S\'ersic indices and ellipticity from parametric fits to COSMOS data, which provide input parameters for galaxies in the F814W band. We employed this approach due to the lack of a comprehensive catalogue of structural parameters measured in the same F606W band we are simulating here, expecting that the impact of the minor differences between these two bands is small (following the results from Sect.\thinspace\ref{sec:light}). Other catalogues which have measured light profiles of galaxies, such as the one presented by \citet{van_der_Wel_2012}, provide structural parameters for CANDELS in the F160W and F125W bands, which are much more different from F606W. This is the reason why we continued using the parametric fits available in \texttt{GALSIM}.
 \begin{figure*}[t!]
  
\centering
 \begin{minipage}{0.49\linewidth}
 
  \includegraphics[width=\linewidth]{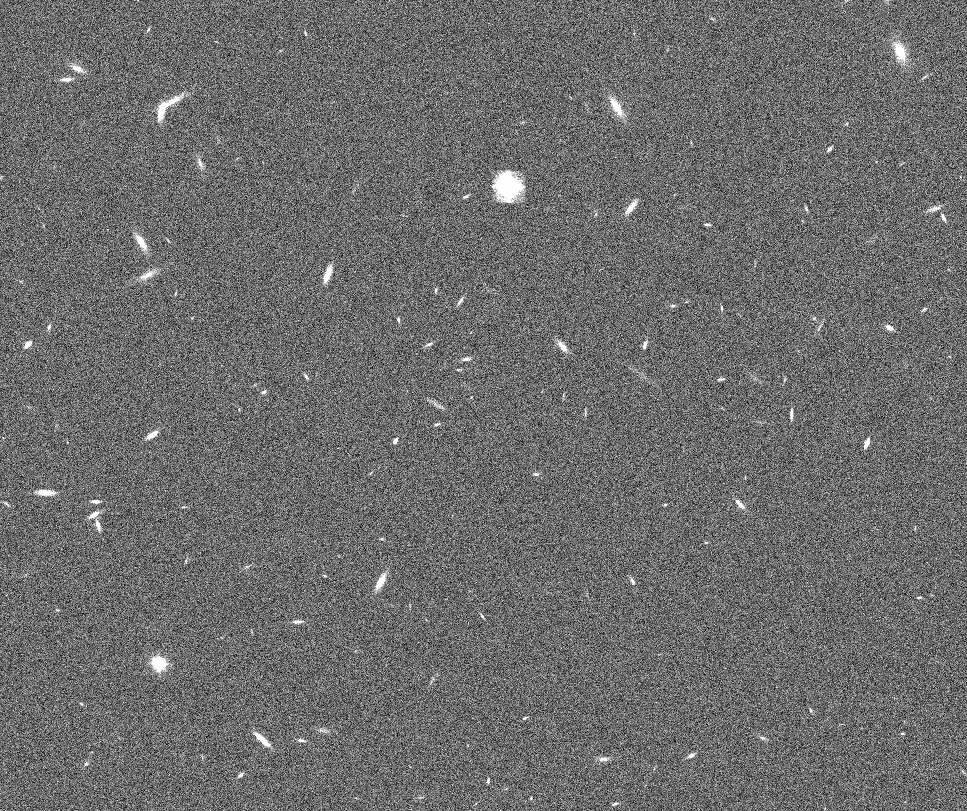}
  \end{minipage}
  \begin{minipage}{0.49\linewidth}
  
    \includegraphics[width=\linewidth]{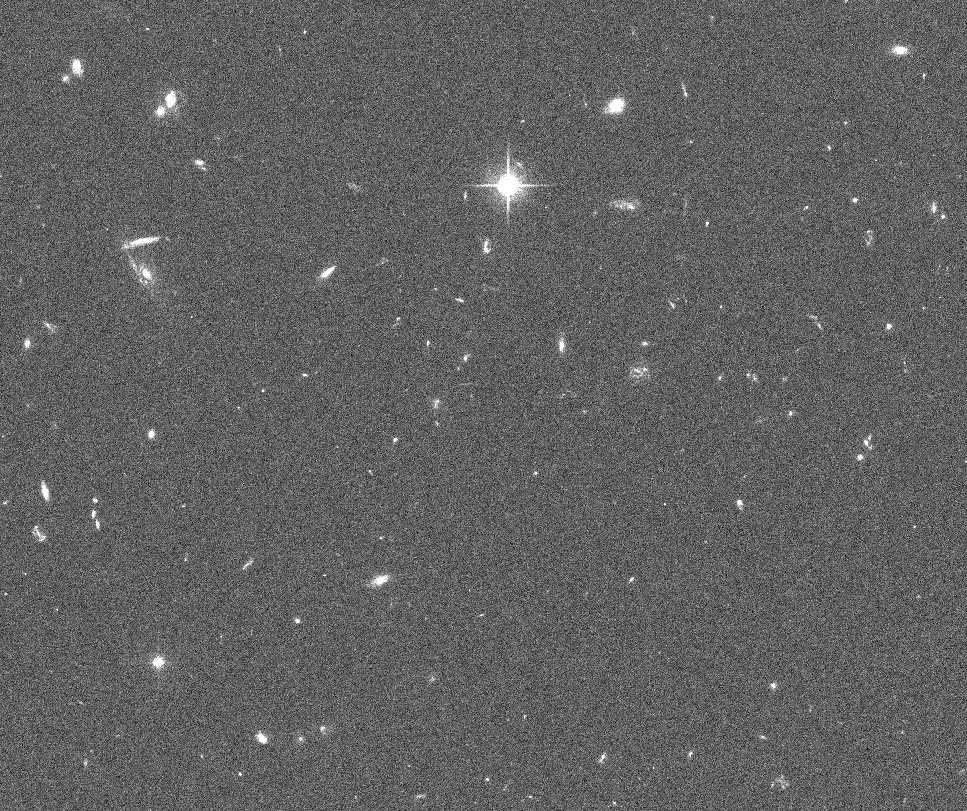}
      \end{minipage}

  \caption{Comparison between the simulated images from GOODS North (left) and the real images (right). The sizes and ellipticities are different between them, but the positions and magnitudes are matched individually.} \label{fig:CANDELS_comp}
 \end{figure*} 

\begin{figure*}[t]
  \centering
 \begin{minipage}{0.45\linewidth}

  \includegraphics[width=\linewidth]{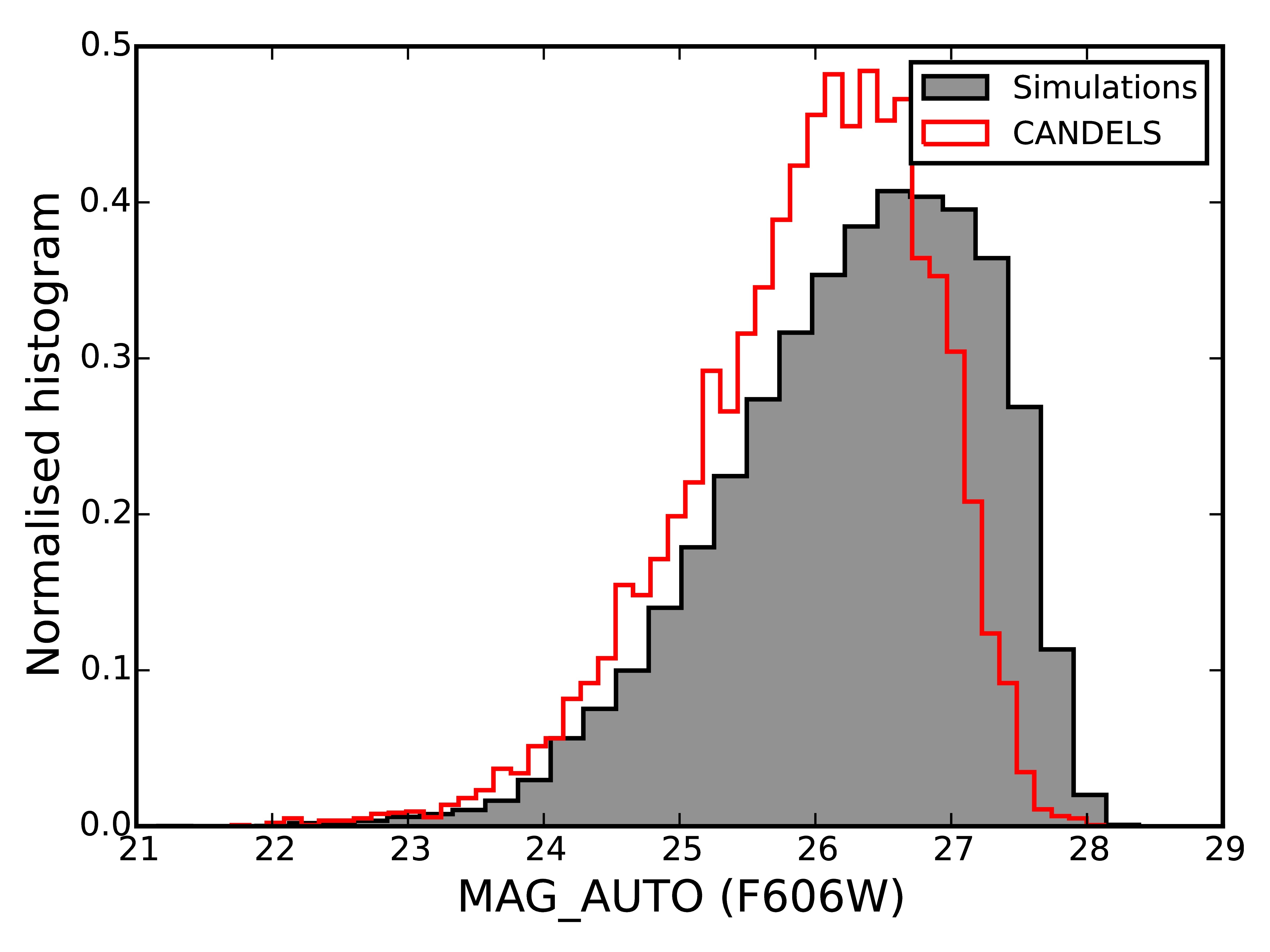}

  \end{minipage}\begin{minipage}{0.45\linewidth}

  \includegraphics[width=\linewidth]{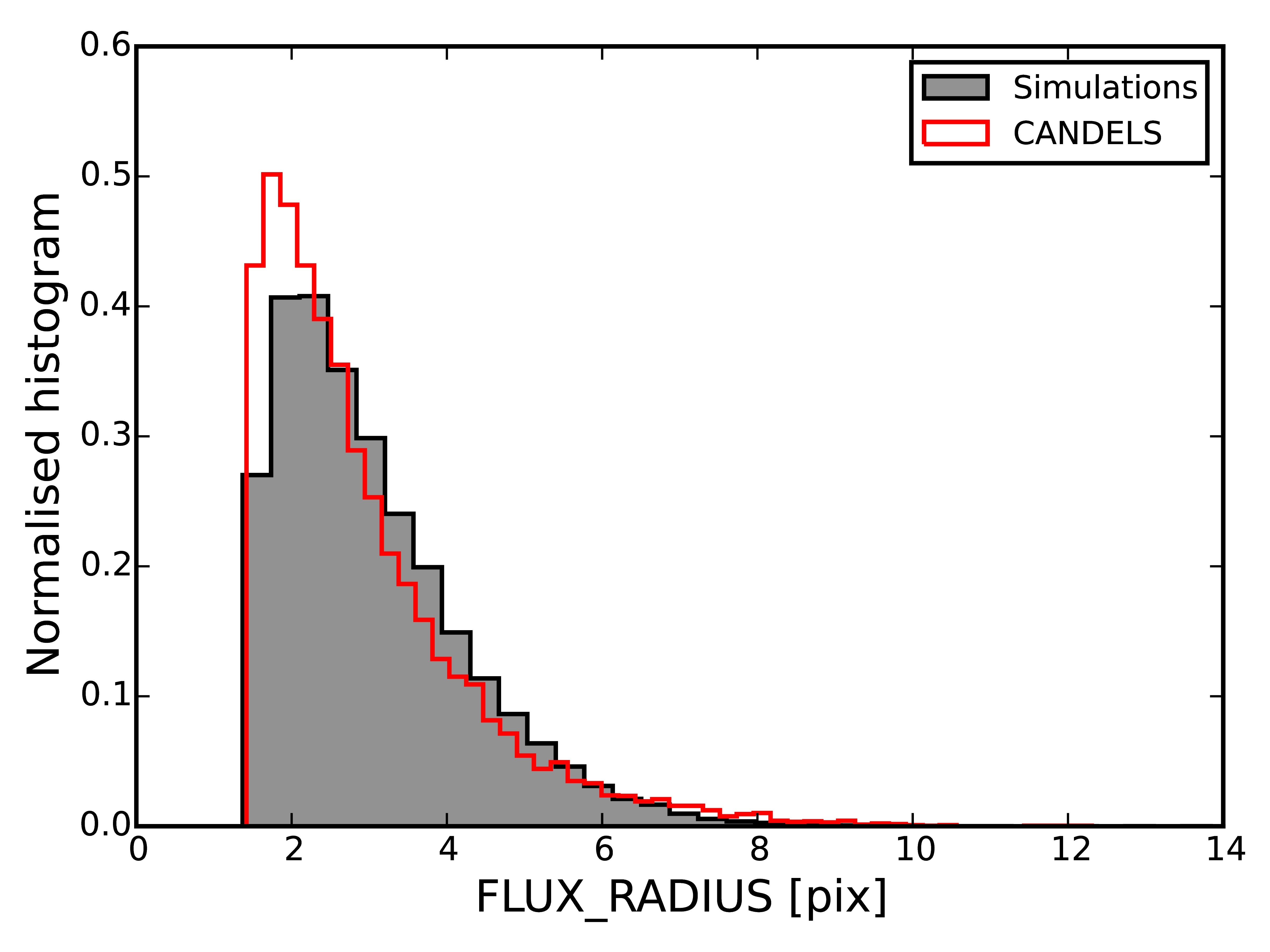}

  \end{minipage}
  \begin{minipage}{0.45\linewidth}

  \includegraphics[width=\linewidth]{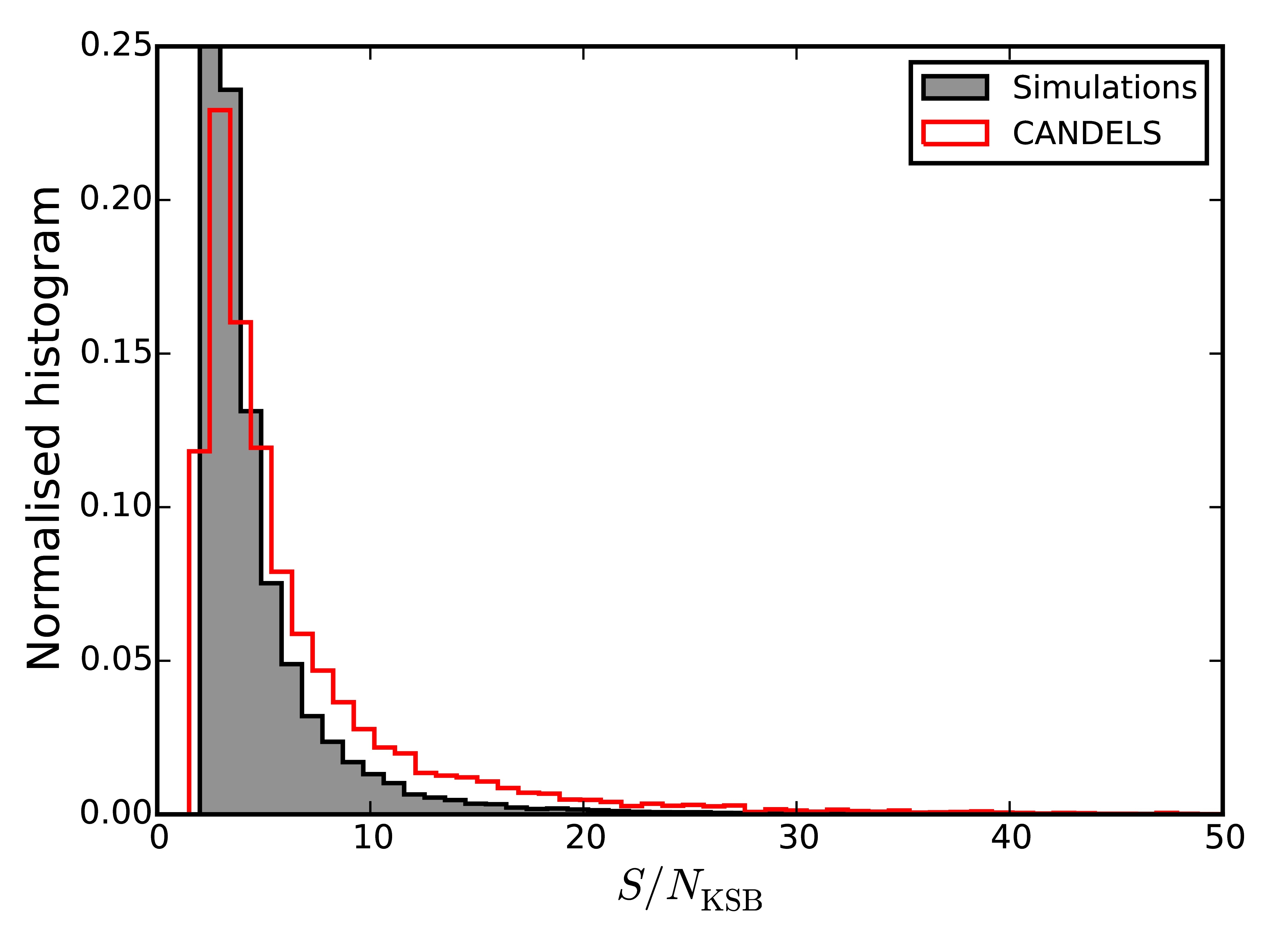}

  \end{minipage}\begin{minipage}{0.45\linewidth}

  \includegraphics[width=\linewidth]{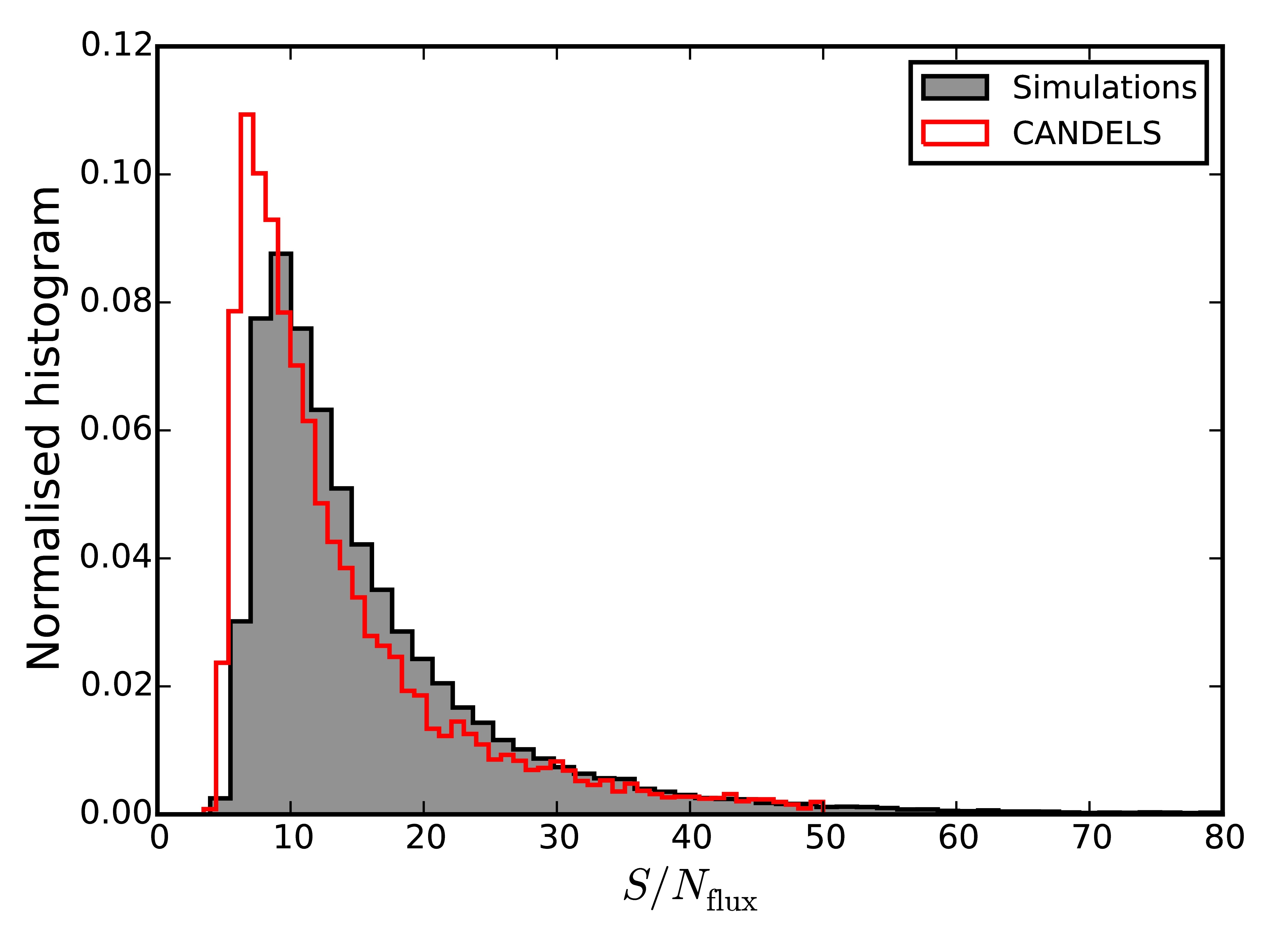}

  \end{minipage}
  \caption{Comparisons between the measured distributions in our CANDELS-like simulations and the KSB+ CANDELS distribution for the F606W magnitudes, the half-light radius, the KSB signal-to-noise ratio $S/N_{\mathrm{KSB}}$, and the SExtractor signal-to-noise ratio $S/N_{\mathrm{flux}}$.} \label{fig:candels_distributions} 
  
 \end{figure*}

We used different patches within the CANDELS fields to provide different realisations of positions and magnitudes. An example of how the simulations look when compared to the real images can be seen in Fig.\thinspace\ref{fig:CANDELS_comp}, where we show a cutout of the GOODS North field in the real images and in our simulation. The sizes and ellipticities differ, but the positions and magnitudes are comparable. Some objects vary since we use a NIR-detected catalogue. When creating the mock images, for the galaxies that are within our colour cuts ($V_{606}-I_{814}<0.4$) and magnitude range ($V_{606}<26.5$), we stored their positions in order to later select them for our analysis. This guarantees that the galaxies used for the shear estimation are indeed similar to what is used in real images. Fainter galaxies were also included in the simulation as present in the \cite{Skelton} catalogue. Importantly this catalogue is deeper than the single-orbit-depth shape catalogue
from \citetalias{Schrabback_2018}, which was used to define the inputs for the simulations described in Sect.\thinspace\ref{sec:grid}. As a result,
this reduces the incompleteness in the input catalogue, leading to a better match in the
recovered distributions between the simulation analysis and the CANDELS analysis, especially in
terms of $S/N_\mathrm{flux}$ (compare Figs.\thinspace \ref{fig:candels} and \ref{fig:candels_distributions}).

With this analysis setup we find a small \mbox{$\sim -0.7\%$} shift in the multiplicative
bias due to the use of realistic positions compared to the grid-based analysis.
This however still lacks the impact of selection bias and very faint \mbox{$V_{606}>27.5$} galaxies, which are accounted for in the following subsections.

While the effect of neighbours has a relevant impact on our analysis, its impact is at a much smaller level than what was found in previous work (\citealp{Mandelbaum_2018}) for ground-based images. This is likely due to the better resolution of the HST images, leading to a weaker impact of the blending and neighbours. In addition, our neighbour rejection may contribute to the lower impact.

\subsection{Selection bias} \label{sec:selection_bias}
Weak lensing is based on the assumption that the orientation of the intrinsic ellipticity of the galaxies is random. When we preferentially select galaxies aligned with, or orthogonally to, the shear direction we introduce a selection bias into our sample. This affects our measured shear \citep{Heymans} and subsequently the mass estimation. The selection bias can come from different sources and in this section we aim to disentangle the different steps in the shape analysis and their impact in the bias determination. In the previous sections the selection bias was neglected as we required matched pairs with opposite intrinsic ellipticities, which artificially removed any preferential selection. However, selection bias can be important (e.g. \citealp{Kannawadi}). For this reason we present here a step-by-step analysis of the selection bias alone before obtaining a joint estimation of the residual bias and the selection bias in Sect.\thinspace\ref{sec:sel_bias_clustering}, as well as their joint signal-to-noise ratio dependence.

In order to measure the selection bias we identified the selected galaxies and computed their average intrinsic ellipticity (from the input catalogues). Any deviation from zero indicates a preferential selection of galaxies depending on their intrinsic ellipticity. We compared the average input ellipticity independently for each input shear and fit a linear relation in order to constrain the shear dependence. Unlike what we did when estimating the residual bias, we did not require detection in both the normal image and the rotated one, since this is not what happens in real images. Both normal and rotated images were still included in order to cancel shape noise for the galaxies detected in both (these galaxies do not contribute to the selection bias), tightening the constraints.

It is important to measure the selection bias using realistic positions and not with the galaxies placed on a grid since the pipeline might introduce a bias by rejecting galaxies from the analysis due to the neighbours. The selection bias that was present after every step is summarised in Table \ref{tab:selection_bias}.

\begin{table}[ht]
\caption{Selection bias after each step in the analysis pipeline.} \label{tab:selection_bias}
\centering
\resizebox{9cm}{!}{
\begin{tabular}{c c c c c}
Step & \textbf{$m_{1,\rm{sel}}$} & \textbf{$m_{2,\rm{sel}}$} \\
\hline
\hline
\texttt{SExtractor} detection & $-0.0291 \pm 0.0015$ & $-0.0266 \pm 0.0018$ \\
+ \mbox{$S/N_{\mathrm{flux}}>10$} cuts & $-0.0259 \pm 0.0020$ & $-0.0259 \pm 0.0014$ \\
+ Neighbour rejection & $-0.0211 \pm 0.0071$ & $-0.0194 \pm 0.0056$ \\
+ Final KSB+ cuts & $-0.0138\pm 0.0020$  & $-0.0174 \pm 0.0013$ \\
\hline
\end{tabular}}
\end{table}

\subsubsection{\texttt{SExtractor} object detection}
\begin{figure}[t]

  \includegraphics[width=\linewidth]{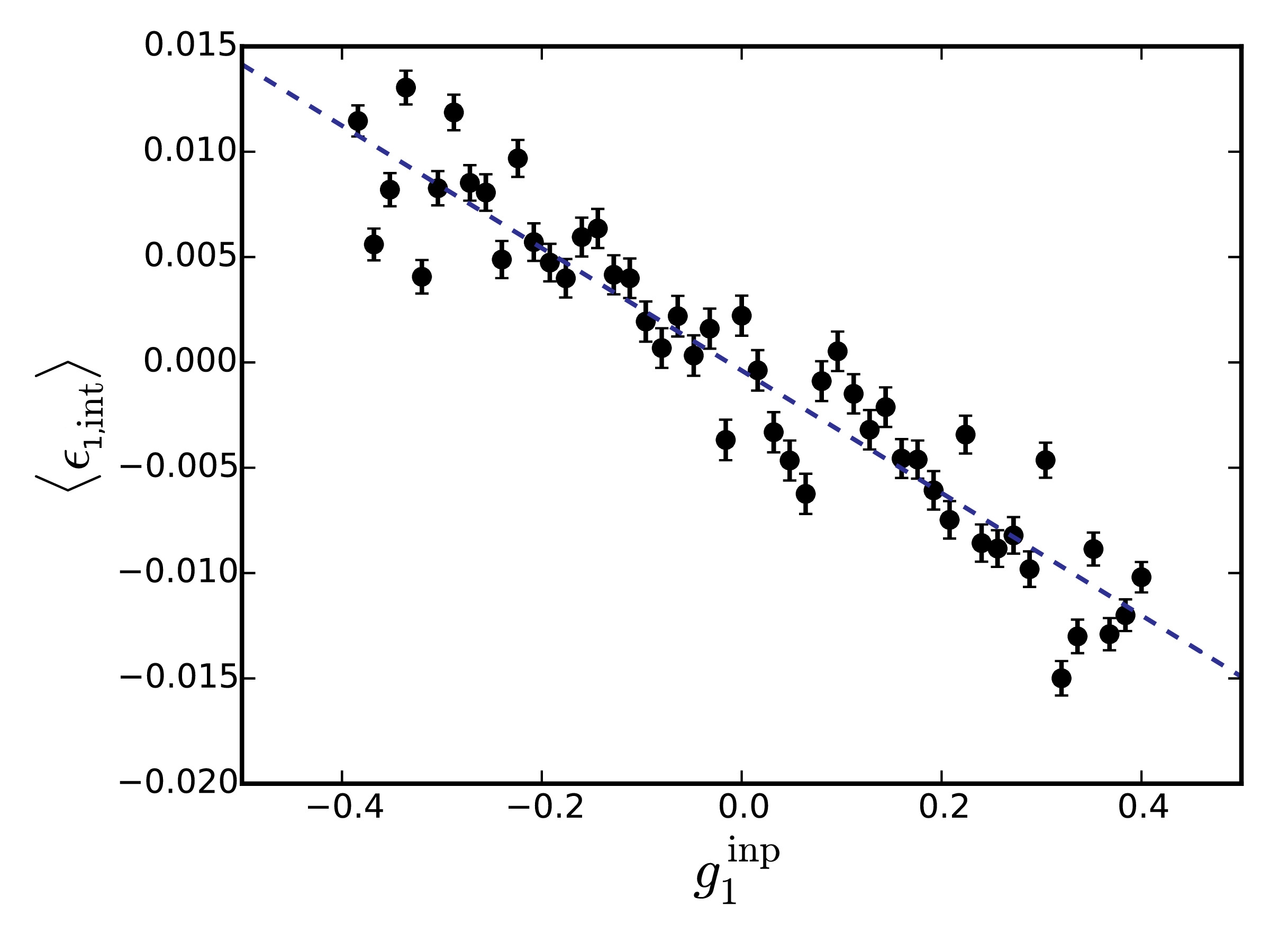}

  \caption{Selection bias on the first component after the \texttt{SExtractor} detection as a function of the input shear.} \label{fig:selection_bias_sex}
 \end{figure}

We used \texttt{SExtractor} to detect the objects and create a catalogue\footnote{Here the most important settings are \texttt{FILTER\_NAME}=gauss\_2.5\_5x5.conv, \texttt{DETECT\_THRESH}=1.4, \texttt{DETECT\_MINAREA}=8, \texttt{DEBLEND\_NTHRESH}=32, and \texttt{DEBLEND\_MINCONT}=0.01.}.
Fig.\thinspace\ref{fig:selection_bias_sex} shows the mean intrinsic ellipticity of the galaxies passing the object detection (and our colour and magnitude selection) as a function of the input shear. 
A linear fit to these points yields the selection bias, which is $m_{1,\rm{sel}}=-0.0291 \pm 0.0015$ and  $m_{2,\rm{sel}} = -0.0266 \pm 0.0018$, similar to the results from \cite{Kannawadi}. This bias does not depend on the actual shear measurement algorithm as it comes directly from the detection software (in our case \texttt{SExtractor}) and therefore it is a more general issue. A main reason for this negative selection bias is likely the isotropic kernel, with which the image is convolved during the \texttt{SExtractor} detection stage. Thus, round galaxies are more likely to be detected. However, further factors such as the deblending may also play a role.

\subsubsection{\texttt{SExtractor} S/N cut}
We selected galaxies for our analysis that had \mbox{$S/N_{\mathrm{flux}}>10$}. This can also introduce additional selection bias. After the cut, the net selection bias was $m_{1,\rm{sel}}=-0.0259 \pm 0.0020$ and 
$m_{2,\rm{sel}} =-0.0259 \pm 0.0014$. We should note that since the cuts were performed after the object detection, the selection bias is cumulative. This indicates that our \mbox{$S/N_{\mathrm{flux}}$} selection actually leads to a slight positive bias which partially corrects for the bias in the \texttt{SExtractor} object detection. This change is marginal, however, which indicates that the exact signal-to-noise cut we used does not have a large impact. In fact, for the alternative cut of galaxies with \mbox{$S/N_{\mathrm{flux}}>7$} we get $m_{1,\rm{sel}}=-0.0279 \pm 0.0010$ and 
$m_{2,\rm{sel}} =-0.0266 \pm 0.0010$.

\subsubsection{Rejection of very close neighbours} \label{sec:sel_neighbours}
For the shape measurement, we performed a selection of objects without bright close neighbours in the detection catalogue. We rejected objects with a brighter neighbour closer than $0\farcs75$. When including this effect we obtained $m_{1,\rm{sel}}=-0.0211 \pm 0.0071$ and $m_{2,\rm{sel}} =-0.0194 \pm 0.0056$. This means that this step also partially compensates for the original bias.

\subsubsection{Final catalogues after KSB+ cuts}
\label{se:selection_bias_after_ksb}
The shape measurement algorithm introduces certain cuts in order to robustly measure the shear which can also introduce artificial biases. The results indicate that there is a net selection bias in the final catalogues of $m_{1,\rm{sel}}=-0.0138\pm 0.0020$ and $m_{2,\rm{sel}}=-0.0174 \pm 0.0013$ which is again smaller than the bias obtained in Sect.\thinspace\ref{sec:sel_neighbours}, suggesting that this step partially compensates for it. A small, but important selection bias is still present in our analysis. 
For the case when \mbox{$S/N_{\mathrm{flux}}>7$} we obtained a final selection bias estimate of $m_{1,\rm{sel}}=-0.0150\pm 0.0018$ and $m_{2,\rm{sel}}=-0.0180 \pm 0.0012$.

\subsection{Joint correction for shape measurement and selection bias based on the CANDELS-like simulations}  \label{sec:realistic}

\begin{figure*}[t]
  \centering
 \begin{minipage}{0.45\linewidth}

  \includegraphics[width=\linewidth]{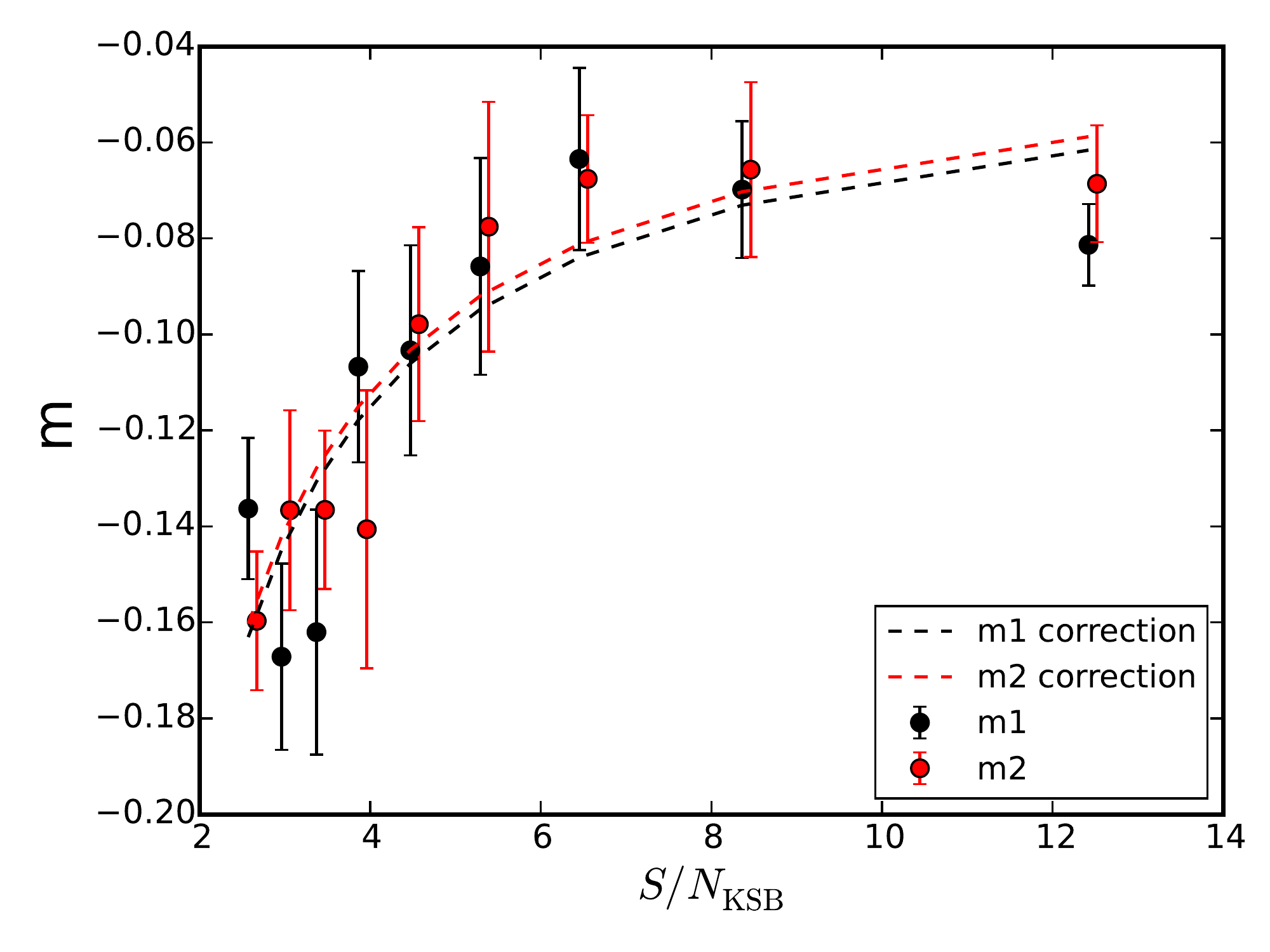}

  \end{minipage}\begin{minipage}{0.45\linewidth}

  \includegraphics[width=\linewidth]{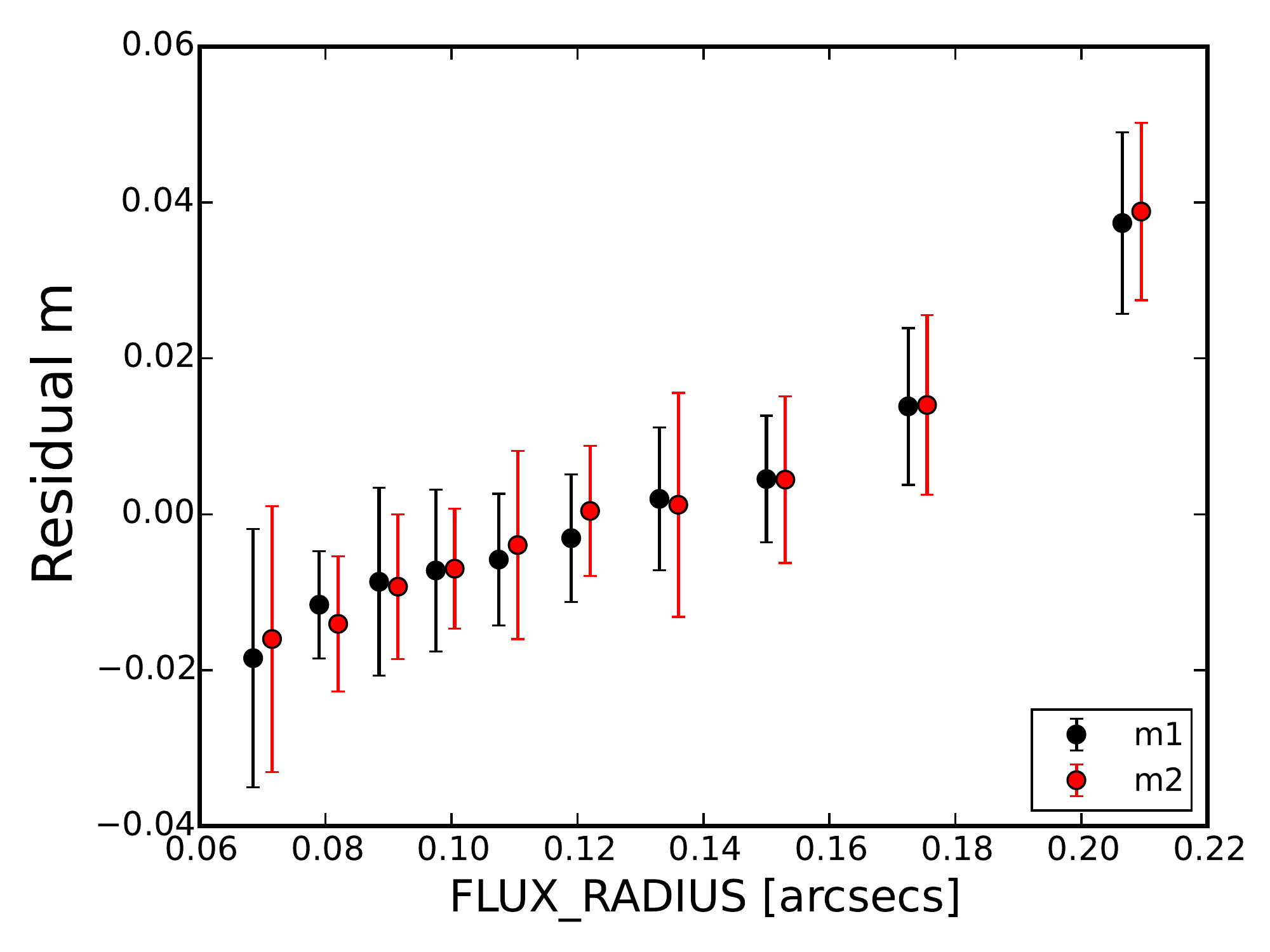}

  \end{minipage}
\caption{\textit{Left}: Dependence of the bias on $S/N_{\mathrm{KSB}}$ computed without correction or matching pairs from the CANDELS-like simulations, which then also includes the impact of selection bias. The dashed lines correspond to the correction from Eq.\thinspace(\ref{eq:sn_corr_final}) for each component. \textit{Right}: Dependence of the residual bias on the FLUX$\_$RADIUS estimated from the CANDELS-like simulations. Here, we have applied the $S/N_{\mathrm{KSB}}$-dependent correction from Eq.\thinspace(\ref{eq:sn_corr_final}). The second component is slightly shifted for visualisation purposes.} \label{fig:faint_dependencies}
 \end{figure*}

In this section we present a joint measurement of the multiplicative bias and the selection bias for our most realistic simulation using the CANDELS positions and magnitudes. These two effects cannot be separated perfectly, which is why it is best to constrain them jointly.
Also, the selection bias may have some dependence on $S/N_\mathrm{KSB}$, which is why we have to
verify if the fit obtain in Sect.\thinspace\ref{sec:noise_bias} still describes the $S/N_\mathrm{KSB}$ dependence well. 

A combined analysis of both effects in 10 $S/N_{\mathrm{KSB}}$ bins is shown in the left panel of Fig.\thinspace\ref{fig:faint_dependencies}. The estimates are more noisy in this case since the shape noise cancellation is less effective given that we no longer require matched pairs.

We find that the dependence of the joint bias on  $S/N_\mathrm{KSB}$ is still well described
by the power-law fit obtained in Sect.\thinspace\ref{sec:noise_bias} if we add a constant bias offset, for which we obtained best-fit parameters
\begin{eqnarray}
\begin{aligned}
m_{1\text{corr}}=-0.358 \left( S/N_{\mathrm{KSB}}\right) ^{-1.145}\,-0.042 \, , \\
m_{2\text{corr}}=-0.357 \left( S/N_{\mathrm{KSB}}\right) ^{-1.298}\,-0.039 \, . \label{eq:sn_corr_final}
\end{aligned}
\end{eqnarray}
Using this correction we obtained a residual bias of
$m_{1,\mathrm{res}}=0.0010 \pm 0.0040$ and $m_{2,\mathrm{res}}=-0.0023 \pm 0.0043$ for  \mbox{$S/N_{\mathrm{flux}}>10$} and $m_{1,\mathrm{res}}=-0.0032 \pm 0.0049$ and $m_{2,\mathrm{res}}=-0.0047 \pm 0.0045$ for  \mbox{$S/N_{\mathrm{flux}}>7$}. This contains the effect of the addition of faint galaxies as present in the \citet{Skelton} catalogues as well as the selection bias.

In the right panel of Fig.\thinspace\ref{fig:faint_dependencies} we show the dependence of the bias on the FLUX\_RADIUS after the $S/N_{\mathrm{KSB}}$-dependent correction from Eq.\thinspace(\ref{eq:sn_corr_final}) is applied. We find that this dependence is very weak when both shape measurement and selection biases are taken into account, confirming that it is sufficient to apply a  $S/N_\mathrm{KSB}$ dependent correction.
As a cross-check we conducted a reweighting analysis, finding that the discrepancy in the FLUX\_RADIUS distributions between the data and simulations affect the bias at a negligible $-0.3\%$ level only.

\subsection{Addition of faint galaxies following \citet{Martinet}} \label{sec:faint}
The CANDELS-like simulations provide a good approximation of the real data. However, \citet{Martinet} finds that galaxies up to 2 magnitudes fainter than the studied source sample need to be included in the simulations in order to fully account for the impact of neighbours. With the CANDELS catalogues we only included galaxies up to approximately 1 magnitude fainter. In the present analysis we simulated images in the F606W band and applied different magnitude and colour cuts, and therefore cannot directly employ the findings from \citet{Martinet}. Instead, we remeasured the clustering and galaxy properties in the HUDF F606W data closely following \citet{Martinet}, but using galaxies with $24.0<V_{606}<26.5$ and  $V_{606}-i_{775}<0.3$\footnote{We note that this is a slightly different cut than used in the rest of this work, but this should not impact the results significantly.} as 'bright' sample; and surrounding sources with $26.5<V_{606}<28.5$ as 'faint' sample. Based on the resulting distributions we injected faint galaxies into our grid-based simulations and remeasured the shape measurement biases. In order to study the increase in the bias due to the addition of galaxies with $27.5<V_{606}<28.5$, we computed the bias in two different setups. The first one includes faint galaxies in the magnitude range $26.5<V_{606}<28.5$, while the other one only includes faint galaxies with $26.5<V_{606}<27.5$. The shift in the bias between both setups needs to be added to the residual bias found in Sect.\thinspace\ref{sec:realistic}. This is a contribution of $\Delta m_1=-0.0037 \pm 0.0055$ and $\Delta m_2=-0.0060 \pm 0.0065$ for \mbox{$S/N_{\mathrm{flux}}>10$} and $\Delta m_1=-0.0061 \pm 0.0036$ and $\Delta m_2=-0.0070 \pm 0.0031$ for \mbox{$S/N_{\mathrm{flux}}>7$}.

\begin{table*}[t!]
 \caption{Summary of the contribution to the bias and its uncertainties from the different effects for the two $S/N_{\mathrm{flux}}$ cuts used in this paper. We present a separated bias for each shear component for all effects except for the light contamination due to bright galaxies, since that is calculated from the tangential shear, which depends on both components. The modelling errors include the uncertainties related to variations in the PSF shape and the galaxy light profile distributions.} \label{tab:final}
\centering
\resizebox{18cm}{!}{
\begin{tabular}{c c c c c }
Contribution & \multicolumn{2}{c}{\mbox{$S/N_{\mathrm{flux}}>10$}} & \multicolumn{2}{c}{\mbox{$S/N_{\mathrm{flux}}>7$}}\\
& $m_1$ & $m_2$ & $m_1$ & $m_2$ \\
\hline
\hline 

Residual bias after CANDELS-like calibration & $0.0010\pm 0.0040$ &$-0.0023\pm 0.0043$ &$-0.0032\pm 0.0049$ &$-0.0047\pm 0.0045$ \\
Bias from extra faint $V_{606}>27.5$ galaxies & $-0.0037 \pm 0.0055$  & $-0.0060\pm 0.0065$ &$-0.0061 \pm 0.0046$  & $-0.0070\pm 0.0041$\\
Bright cluster galaxy contribution ($z>0.7$): Shape measurement bias  & \multicolumn{2}{c}{$0.0048\pm 0.0038$} &  \multicolumn{2}{c}{$0.0025\pm 0.0040$}\\
Bright cluster galaxy contribution ($z>0.7$): Selection bias  & \multicolumn{2}{c}{$0.0053\pm 0.0019$} &  \multicolumn{2}{c}{$0.0060\pm 0.0029$}\\
Other modelling uncertainties (simulation mismatch effects) & $\pm 0.0128$ & $\pm 0.0130$ & $\pm 0.0126$ & $\pm 0.0131$ \\
\hline
Total residual bias + uncertainty & $0.0074\pm 0.0151$& $0.0017\pm 0.0157$ &$-0.0008\pm 0.0151$& $-0.0032\pm 0.0153$\\

 \hline
\end{tabular}}
\end{table*}
 
As a consistency check we also verified that the setup of a grid-based simulation with added \mbox{$26.5<V_{606}<27.5$} HUDF-like galaxies
yields shape measurement biases that are
consistent with the results obtained using the CANDELS-like simulations. Also note that, in total, the full set of faint galaxies (\mbox{$26.5<V_{606}<28.5$}) causes a shift in the shape measurement bias by \mbox{$\sim -1.3\%$}, as estimated with the HUDF-informed analysis.

\subsection{Selection bias due to the addition of cluster members.} \label{sec:sel_bias_clustering}

\begin{figure}[t!]
 
  \includegraphics[width=\linewidth]{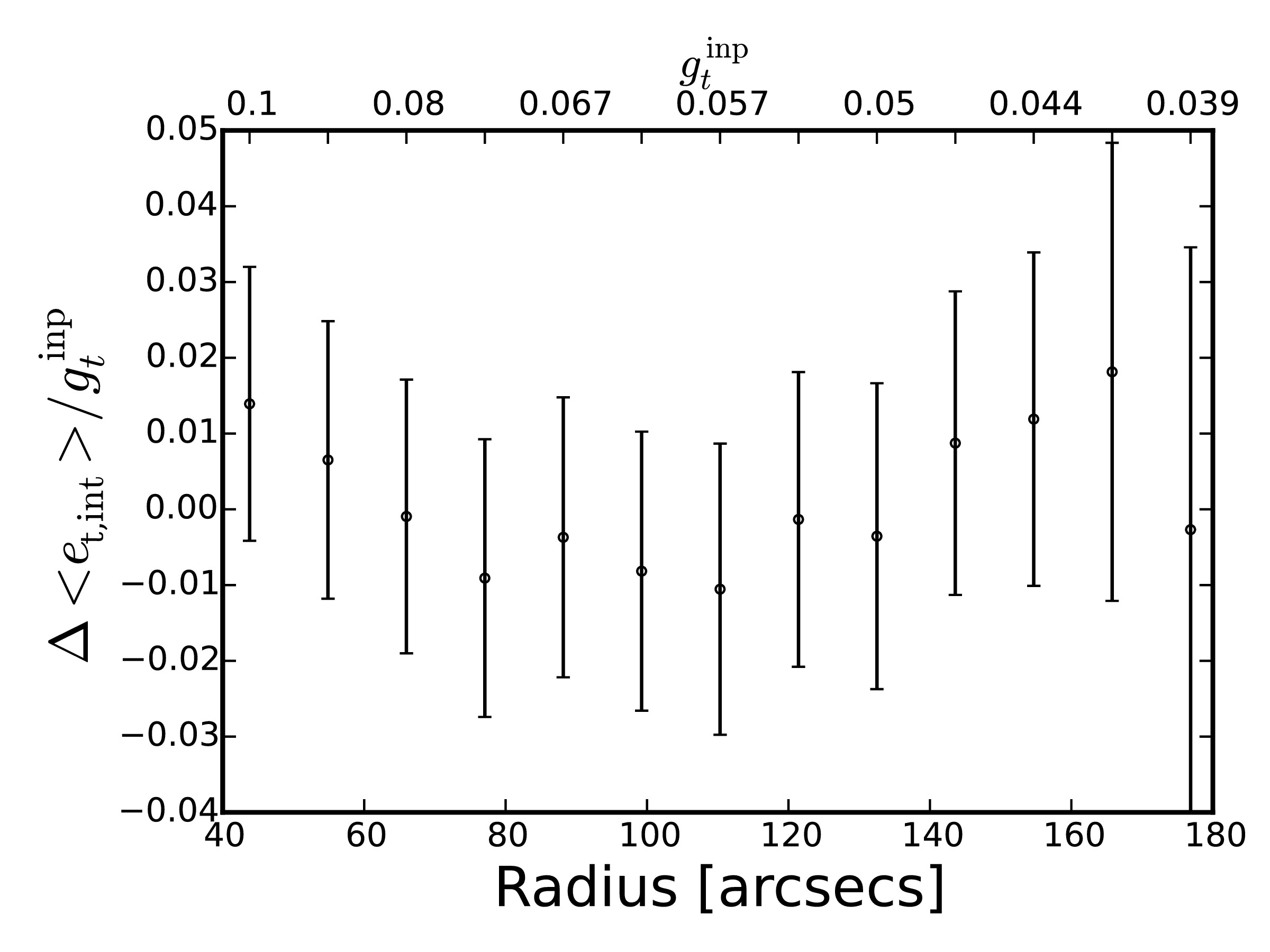}

   \caption{Relative difference in the selection bias estimates for the background only and the background+cluster members simulations for clusters at $z>0.7$  and \mbox{$S/N_{\mathrm{flux}}>7$} galaxies.} \label{fig:sel_cluster_members}
 \end{figure}

Employing the simulations analysed in Sect.\thinspace\ref{sec:clustering} we can estimate how much the addition of cluster members affects the selection bias.  Fig.\thinspace\ref{fig:sel_cluster_members} shows the difference in the selection bias for the simulations with cluster members and the background-only simulations as a function of cluster-centric distance. Averaged over all scales we find that the addition of cluster members yields a slightly positive selection bias of \mbox{$0.0085\pm 0.0024$} if all cluster redshifts are considered, and \mbox{$0.0060\pm 0.0029$} for the clusters at \mbox{$z>0.7$} and \mbox{$S/N_{\mathrm{flux}}>7$} galaxies, which we add to our total residual bias estimation in Sect.\thinspace\ref{sec:final}.

\section{Summary of the bias estimates} \label{sec:final}
The bias estimate from Sect.\thinspace\ref{sec:neighbours} was obtained from the CANDELS-like simulations that closely resemble the real images. However, as mentioned before, this lacks bright cluster members, as well as faint ($V_{606}>27.5$) neighbours. In this section we present a summary of the different contributions and an estimate of the final remaining biases and uncertainties after the correction in Eq.\thinspace(\ref{eq:sn_corr_final}) is applied. The contribution of the general selection bias is not separately reported in the table, but it is included in the bias calibration and residual bias estimates from the CANDELS-like simulations (Sect.\thinspace\ref{sec:realistic}). We computed the bias in all cases for $|g|<0.2$ and for two different signal-to-noise ratio cuts (\mbox{$S/N_{\mathrm{flux}}>10$} and \mbox{$S/N_{\mathrm{flux}}>7$}). 
In order to obtain robust estimates of the uncertainties caused by limitations of our simulations we summarised the most important contributions (explained below) as 'Other modelling uncertainties' in Table \ref{tab:final} (all added in quadrature). These include the differences in the multiplicative bias from a flat S\'ersic index distribution and the more realistic parametric fits (see Sect.\thinspace\ref{sec:light}) as a conservative estimate for the uncertainties in our galaxy models. The Tiny Tim PSF model should provide a good match to the actual PSF shapes in HST data. However, as the bias shows some dependence on the PSF shape (see Sect.\thinspace\ref{sec:PSF}), we additionally included half of the bias difference from the analysis using a Moffat PSF (which is clearly not a good model for HST) as an additional error contribution. To account for the slight dependence of the multiplicative bias on the PSF ellipticity (see Sect.\thinspace\ref{sec:PSF}) we also added half of the bias difference between the setups of a circular and an elliptical (\mbox{$|e|=0.1$}) Moffat PSF to the systematic error budget. 

The final residual bias estimate and uncertainty is $m_1=0.0074\pm 0.0151$, $m_2=0.0017\pm 0.0157$ for \mbox{$S/N_{\mathrm{flux}}>10$} galaxies and $m_1=-0.0008\pm 0.0151$, $m_2=-0.032\pm 0.0153$ for \mbox{$S/N_{\mathrm{flux}}>7$} galaxies. These residual biases need to be corrected for when analysing real data, as done in Sect.\thinspace\ref{sec:mass}. We find that after applying these corrections, multiplicative shape measurement biases should be accurately calibrated to the $1.5\%$ level.

\section{HST weak lensing measurements of three highly relaxed clusters from the South Pole Telescope Sunyaev-Zel'dovich Survey} \label{sec:mass}
To demonstrate  the use of
weak lensing measurements in the inner cluster and stronger shear regime we studied three clusters from the 2,500 deg$^2$ South Pole Telescope  Sunyaev-Zel'dovich (SPT-SZ) Survey \citep{Bleem}.
According to their X-ray properties these three clusters are among the most relaxed clusters found in the SPT-SZ sample  \citep{mcdonald19}.
These clusters are part of a larger sample of distant SPT-SZ clusters  analysed using HST weak lensing data (\citetalias{Schrabback_2018} and \citetalias{Schrabback_2019}).
Since general cluster samples are more strongly affected by miscentring and substructure in the inner cluster regions,  \citetalias{Schrabback_2018} and \citetalias{Schrabback_2019} excluded the cluster cores from their analysis and only incorporated scales \mbox{$r>500$ kpc} when fitting radial shear profiles.
As these astrophysical limitations are less severe for relaxed clusters, we can use these clusters as a test case
to investigate how much constraints can be tightened by including weak lensing measurements from the cluster cores.

SPT-CL{\thinspace}$J$0000$-$5748 (\mbox{$z=0.702$}) and SPT-CL{\thinspace}$J$2331$-$5051 (\mbox{$z=0.576$}) were initially studied by \citetalias{Schrabback_2018}, who measured  weak lensing shapes in \mbox{$2\times 2$} HST/ACS mosaic $V_{606}$ images and selected mostly background galaxies using  $V-I$ colour.
For the source selection they employed HST  $I_{814}$ imaging in the cluster core and VLT $I$-band imaging in the cluster outskirts. \citetalias{Schrabback_2019} recently updated these measurements using a revised reference sample for the calibration of the source redshift distribution \citep{Raihan_2019} and employing our revised shear calibration as summarised in Sect.\thinspace\ref{sec:final}. \citetalias{Schrabback_2019} also incorporate deeper VLT $I$-band imaging for the source selection for SPT-CL{\thinspace}$J$0000$-$5748.
SPT-CL{\thinspace}$J$2043$-$5035 (\mbox{$z=0.723$}) was studied in  \citetalias{Schrabback_2019} using the same calibrations, employing shape measurements from   mosaic ACS $V_{606}$ images and a source selection that incorporates mosaic ACS $I_{814}$ imaging for the full cluster field. We refer the reader to these publications for further details on the data, shape measurements, source selection,   calibration of the source redshift distribution, and fitting procedure.

\begin{figure*}
 \includegraphics[width=0.69\columnwidth]{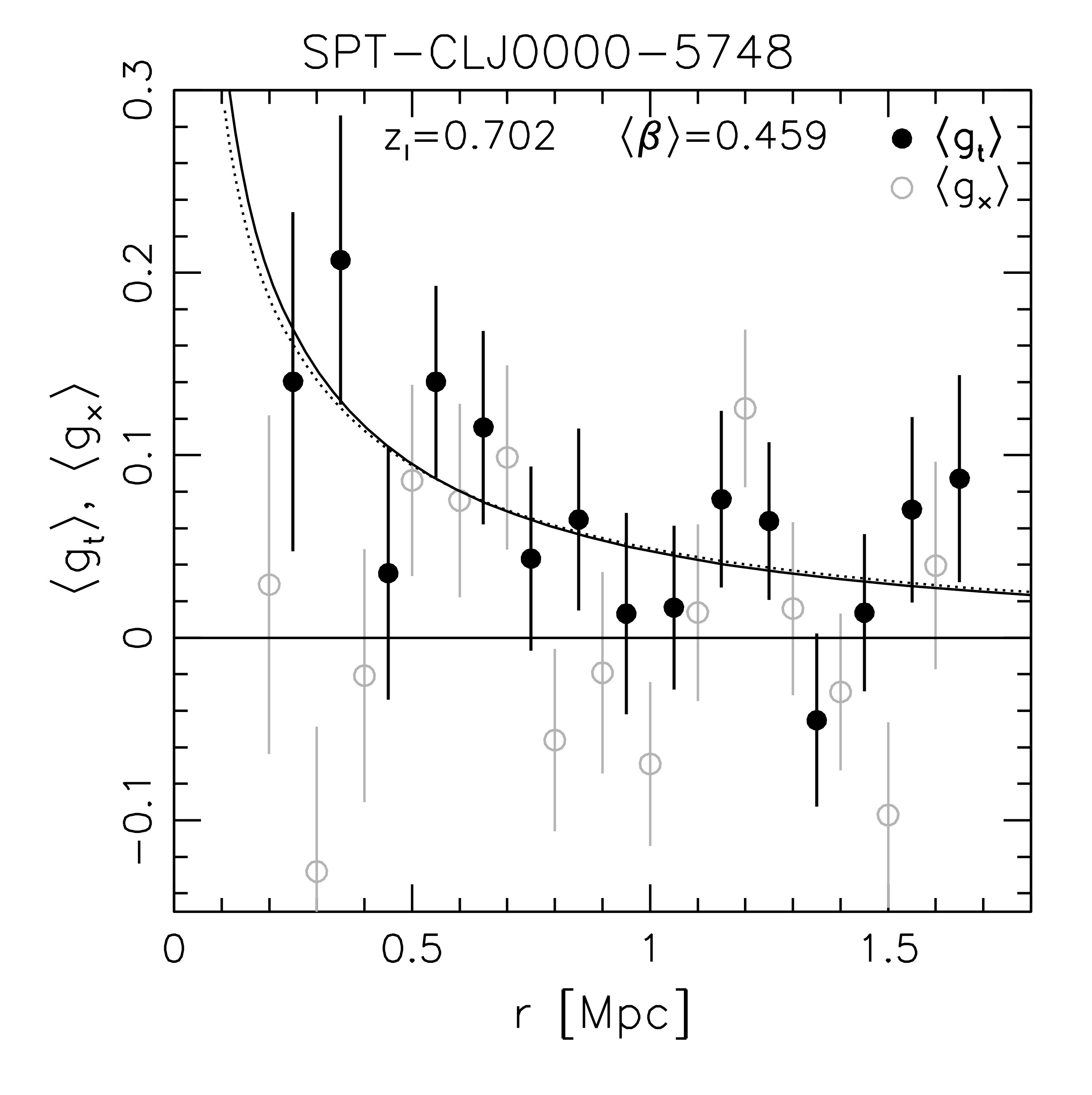}
 \includegraphics[width=0.69\columnwidth]{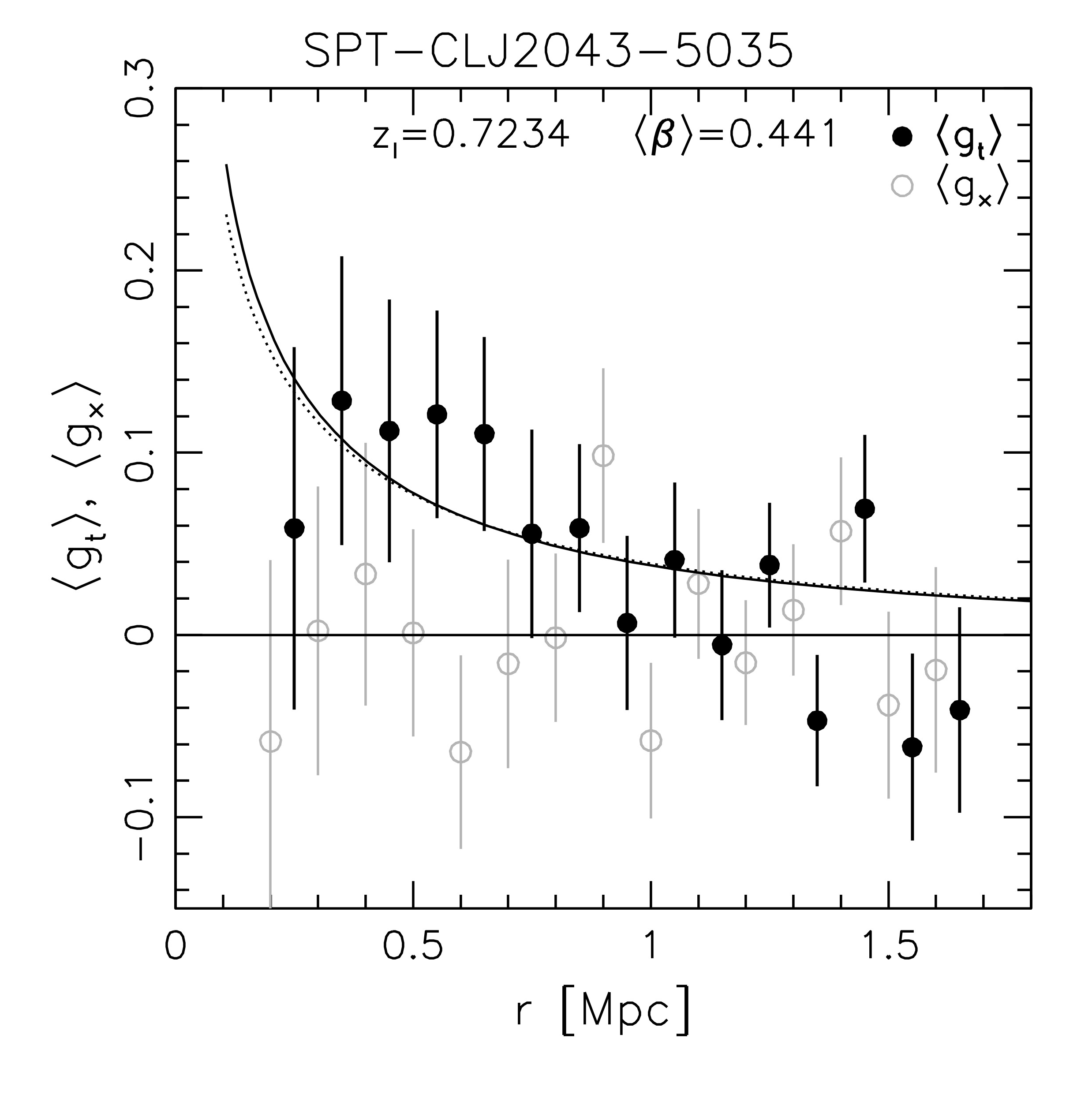}
 \includegraphics[width=0.69\columnwidth]{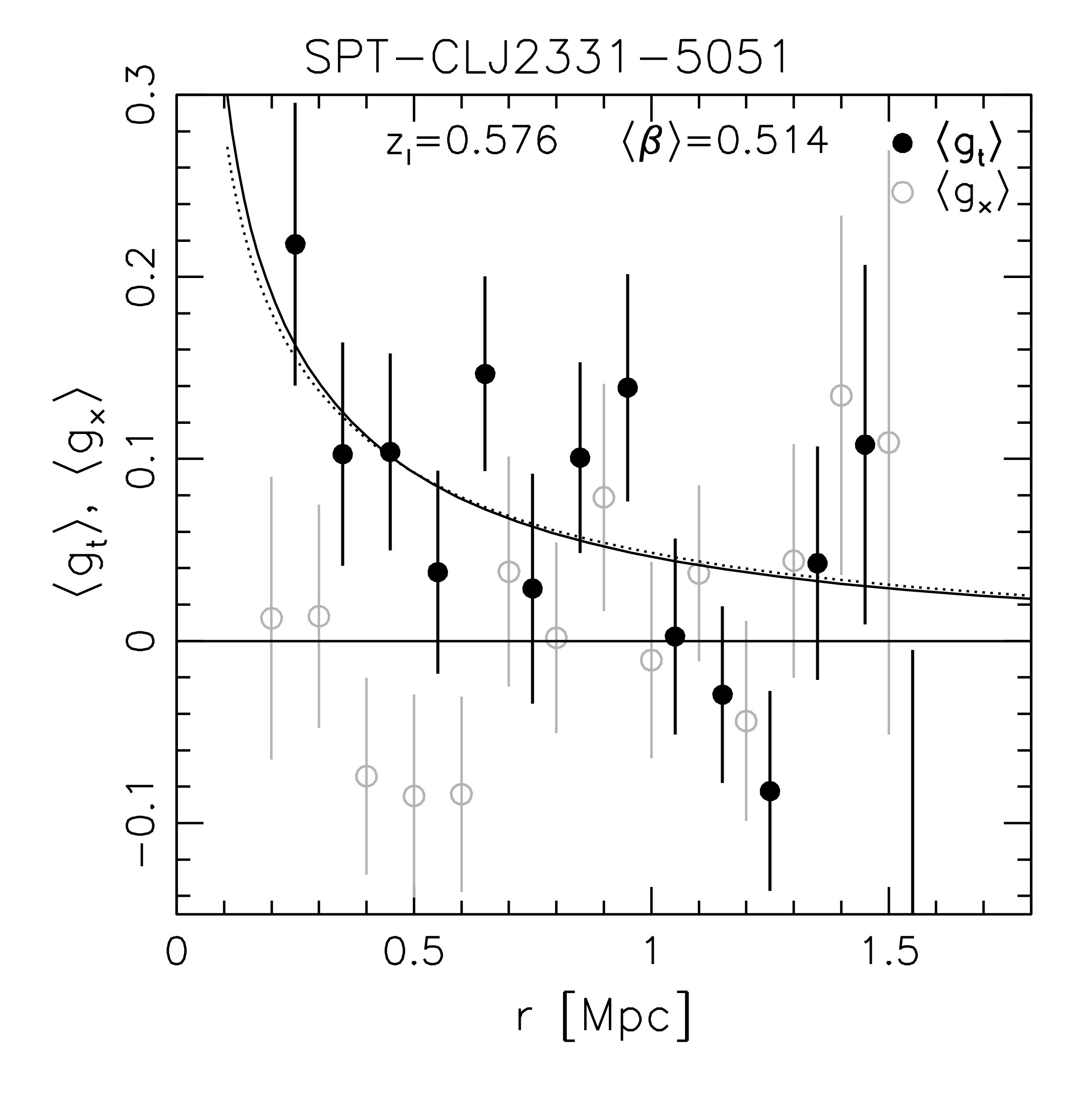}
 \caption{Reduced shear profiles around the X-ray centres 
   of the three relaxed clusters analysed in this study,
   showing the tangential (black solid circles) and cross (grey open circles) components.
   The  curves correspond to the best-fitting NFW models assuming the \citetalias{Diemer_2019} $c(M)$ relation (dotted) and increased concentrations  \mbox{$c_\mathrm{200c}=1.14 c_\mathrm{200c}^{D19}$} (solid), which are expected to more accurately represent relaxed clusters.
   \label{fi:bea:shearprofiles}}
\end{figure*}

As primary difference to the previous work we included smaller scales  (\mbox{$r>200$ kpc})\footnote{This limit ensures that only reduced shear estimates in the robustly calibrated \mbox{$|g|\lesssim 0.2$} regime are included in the fit (compare Fig.\thinspace\ref{fi:bea:shearprofiles}).} in the analysis of the reduced shear profiles of these clusters (see Fig.\thinspace\ref{fi:bea:shearprofiles}). 
Following \citetalias{Schrabback_2018} and \citetalias{Schrabback_2019} we fitted NFW shear profile models \citep{Brainerd} to the tangential and cross shear 
\begin{eqnarray}
\begin{aligned}
g_{\mathrm{t}}&=&-g_1 \cos 2 \Phi -g_2 \sin 2\Phi \, , \\
g_{\times}&=&g_1 \sin 2 \Phi -g_2 \cos 2\Phi \, ,
\end{aligned}
\end{eqnarray}
where $\Phi$ is the azimuthal angle with respect to the centre. The cross shear component serves as a test for systematics. For each of our targets it is consistent with zero within $1\sigma$ when averaged over all radial bins (compare Fig.\thinspace\ref{fi:bea:shearprofiles}). We accounted for the magnitude dependence of the mean geometric lensing efficiency of the sources and the impact of weak lensing magnification on the source redshift distribution. \citetalias{Schrabback_2019} employ a fixed concentration--mass relation from \citet[][\citetalias{Diemer_2019} henceforth]{Diemer_2019}.
Assuming the same relation in a first step, we can estimate by how much the cluster mass constraints
tighten when including the information from the inner cluster regions.
We list the mass signal-to-noise ratio \mbox{$S/N_\mathrm{mass,D19}=M_\mathrm{200c}/\Delta M_\mathrm{200c}$} (considering only shape-noise uncertainties) for both fit ranges and all three clusters  in Table \ref{tab:bea_results_ratios}, finding that it improves on average by a factor of 1.38 when scales \mbox{$r>200$ kpc} are used.

\begin{table*}  
  \caption{Cluster properties and achieved weak lensing mass signal-to-noise ratios.
   \label{tab:bea_results_ratios}}
\begin{center}
 
\begin{tabular}{lcccc}
  \hline
  \hline
  Cluster & $z$ & $S/N_\mathrm{mass,D19}^{>500\mathrm{kpc}}$ &  $S/N_\mathrm{mass,D19}^{>200\mathrm{kpc}}$ & $(S/N_\mathrm{mass,D19}^{>200\mathrm{kpc}}) /  (S/N_\mathrm{mass,D19}^{>500\mathrm{kpc}})$ \\
  \hline

SPT-CL{\thinspace}$J$0000$-$5748 & 0.702 & 2.63 & 3.25 & 1.24 \\
  SPT-CL{\thinspace}$J$2043$-$5035 & 0.723 & 2.15 & 2.61 & 1.22 \\
  SPT-CL{\thinspace}$J$2331$-$5051 & 0.576 & 1.90 & 3.19 & 1.67 \\
 \hline
\end{tabular}
 
\end{center}
{\flushleft
Note. --- 
{\it Column 1:} Cluster designation.
{\it Column 2:} Cluster redshift from \citet{Bleem}.
{\it Column 3:} Mass signal-to-noise ratio \mbox{$S/N_\mathrm{mass,D19}=M_\mathrm{200c}/\Delta M_\mathrm{200c}$} (considering only shape-noise uncertainties) obtained from  NFW reduced shear profile fits using scales  \mbox{$0.5 \mathrm{Mpc}<r<1.5 \mathrm{Mpc}$} and assuming the  \citetalias{Diemer_2019} $c(M)$ relation.
{\it Column 4:} Mass signal-to-noise ratio \mbox{$S/N_\mathrm{mass,D19}=M_\mathrm{200c}/\Delta M_\mathrm{200c}$} (considering only shape-noise uncertainties) obtained from  NFW reduced shear profile fits using scales  \mbox{$0.2 \mathrm{Mpc}<r<1.5 \mathrm{Mpc}$} and assuming the  \citetalias{Diemer_2019} $c(M)$ relation.
{\it Column 5:} Ratio of the values from columns 4 and 3.\\
}
\end{table*}

Table \ref{tab:bea_results_mass} lists the mass constraints obtained when including the inner cluster regions (\mbox{$r>200$ kpc})
and assuming the  \citetalias{Diemer_2019} concentration--mass relation.
Here we not only list the fit uncertainties caused by shape noise, but also the additional minor noise contributions from  projections of uncorrelated large-scale structure and line-of-sight variations in the redshift distribution (see \citetalias{Schrabback_2018} for details).

The \citetalias{Diemer_2019} concentration--mass relation provides an estimate for the expected average concentration as a function of mass for an average (approximately mass-selected) cluster population, as adequate for the overall sample
studied in \citetalias{Schrabback_2018} and \citetalias{Schrabback_2019}.
Given their relaxed nature, we would however expect that the three clusters studied here should -- on average -- have higher concentrations.
For simulated haloes with masses comparable to the masses of our clusters, \citet{Neto} find that the median concentration of relaxed haloes is higher by a factor $\times 1.14$  compared to the median concentration of the full halo population.
Following \citetalias{Schrabback_2018b} we therefore refitted the clusters using an increased concentration \mbox{$c_\mathrm{200c}=1.14 c_\mathrm{200c}^{D19}$}, where $c_\mathrm{200c}^{D19}$ is the concentration that  corresponds to the best-fit mass from the initial fit when assuming the \citetalias{Diemer_2019} $c(M)$ relation.
The resulting mass constraints, which shift noticeably for $M_\mathrm{200c}$ but only little for $M_\mathrm{500c}$, are also listed in Table \ref{tab:bea_results_mass}. 
Our derived constraints on $M_\mathrm{500c}$ agree within \mbox{$\sim 1\sigma$} with  {\it Chandra}
X-ray estimates computed by \citet{mcdonald19} assuming
the $Y_\mathrm{X}$--$M$ scaling relation from \citet[][compare Table \ref{tab:bea_results_mass}]{vikhlinin09b}.

Using simulations, \citetalias{Schrabback_2019} compute a mass modelling correction as a function of mass and redshift for approximately mass-selected cluster populations. Thus, their correction should on average be accurate for the full cluster sample studied by \citetalias{Schrabback_2019}.
However, since we are studying a subset consisting of particularly relaxed clusters, which should not suffer significantly 
from  miscentring or substructure, it is more adequate to not apply these mass modelling corrections to our results. For this reason, and also given that we employed the increased concentrations applicable for relaxed clusters, our masses have lower absolute values compared to the results from \citetalias{Schrabback_2019}.
We expect that the mass constraints reported for these particular clusters in  \citetalias{Schrabback_2019} are likely scattered up given that their analysis is blind to cluster morphology. We note that this does not affect the population-averaged results derived by   \citetalias{Schrabback_2019}, as particularly disturbed clusters will likely scatter down in their analysis. In the future it might be possible to reduce this scatter by deriving morphology-dependent corrections for mass-modelling biases  from hydrodynamical simulations.

Systematic mass uncertainties are small compared to the statistical uncertainties reported in Table \ref{tab:bea_results_mass}. The \mbox{$1.5\%$} uncertainty in the shear calibration (see Table \ref{tab:final}) translates to a 2.3\% mass uncertainty. The mass uncertainty associated with the calibration of the source redshift distribution amounts to 4.7\% (\citetalias{Schrabback_2019}). While we did not apply a mass modelling correction (as discussed above), we conservatively consider the same residual (post-correction) mass modelling uncertainty as \citetalias{Schrabback_2019} to reflect the lack of a  morphology-dependent mass modelling correction (5.3\%). Added in quadrature, the total systematic mass uncertainty amounts to 7.4\%.

\begin{table*}  
  \caption{Weak lensing constraints derived for the fit range \mbox{$0.2 \mathrm{Mpc}<r<1.5 \mathrm{Mpc}$}. Masses are in  $10^{14}\mathrm{M}_\odot$ and are reported   for two different over-densities
  \mbox{$\Delta \in \{200\mathrm{c}, 500\mathrm{c}\}$}.
  For comparison we also list {\it Chandra} X-ray mass estimates $M_{500\mathrm{c},Y_\mathrm{X}}$ from \citet{mcdonald19}.
   \label{tab:bea_results_mass}}
\begin{center}
\begin{tabular}{lcccccc}
  \hline
  \hline
  &  \multicolumn{3}{c}{Using $c(M)$ from \citetalias{Diemer_2019}} &  \multicolumn{2}{c}{Using \mbox{$c_\mathrm{200c}=1.14 c_\mathrm{200c}^{D19}$}}&\\
 Cluster & $M_\mathrm{200c}$ & $M_\mathrm{500c}$ & $c_\mathrm{200c}^{D19}$  & $M_\mathrm{200c}$ & $M_\mathrm{500c}$  &  $M_{500\mathrm{c},Y_\mathrm{X}}$  \\
 \hline
SPT-CL$J${\thinspace}0000$-$5748 & $5.6_{-1.7}^{+1.8}\pm 0.7\pm 0.3 $ & $3.8_{-1.1}^{+1.3} \pm 0.5\pm 0.2 $    & 3.72 & $5.2_{-1.5}^{+1.7}\pm 0.7\pm 0.3 $ & $3.7_{-1.1}^{+1.2} \pm 0.5\pm 0.2 $ & $4.1^{+0.7}_{-0.6}$\\
SPT-CL$J${\thinspace}2043$-$5035 & $4.2_{-1.5}^{+1.7}\pm 0.6\pm 0.3 $ & $2.8_{-1.0}^{+1.2} \pm 0.4\pm 0.2 $ & 3.66 &  $3.9_{-1.4}^{+1.6}\pm 0.7\pm 0.3 $ & $2.8_{-1.0}^{+1.1} \pm 0.5\pm 0.2 $ & $4.2^{+0.1}_{-0.2}$\\
SPT-CL$J${\thinspace}2331$-$5051 & $5.7_{-1.7}^{+1.8}\pm 0.8\pm 0.4 $ & $3.9_{-1.2}^{+1.3} \pm 0.5\pm 0.3 $ & 3.69 & $5.2_{-1.6}^{+1.7}\pm 0.7\pm 0.3 $ & $3.6_{-1.1}^{+1.2} \pm 0.5\pm 0.2 $ & $4.3^{+0.3}_{-0.4}$\\
  \hline
\end{tabular}
 
\end{center}
{\flushleft
Note. --- 
{\it Column 1:} Cluster designation.
{\it Column 2:}  $M_\mathrm{200c}$ constraints obtained when assuming the $c(M)$ relation from \citetalias{Diemer_2019}.
{\it Column 3:}  $M_\mathrm{500c}$ constraints obtained when assuming the $c(M)$ relation from \citetalias{Diemer_2019}.
{\it Column 4:}  Concentration derived from  the \citetalias{Diemer_2019} $c(M)$ relation for the best-fit $M_\mathrm{200c}$.
{\it Column 5:}  $M_\mathrm{200c}$ constraints obtained for a fixed concentration \mbox{$c_\mathrm{200c}=1.14 c_\mathrm{200c}^{D19}$}. 
{\it Column 6:}  $M_\mathrm{500c}$ constraints obtained for a fixed concentration \mbox{$c_\mathrm{200c}=1.14 c_\mathrm{200c}^{D19}$}. 
The statistical uncertainties listed for the mass constraints correspond to shape noise (asymmetric), uncorrelated large-scale structure projections, and line-of-sight variations in the source redshift distribution. The systematic mass uncertainty amounts to 7.4\%. 
{\it Column 7:} {\it Chandra} X-ray mass estimates computed by  \citet{mcdonald19} assuming the $Y_\mathrm{X}$--$M$ scaling relation from \citet[][]{vikhlinin09b}.\\
}
\end{table*}

\section{Conclusions}
We created sets of simulated images mimicking real HST/ACS cluster observations in order to test and better calibrate our implementation of the KSB+ algorithm which is used to measure the WL signals of clusters. As a first simplified setup we placed galaxies on a grid, using real catalogues analysed with KSB+ as comparison to determine the input parameters of the mock galaxies. With these grid simulations we studied the influence of choices for the input parameters, focussing on the galaxy properties while excluding the impact of neighbours. We found that the strongest impact comes from the distributions describing the synthetic galaxy light profiles and the details of the PSF shape. We gathered that the method yields results that are robust to $\lesssim 1\%$ for typical source populations, but note that extreme populations such as pure De Vaucouleurs profiles yield significant bias shifts. Similarly, we find a stability of the results to \mbox{$\Delta m \sim 0.01 $} when varying the PSF model from fairly realistic Tiny Tim ACS PSF models (including charge diffusion) to a Moffat PSF with similar half-light radius. Here we note that the resulting multiplicative bias is weakly dependent on the PSF ellipticity, likely reflecting limitations of the KSB+ algorithm. We checked that for the stronger shear regime ($|g|<0.4$) we do not find any non-linear deviations of the multiplicative bias, but we note a shear dependence of the additive bias at $|g|>0.2$. We studied the $S/N_{\mathrm{KSB}}$ and magnitude dependence of this bias, and confirmed the need for a $S/N_{\mathrm{KSB}}$-dependent correction (for which we provide updated fit parameters) in order to keep biases small at low signal-to-noise ratios. This correction reduces the bias to the $\lesssim 0.01$ level, also when pushing our signal-to-noise cuts to \mbox{$S/N_{\mathrm{flux}}>7$}, which opens up the possibility to include a larger number of galaxies in the analysis. Differences in the intrinsic ellipticity distribution were shown to have negligible impact for current cluster applications. 

We studied the influence of light contamination from cluster galaxies by creating two sets of simulations, one with background galaxies only and one which also includes cluster members and foreground galaxies. We generally find that the impact of such bright galaxies is small (\mbox{$\Delta m \lesssim 0.005$}), except for clusters at lower redshifts \mbox{$z<0.7$} and cluster centric distances $70^{\prime\prime}$--$100^{\prime\prime}$, for which we obtain
\mbox{$\Delta m = (0.0113\pm 0.0033)$}. The impact of selection bias with the addition of cluster members is also low for (\mbox{$z>0.7$}) clusters (\mbox{$\Delta m \sim 0.006$}).

Aiming to create more realistic scenarios which include the impact of neighbouring galaxies we generated mock CANDELS observations, which used actual galaxy positions and magnitudes to provide realistic clustering. This setup also allows us to constrain selection bias, for which we obtained a first estimate by measuring the average intrinsic input ellipticity of the detected and selected galaxies. Confirming results from \citet{Kannawadi}, we find that the object detection in \texttt{SExtractor} leads to a selection bias $m_\mathrm{sel}=-0.026$, which is however reduced to a net selection bias of $m_\mathrm{sel}=-0.014$ once additional cuts applied in our pipeline are taken into account. Since both effects do not separate perfectly, we obtained a joint correction for shape measurement and selection biases from the CANDELS-like simulations (not requiring matched galaxy pairs). Here we found that the dependence of the bias on \mbox{$S/N_\mathrm{KSB}$} is still well described by the calibration that was initially obtained for shape measurement bias using the grid-based simulations, if a constant bias offset is added. After applying this revised calibration the dependence of residual shape measurement bias on the galaxy flux radius is sufficiently small that it can be neglected for background-selected cluster weak lensing analyses. As final ingredient for our bias correction we obtained an estimate for the shape measurement
bias caused by very faint (\mbox{$V_{606}>27.5$}) neighbouring galaxies, following the analysis from \citet{Martinet}.

In summary,  we derived a conservative updated shear calibration for our HST-like cluster weak lensing analyses, considering many different sources of bias. 
The small remaining residual biases can be corrected for in the analysis of real data depending on the employed \mbox{$S/N_\mathrm{flux}$} cut (see Table \ref{tab:final}).
Our analysis suggests that the corrected shear estimates are accurate to 1.5\%, which is fully sufficient for current cluster studies. Our results also give us confidence to extend our current shear estimates to the inner cluster regions, where the lensing effect approaches the non-weak regime. The achieved accuracy does, however, not meet the stringent requirements of next generation surveys such as Euclid (see \citealp{Kohlinger}), which will require more careful matching of the simulation parameters and likely the use of more advanced shape measurement methods. Shear calibration campaigns of next generation WL surveys will need to include the cluster regime probed in this work in their image simulations, in particular accounting for cluster galaxy blending and non-weak shears, in order to achieve the accuracy requirements for their mass estimates.

We have not included magnification or flexion effects in this work, which could further impact the results. However, the impact of flexion should be small in the more relevant \mbox{$|g|<0.2$} regime.
Magnification affects the observed distributions of faint background sources, which can impact cluster weak lensing studies by changing the source redshift distribution (\citetalias{Schrabback_2018}).
The impact on shear estimation biases is however expected to be negligible at the accuracy level of our study (see also \citealp{Martinet}).

A set of HAWK-I-like image simulations of high-$z$ sources following the same approach showed consistent results with the use of the same $S/N_{\mathrm{KSB}}$-dependent correction and a residual bias below $0.5 \%$ (see Appendix \ref{sec:HAWKI}). This indicates that the results presented here can be applied to other high-$z$ cluster weak lensing observations and not only to HST data. In particular they can be implemented in future high-$z$ cluster studies using $K_\mathrm{s}$-band shape measurements from HAWK-I as pioneered by \citetalias{Schrabback_2018b}.

Finally, we used the corrections derived in this paper and the confirmation that systematics are well controlled also in the non-weak regime to obtain updated mass constraints for three relaxed clusters. In particular, we demonstrated that the inclusion of smaller scales (\mbox{$r>200$ kpc}, as enabled by our work), increases the mass signal-to-noise ratio \mbox{$S/N_\mathrm{mass,D19}=M_\mathrm{200c}/\Delta M_\mathrm{200c}$} by a factor of 1.38 on average compared to the default scale \mbox{$r>500$ kpc} used by \citetalias{Schrabback_2019}.

\section*{Acknowledgements}
This work is based on observations made with the NASA/ESA {\it Hubble Space
  Telescope}, using imaging data from the SPT follow-up GO programmes
12246 (PI: C.~Stubbs), 14352 (PI: J.~Hlavacek-Larrondo), and 
13412 (PI: Schrabback).
STScI is operated by the Association of Universities for Research in Astronomy, Inc. under NASA contract NAS 5-26555.
This work is also based on observations made with ESO Telescopes at the La Silla Paranal Observatory under programmes
088.A-0889 (PI: J.~Mohr) and  0100.A-0217 (PI: B.~Hern\'andez-Mart\'in).
PISCO observations are supported by NSF AST-1814719.  Argonne National Lab, a U.S. Department of Energy Office of Science Laboratory, is operated by UChicago Argonne LLC under contract no. DE-AC02-06CH11357.

The Bonn group acknowledges support from  the German Federal Ministry of Economics and Technology (BMWi) provided through DLR under projects  50 OR 1407,  50 OR 1610, 50 OR 1803, and 50 QE 1103, as well as support provided  by the Deutsche Forschungsgemeinschaft under project  SCHR 1400/3-1. TS acknowledges support through ERC H2020-COMPET-2017 project \#776247. HH acknowledges support from the Netherlands Organisation for Scientific Research (NWO) under grant numbers 639.043.512 and NM acknowledges support from a CNES fellowship. We want to thank Fatimah Raihan, Hannah Zohren, Diana Scognamiglio, Martin Sommer and Florian Kleinebreil for valuable discussions of this work.
We also thank Bryan Gillis and collaborators for providing updated Tiny Tim parameters prior to the submission of \citet{Gillis}. We thank Peter Schneider for useful comments on this manuscript.

\bibliographystyle{apj}
\bibliography{mybib}

\begin{thebibliography}{}
\expandafter\ifx\csname natexlab\endcsname\relax\def\natexlab#1{#1}\fi

\bibitem[{{Bacon} {et~al.}(2006){Bacon}, {Goldberg}, {Rowe}, \&
  {Taylor}}]{Bacon}
{Bacon}, D.~J., {Goldberg}, D.~M., {Rowe}, B.~T.~P., \& {Taylor}, A.~N. 2006,
  \mnras, 365, 414

\bibitem[{{Bartelmann} \& {Schneider}(2001)}]{Bartelmann}
{Bartelmann}, M., \& {Schneider}, P. 2001, \physrep, 340, 291

\bibitem[{{Beckwith} {et~al.}(2006){Beckwith}, {Stiavelli}, {Koekemoer},
  {Caldwell}, {Ferguson}, {Hook}, {Lucas}, {Bergeron}, {Corbin}, {Jogee},
  {Panagia}, {Robberto}, {Royle}, {Somerville}, \& {Sosey}}]{Beckwith}
{Beckwith}, S. V.~W., {Stiavelli}, M., {Koekemoer}, A.~M., {et~al.} 2006, \aj,
  132, 1729

\bibitem[{{Bernstein} {et~al.}(2016){Bernstein}, {Armstrong}, {Krawiec}, \&
  {March}}]{Bernstein}
{Bernstein}, G.~M., {Armstrong}, R., {Krawiec}, C., \& {March}, M.~C. 2016,
  \mnras, 459, 4467

\bibitem[{{Bertin} \& {Arnouts}(1996)}]{Bertin}
{Bertin}, E., \& {Arnouts}, S. 1996, AAP, 117, 393

\bibitem[{{Bleem} {et~al.}(2015){Bleem}, {Stalder}, {de Haan}, {Aird}, {Allen},
  {Applegate}, {Ashby}, {Bautz}, {Bayliss}, {Benson}, {Bocquet}, {Brodwin},
  {Carlstrom}, {Chang}, {Chiu}, {Cho}, {Clocchiatti}, {Crawford}, {Crites},
  {Desai}, {Dietrich}, {Dobbs}, {Foley}, {Forman}, {George}, {Gladders},
  {Gonzalez}, {Halverson}, {Hennig}, {Hoekstra}, {Holder}, {Holzapfel},
  {Hrubes}, {Jones}, {Keisler}, {Knox}, {Lee}, {Leitch}, {Liu}, {Lueker},
  {Luong-Van}, {Mantz}, {Marrone}, {McDonald}, {McMahon}, {Meyer}, {Mocanu},
  {Mohr}, {Murray}, {Padin}, {Pryke}, {Reichardt}, {Rest}, {Ruel}, {Ruhl},
  {Saliwanchik}, {Saro}, {Sayre}, {Schaffer}, {Schrabback}, {Shirokoff},
  {Song}, {Spieler}, {Stanford}, {Staniszewski}, {Stark}, {Story}, {Stubbs},
  {Vanderlinde}, {Vieira}, {Vikhlinin}, {Williamson}, {Zahn}, \&
  {Zenteno}}]{Bleem}
{Bleem}, L.~E., {Stalder}, B., {de Haan}, T., {et~al.} 2015, The Astrophysical
  Journal Supplement Series, 216, 27

\bibitem[{{Bleem} {et~al.}(2019){Bleem}, {Bocquet}, {Stalder}, {Gladders},
  {Ade}, {Allen}, {Anderson}, {Annis}, {Ashby}, {Austermann}, {Avila}, {Avva},
  {Bayliss}, {Beall}, {Bechtol}, {Bender}, {Benson}, {Bertin}, {Bianchini},
  {Blake}, {Brodwin}, {Brooks}, {Buckley-Geer}, {Burke}, {Carlstrom}, {Carnero
  Rosell}, {Carrasco Kind}, {Carretero}, {Chang}, {Chiang}, {Citron}, {Corbett
  Moran}, {Costanzi}, {Crawford}, {Crites}, {da Costa}, {de Haan}, {De
  Vicente}, {Desai}, {Diehl}, {Dietrich}, {Dobbs}, {Eifler}, {Everett},
  {Flaugher}, {Floyd}, {Frieman}, {Gallicchio}, {Garc{\'\i}a-Bellido},
  {George}, {Gerdes}, {Gilbert}, {Gruen}, {Gruendl}, {Gschwend}, {Gupta},
  {Gutierrez}, {Halverson}, {Harrington}, {Henning}, {Heymans}, {Holder},
  {Hollowood}, {Holzapfel}, {Honscheid}, {Hrubes}, {Huang}, {Hubmayr}, {Irwin},
  {James}, {Jeltema}, {Joudaki}, {Khullar}, {Klein}, {Knox}, {Kuropatkin},
  {Lee}, {Li}, {Lidman}, {Lowitz}, {MacCrann}, {Mahler}, {Maia}, {Marshall},
  {McDonald}, {McMahon}, {Melchior}, {Menanteau}, {Meyer}, {Miquel}, {Mocanu},
  {Mohr}, {Montgomery}, {Nadolski}, {Natoli}, {Nibarger}, {Noble}, {Novosad},
  {Padin}, {Palmese}, {Parkinson}, {Patil}, {Paz-Chinch{\'o}n}, {Plazas},
  {Pryke}, {Ramachandra}, {Reichardt}, {Remolina Gonz{\'a}lez}, {Romer},
  {Roodman}, {Ruhl}, {Rykoff}, {Saliwanchik}, {Sanchez}, {Saro}, {Sayre},
  {Schaffer}, {Schrabback}, {Serrano}, {Sharon}, {Sievers}, {Smecher}, {Smith},
  {Soares-Santos}, {Stark}, {Story}, {Suchyta}, {Tarle}, {Tucker},
  {Vanderlinde}, {Veach}, {Vieira}, {Wang}, {Weller}, {Whitehorn}, {Wu},
  {Yefremenko}, \& {Zhang}}]{Bleem_2019}
{Bleem}, L.~E., {Bocquet}, S., {Stalder}, B., {et~al.} 2019, \apj, submitted,
  arXiv:1910.04121

\bibitem[{{Bocquet} {et~al.}(2019){Bocquet}, {Dietrich}, {Schrabback}, {Bleem},
  {Klein}, {Allen}, {Applegate}, {Ashby}, {Bautz}, {Bayliss}, {Benson},
  {Brodwin}, {Bulbul}, {Canning}, {Capasso}, {Carlstrom}, {Chang}, {Chiu},
  {Cho}, {Clocchiatti}, {Crawford}, {Crites}, {de Haan}, {Desai}, {Dobbs},
  {Foley}, {Forman}, {Garmire}, {George}, {Gladders}, {Gonzalez}, {Grandis},
  {Gupta}, {Halverson}, {Hlavacek-Larrondo}, {Hoekstra}, {Holder}, {Holzapfel},
  {Hou}, {Hrubes}, {Huang}, {Jones}, {Khullar}, {Knox}, {Kraft}, {Lee}, {von
  der Linden}, {Luong-Van}, {Mantz}, {Marrone}, {McDonald}, {McMahon}, {Meyer},
  {Mocanu}, {Mohr}, {Morris}, {Padin}, {Patil}, {Pryke}, {Rapetti},
  {Reichardt}, {Rest}, {Ruhl}, {Saliwanchik}, {Saro}, {Sayre}, {Schaffer},
  {Shirokoff}, {Stalder}, {Stanford}, {Staniszewski}, {Stark}, {Story},
  {Strazzullo}, {Stubbs}, {Vanderlinde}, {Vieira}, {Vikhlinin}, {Williamson},
  \& {Zenteno}}]{Bocquet}
{Bocquet}, S., {Dietrich}, J.~P., {Schrabback}, T., {et~al.} 2019, \apj, 878,
  55

\bibitem[{{Bridle} {et~al.}(2010){Bridle}, {Balan}, {Bethge}, {Gentile},
  {Harmeling}, {Heymans}, {Hirsch}, {Hosseini}, {Jarvis}, {Kirk}, {Kitching},
  {Kuijken}, {Lewis}, {Paulin-Henriksson}, {Sch{\"o}lkopf}, {Velander},
  {Voigt}, {Witherick}, {Amara}, {Bernstein}, {Courbin}, {Gill}, {Heavens},
  {Mandelbaum}, {Massey}, {Moghaddam}, {Rassat}, {R{\'e}fr{\'e}gier}, {Rhodes},
  {Schrabback}, {Shawe-Taylor}, {Shmakova}, {van Waerbeke}, \&
  {Wittman}}]{Bridle_2010}
{Bridle}, S., {Balan}, S.~T., {Bethge}, M., {et~al.} 2010, \mnras, 405, 2044

\bibitem[{{Bridle} {et~al.}(2002){Bridle}, {Kneib}, {Bardeau}, {Gull}, \&
  {Natarajan}}]{Bridle}
{Bridle}, S.~L., {Kneib}, J.~P., {Bardeau}, S., {Gull}, S.~F., \& {Natarajan},
  P. 2002, in The Shapes of Galaxies and their Dark Halos, 38--46

\bibitem[{{Diemer} \& {Joyce}(2019)}]{Diemer_2019}
{Diemer}, B., \& {Joyce}, M. 2019, \apj, 871, 168

\bibitem[{{Dietrich} {et~al.}(2019){Dietrich}, {Bocquet}, {Schrabback},
  {Applegate}, {Hoekstra}, {Grandis}, {Mohr}, {Allen}, {Bayliss}, {Benson},
  {Bleem}, {Brodwin}, {Bulbul}, {Capasso}, {Chiu}, {Crawford}, {Gonzalez}, {de
  Haan}, {Klein}, {von der Linden}, {Mantz}, {Marrone}, {McDonald},
  {Raghunathan}, {Rapetti}, {Reichardt}, {Saro}, {Stalder}, {Stark}, {Stern},
  \& {Stubbs}}]{Dietrich}
{Dietrich}, J.~P., {Bocquet}, S., {Schrabback}, T., {et~al.} 2019, \mnras, 483,
  2871

\bibitem[{{Erben} {et~al.}(2001){Erben}, {Van Waerbeke}, {Bertin}, {Mellier},
  \& {Schneider}}]{Erben}
{Erben}, T., {Van Waerbeke}, L., {Bertin}, E., {Mellier}, Y., \& {Schneider},
  P. 2001, \aap, 366, 717

\bibitem[{{Euclid Collaboration: Martinet} {et~al.}(2019){Euclid Collaboration:
  Martinet}, {Schrabback}, {Hoekstra}, {Tewes}, {Herbonnet}, {Schneider},
  {Hernandez-Martin}, {Taylor}, {Brinchmann}, {Carvalho}, {Castellano},
  {Congedo}, {Gillis}, {Jullo}, {K{\"u}mmel}, {Ligori}, {Lilje}, {Padilla},
  {Paris}, {Peacock}, {Pilo}, {Pujol}, {Scott}, \& {Toledo-Moreo}}]{Martinet}
{Euclid Collaboration: Martinet}, N., {Schrabback}, T., {Hoekstra}, H.,
  {et~al.} 2019, \aap, 627, A59

\bibitem[{{Fenech Conti} {et~al.}(2017){Fenech Conti}, {Herbonnet}, {Hoekstra},
  {Merten}, {Miller}, \& {Viola}}]{Fenech-Conti}
{Fenech Conti}, I., {Herbonnet}, R., {Hoekstra}, H., {et~al.} 2017, \mnras,
  467, 1627

\bibitem[{{Gillis} {et~al.}(2019){Gillis}, {Schrabback}, {Maggraf},
  {Mandelbaum}, {Massey}, {Rhodes}, \& {Taylor}}]{Gillis}
{Gillis}, B., {Schrabback}, T., {Maggraf}, O., {et~al.} 2019, \mnras, submitted

\bibitem[{{Goldberg} \& {Bacon}(2005)}]{Goldberg}
{Goldberg}, D.~M., \& {Bacon}, D.~J. 2005, \apj, 619, 741

\bibitem[{{Goldberg} \& {Natarajan}(2002)}]{Goldberg_2002}
{Goldberg}, D.~M., \& {Natarajan}, P. 2002, \apj, 564, 65

\bibitem[{{Grogin} {et~al.}(2011){Grogin}, {Kocevski}, {Faber}, {Ferguson},
  {Koekemoer}, {Riess}, {Acquaviva}, {Alexander}, {Almaini}, {Ashby}, {Barden},
  {Bell}, {Bournaud}, {Brown}, {Caputi}, {Casertano}, {Cassata}, {Castellano},
  {Challis}, {Chary}, {Cheung}, {Cirasuolo}, {Conselice}, {Roshan Cooray},
  {Croton}, {Daddi}, {Dahlen}, {Dav{\'e}}, {de Mello}, {Dekel}, {Dickinson},
  {Dolch}, {Donley}, {Dunlop}, {Dutton}, {Elbaz}, {Fazio}, {Filippenko},
  {Finkelstein}, {Fontana}, {Gardner}, {Garnavich}, {Gawiser}, {Giavalisco},
  {Grazian}, {Guo}, {Hathi}, {H{\"a}ussler}, {Hopkins}, {Huang}, {Huang},
  {Jha}, {Kartaltepe}, {Kirshner}, {Koo}, {Lai}, {Lee}, {Li}, {Lotz}, {Lucas},
  {Madau}, {McCarthy}, {McGrath}, {McIntosh}, {McLure}, {Mobasher},
  {Moustakas}, {Mozena}, {Nandra}, {Newman}, {Niemi}, {Noeske}, {Papovich},
  {Pentericci}, {Pope}, {Primack}, {Rajan}, {Ravindranath}, {Reddy}, {Renzini},
  {Rix}, {Robaina}, {Rodney}, {Rosario}, {Rosati}, {Salimbeni}, {Scarlata},
  {Siana}, {Simard}, {Smidt}, {Somerville}, {Spinrad}, {Straughn}, {Strolger},
  {Telford}, {Teplitz}, {Trump}, {van der Wel}, {Villforth}, {Wechsler},
  {Weiner}, {Wiklind}, {Wild}, {Wilson}, {Wuyts}, {Yan}, \& {Yun}}]{Grogin}
{Grogin}, N.~A., {Kocevski}, D.~D., {Faber}, S.~M., {et~al.} 2011, \apjs, 197,
  35

\bibitem[{{Herbonnet} {et~al.}(2019){Herbonnet}, {von der Linden}, {Allen},
  {Mantz}, {Modumudi}, {Morris}, \& {Kelly}}]{Herbonnet}
{Herbonnet}, R., {von der Linden}, A., {Allen}, S.~W., {et~al.} 2019, \mnras,
  490, 4889

\bibitem[{{Heymans} {et~al.}(2005){Heymans}, {Brown}, {Barden}, {Caldwell},
  {Jahnke}, {Peng}, {Rix}, {Taylor}, {Beckwith}, {Bell}, {Borch},
  {H{\"a}u{\ss}ler}, {Jogee}, {McIntosh}, {Meisenheimer}, {S{\'a}nchez},
  {Somerville}, {Wisotzki}, \& {Wolf}}]{Heymans_2005}
{Heymans}, C., {Brown}, M.~L., {Barden}, M., {et~al.} 2005, \mnras, 361, 160

\bibitem[{{Heymans} {et~al.}(2006){Heymans}, {Van Waerbeke}, {Bacon}, {Berge},
  {Bernstein}, {Bertin}, {Bridle}, {Brown}, {Clowe}, {Dahle}, {Erben}, {Gray},
  {Hetterscheidt}, {Hoekstra}, {Hudelot}, {Jarvis}, {Kuijken}, {Margoniner},
  {Massey}, {Mellier}, {Nakajima}, {Refregier}, {Rhodes}, {Schrabback}, \&
  {Wittman}}]{Heymans}
{Heymans}, C., {Van Waerbeke}, L., {Bacon}, D., {et~al.} 2006, \mnras, 368,
  1323

\bibitem[{{Hoekstra} {et~al.}(1998){Hoekstra}, {Franx}, {Kuijken}, \&
  {Squires}}]{Hoekstra}
{Hoekstra}, H., {Franx}, M., {Kuijken}, K., \& {Squires}, G. 1998, ApJ, 504,
  636

\bibitem[{{Hoekstra} {et~al.}(2015){Hoekstra}, {Herbonnet}, {Muzzin}, {Babul},
  {Mahdavi}, {Viola}, \& {Cacciato}}]{Hoekstra_2015}
{Hoekstra}, H., {Herbonnet}, R., {Muzzin}, A., {et~al.} 2015, \mnras, 449, 685

\bibitem[{{Hoekstra} {et~al.}(2017){Hoekstra}, {Viola}, \&
  {Herbonnet}}]{Hoekstra_2017}
{Hoekstra}, H., {Viola}, M., \& {Herbonnet}, R. 2017, \mnras, 468, 3295

\bibitem[{{Kaiser} {et~al.}(1995){Kaiser}, {Squires}, \& {Broadhurst}}]{Kaiser}
{Kaiser}, N., {Squires}, G., \& {Broadhurst}, T. 1995, ApJ, 449, 460

\bibitem[{{Kannawadi} {et~al.}(2019){Kannawadi}, {Hoekstra}, {Miller}, {Viola},
  {Fenech Conti}, {Herbonnet}, {Erben}, {Heymans}, {Hildebrandt}, {Kuijken},
  {Vakili}, \& {Wright}}]{Kannawadi}
{Kannawadi}, A., {Hoekstra}, H., {Miller}, L., {et~al.} 2019, \aap, 624, A92

\bibitem[{{Koekemoer} {et~al.}(2011){Koekemoer}, {Faber}, {Ferguson}, {Grogin},
  {Kocevski}, {Koo}, {Lai}, {Lotz}, {Lucas}, {McGrath}, {Ogaz}, {Rajan},
  {Riess}, {Rodney}, {Strolger}, {Casertano}, {Castellano}, {Dahlen},
  {Dickinson}, {Dolch}, {Fontana}, {Giavalisco}, {Grazian}, {Guo}, {Hathi},
  {Huang}, {van der Wel}, {Yan}, {Acquaviva}, {Alexander}, {Almaini}, {Ashby},
  {Barden}, {Bell}, {Bournaud}, {Brown}, {Caputi}, {Cassata}, {Challis},
  {Chary}, {Cheung}, {Cirasuolo}, {Conselice}, {Roshan Cooray}, {Croton},
  {Daddi}, {Dav{\'e}}, {de Mello}, {de Ravel}, {Dekel}, {Donley}, {Dunlop},
  {Dutton}, {Elbaz}, {Fazio}, {Filippenko}, {Finkelstein}, {Frazer}, {Gardner},
  {Garnavich}, {Gawiser}, {Gruetzbauch}, {Hartley}, {H{\"a}ussler},
  {Herrington}, {Hopkins}, {Huang}, {Jha}, {Johnson}, {Kartaltepe},
  {Khostovan}, {Kirshner}, {Lani}, {Lee}, {Li}, {Madau}, {McCarthy},
  {McIntosh}, {McLure}, {McPartland}, {Mobasher}, {Moreira}, {Mortlock},
  {Moustakas}, {Mozena}, {Nandra}, {Newman}, {Nielsen}, {Niemi}, {Noeske},
  {Papovich}, {Pentericci}, {Pope}, {Primack}, {Ravindranath}, {Reddy},
  {Renzini}, {Rix}, {Robaina}, {Rosario}, {Rosati}, {Salimbeni}, {Scarlata},
  {Siana}, {Simard}, {Smidt}, {Snyder}, {Somerville}, {Spinrad}, {Straughn},
  {Telford}, {Teplitz}, {Trump}, {Vargas}, {Villforth}, {Wagner}, {Wandro},
  {Wechsler}, {Weiner}, {Wiklind}, {Wild}, {Wilson}, {Wuyts}, \&
  {Yun}}]{Koekemoer}
{Koekemoer}, A.~M., {Faber}, S.~M., {Ferguson}, H.~C., {et~al.} 2011, \apjs,
  197, 36

\bibitem[{{K{\"o}hlinger} {et~al.}(2015){K{\"o}hlinger}, {Hoekstra}, \&
  {Eriksen}}]{Kohlinger}
{K{\"o}hlinger}, F., {Hoekstra}, H., \& {Eriksen}, M. 2015, \mnras, 453, 3107

\bibitem[{{Krist}(2003)}]{Krist_2003}
{Krist}, J. 2003, {ACS WFC; HRC field dependent PSF variations due to optical
  and charge diffusion effects}, Tech. rep.

\bibitem[{{Krist} {et~al.}(2011){Krist}, {Hook}, \& {Stoehr}}]{Krist}
{Krist}, J.~E., {Hook}, R.~N., \& {Stoehr}, F. 2011, in Optical Modeling and
  Performance Predictions V, Vol. 8127, 81270J

\bibitem[{{Luppino} \& {Kaiser}(1997)}]{Luppino}
{Luppino}, G.~A., \& {Kaiser}, N. 1997, ApJ, 475, 20

\bibitem[{{Mandelbaum} {et~al.}(2012){Mandelbaum}, {Hirata}, {Leauthaud},
  {Massey}, \& {Rhodes}}]{Mandelbaum_2012}
{Mandelbaum}, R., {Hirata}, C.~M., {Leauthaud}, A., {Massey}, R.~J., \&
  {Rhodes}, J. 2012, \mnras, 420, 1518

\bibitem[{{Mandelbaum} {et~al.}(2015){Mandelbaum}, {Rowe}, {Armstrong}, {Bard},
  {Bertin}, {Bosch}, {Boutigny}, {Courbin}, {Dawson}, {Donnarumma}, {Fenech
  Conti}, {Gavazzi}, {Gentile}, {Gill}, {Hogg}, {Huff}, {Jee}, {Kacprzak},
  {Kilbinger}, {Kuntzer}, {Lang}, {Luo}, {March}, {Marshall}, {Meyers},
  {Miller}, {Miyatake}, {Nakajima}, {Ngol{\'e} Mboula}, {Nurbaeva}, {Okura},
  {Paulin-Henriksson}, {Rhodes}, {Schneider}, {Shan}, {Sheldon}, {Simet},
  {Starck}, {Sureau}, {Tewes}, {Zarb Adami}, {Zhang}, \&
  {Zuntz}}]{Mandelbaum_2015}
{Mandelbaum}, R., {Rowe}, B., {Armstrong}, R., {et~al.} 2015, \mnras, 450, 2963

\bibitem[{{Mandelbaum} {et~al.}(2018){Mandelbaum}, {Lanusse}, {Leauthaud},
  {Armstrong}, {Simet}, {Miyatake}, {Meyers}, {Bosch}, {Murata}, {Miyazaki}, \&
  {Tanaka}}]{Mandelbaum_2018}
{Mandelbaum}, R., {Lanusse}, F., {Leauthaud}, A., {et~al.} 2018, \mnras, 481,
  3170

\bibitem[{{Massey} {et~al.}(2007){Massey}, {Heymans}, {Berg{\'e}}, {Bernstein},
  {Bridle}, {Clowe}, {Dahle}, {Ellis}, {Erben}, {Hetterscheidt}, {High},
  {Hirata}, {Hoekstra}, {Hudelot}, {Jarvis}, {Johnston}, {Kuijken},
  {Margoniner}, {Mandelbaum}, {Mellier}, {Nakajima}, {Paulin-Henriksson},
  {Peeples}, {Roat}, {Refregier}, {Rhodes}, {Schrabback}, {Schirmer}, {Seljak},
  {Semboloni}, \& {van Waerbeke}}]{Massey}
{Massey}, R., {Heymans}, C., {Berg{\'e}}, J., {et~al.} 2007, \mnras, 376, 13

\bibitem[{{McDonald} {et~al.}(2019){McDonald}, {Allen}, {Hlavacek-Larrondo},
  {Mantz}, {Bayliss}, {Benson}, {Brodwin}, {Bulbul}, {Canning}, {Chiu},
  {Forman}, {Garmire}, {Gupta}, {Khullar}, {Mohr}, {Reichardt}, \&
  {Schrabback}}]{mcdonald19}
{McDonald}, M., {Allen}, S.~W., {Hlavacek-Larrondo}, J., {et~al.} 2019, \apj,
  870, 85

\bibitem[{{Miller} {et~al.}(2007){Miller}, {Kitching}, {Heymans}, {Heavens}, \&
  {van Waerbeke}}]{Miller}
{Miller}, L., {Kitching}, T.~D., {Heymans}, C., {Heavens}, A.~F., \& {van
  Waerbeke}, L. 2007, \mnras, 382, 315

\bibitem[{{Navarro} {et~al.}(1997){Navarro}, {Frenk}, \& {White}}]{Navarro}
{Navarro}, J.~F., {Frenk}, C.~S., \& {White}, S. D.~M. 1997, \apj, 490, 493

\bibitem[{{Neto} {et~al.}(2007){Neto}, {Gao}, {Bett}, {Cole}, {Navarro},
  {Frenk}, {White}, {Springel}, \& {Jenkins}}]{Neto}
{Neto}, A.~F., {Gao}, L., {Bett}, P., {et~al.} 2007, \mnras, 381, 1450

\bibitem[{{Pujol} {et~al.}(2019){Pujol}, {Kilbinger}, {Sureau}, \&
  {Bobin}}]{Pujol}
{Pujol}, A., {Kilbinger}, M., {Sureau}, F., \& {Bobin}, J. 2019, \aap, 621, A2

\bibitem[{{Raihan} {et~al.}({2020}){Raihan}, {Schrabback}, {Hildebrandt},
  {Applegate}, \& {Mahler}}]{Raihan_2019}
{Raihan}, S.~F., {Schrabback}, T., {Hildebrandt}, H., {Applegate}, D., \&
  {Mahler}, G. {2020}, \mnras, submitted

\bibitem[{{Refregier} {et~al.}(2012){Refregier}, {Kacprzak}, {Amara}, {Bridle},
  \& {Rowe}}]{Refregier}
{Refregier}, A., {Kacprzak}, T., {Amara}, A., {Bridle}, S., \& {Rowe}, B. 2012,
  \mnras, 425, 1951

\bibitem[{{Rhodes} {et~al.}(2006){Rhodes}, {Massey}, {Albert}, {Taylor},
  {Koekemoer}, \& {Leauthaud}}]{Rhodes}
{Rhodes}, J.~D., {Massey}, R., {Albert}, J., {et~al.} 2006, in The 2005 HST
  Calibration Workshop: Hubble After the Transition to Two-Gyro Mode, ed. A.~M.
  {Koekemoer}, P.~{Goudfrooij}, \& L.~L. {Dressel}, 21

\bibitem[{{Rowe} {et~al.}(2015){Rowe}, {Jarvis}, {Mandelbaum}, {Bernstein},
  {Bosch}, {Simet}, {Meyers}, {Kacprzak}, {Nakajima}, {Zuntz}, {Miyatake},
  {Dietrich}, {Armstrong}, {Melchior}, \& {Gill}}]{Rowe}
{Rowe}, B.~T.~P., {Jarvis}, M., {Mandelbaum}, R., {et~al.} 2015, Astronomy and
  Computing, 10, 121

\bibitem[{{Samuroff} {et~al.}(2018){Samuroff}, {Bridle}, {Zuntz}, {Troxel},
  {Gruen}, {Rollins}, {Bernstein}, {Eifler}, {Huff}, {Kacprzak}, {Krause},
  {MacCrann}, {Abdalla}, {Allam}, {Annis}, {Bechtol}, {Benoit-L{\'e}vy},
  {Bertin}, {Brooks}, {Buckley-Geer}, {Carnero Rosell}, {Carrasco Kind},
  {Carretero}, {Crocce}, {D'Andrea}, {da Costa}, {Davis}, {Desai}, {Doel},
  {Fausti Neto}, {Flaugher}, {Fosalba}, {Frieman}, {Garc{\'\i}a-Bellido},
  {Gerdes}, {Gruendl}, {Gschwend}, {Gutierrez}, {Honscheid}, {James}, {Jarvis},
  {Jeltema}, {Kirk}, {Kuehn}, {Kuhlmann}, {Li}, {Lima}, {Maia}, {March},
  {Marshall}, {Martini}, {Melchior}, {Menanteau}, {Miquel}, {Nord}, {Ogando},
  {Plazas}, {Roodman}, {Sanchez}, {Scarpine}, {Schindler}, {Schubnell},
  {Sevilla-Noarbe}, {Sheldon}, {Smith}, {Soares-Santos}, {Sobreira}, {Suchyta},
  {Tarle}, {Thomas}, {Tucker}, \& {DES Collaboration}}]{Samuroff}
{Samuroff}, S., {Bridle}, S.~L., {Zuntz}, J., {et~al.} 2018, \mnras, 475, 4524

\bibitem[{{Schrabback}({et al.~in prep.})}]{Schrabback_2019}
{Schrabback}, T. {et al.~in prep.}

\bibitem[{{Schrabback} {et~al.}(2007){Schrabback}, {Erben}, {Simon},
  {Miralles}, {Schneider}, {Heymans}, {Eifler}, {Fosbury}, {Freudling},
  {Hetterscheidt}, {Hildebrandt}, \& {Pirzkal}}]{Schrabback_2007}
{Schrabback}, T., {Erben}, T., {Simon}, P., {et~al.} 2007, \aap, 468, 823

\bibitem[{{Schrabback} {et~al.}(2010){Schrabback}, {Hartlap}, {Joachimi},
  {Kilbinger}, {Simon}, {Benabed}, {Brada{\v c}}, {Eifler}, {Erben},
  {Fassnacht}, {High}, {Hilbert}, {Hildebrandt}, {Hoekstra}, {Kuijken},
  {Marshall}, {Mellier}, {Morganson}, {Schneider}, {Semboloni}, {van Waerbeke},
  \& {Velander}}]{Schrabback}
{Schrabback}, T., {Hartlap}, J., {Joachimi}, B., {et~al.} 2010, AAP, 516, A63

\bibitem[{{Schrabback} {et~al.}(2018{\natexlab{a}}){Schrabback}, {Applegate},
  {Dietrich}, {Hoekstra}, {Bocquet}, {Gonzalez}, {von der Linden}, {McDonald},
  {Morrison}, {Raihan}, {Allen}, {Bayliss}, {Benson}, {Bleem}, {Chiu}, {Desai},
  {Foley}, {de Haan}, {High}, {Hilbert}, {Mantz}, {Massey}, {Mohr},
  {Reichardt}, {Saro}, {Simon}, {Stern}, {Stubbs}, \&
  {Zenteno}}]{Schrabback_2018}
{Schrabback}, T., {Applegate}, D., {Dietrich}, J.~P., {et~al.}
  2018{\natexlab{a}}, \mnras, 474, 2635

\bibitem[{{Schrabback} {et~al.}(2018{\natexlab{b}}){Schrabback}, {Schirmer},
  {van der Burg}, {Hoekstra}, {Buddendiek}, {Applegate}, {Brada{\v{c}}},
  {Eifler}, {Erben}, {Gladders}, {Hern{\'a}ndez- Mart{\'\i}n}, {Hildebrandt},
  {Hoag}, {Klaes}, {von der Linden}, {Marchesini}, {Muzzin}, {Sharon}, \&
  {Stefanon}}]{Schrabback_2018b}
{Schrabback}, T., {Schirmer}, M., {van der Burg}, R. F.~J., {et~al.}
  2018{\natexlab{b}}, \aap, 610, A85

\bibitem[{{Sheldon} \& {Huff}(2017)}]{Sheldon}
{Sheldon}, E.~S., \& {Huff}, E.~M. 2017, \apj, 841, 24

\bibitem[{{Skelton} {et~al.}(2014){Skelton}, {Whitaker}, {Momcheva}, {Brammer},
  {van Dokkum}, {Labb{\'e}}, {Franx}, {van der Wel}, {Bezanson}, {Da Cunha},
  {Fumagalli}, {F{\"o}rster Schreiber}, {Kriek}, {Leja}, {Lundgren}, {Magee},
  {Marchesini}, {Maseda}, {Nelson}, {Oesch}, {Pacifici}, {Patel}, {Price},
  {Rix}, {Tal}, {Wake}, \& {Wuyts}}]{Skelton}
{Skelton}, R.~E., {Whitaker}, K.~E., {Momcheva}, I.~G., {et~al.} 2014, \apjs,
  214, 24

\bibitem[{{Stalder} {et~al.}(2014){Stalder}, {Stark}, {Amato}, {Geary},
  {Shectman}, {Stubbs}, \& {Szentgyorgyi}}]{Stalder}
{Stalder}, B., {Stark}, A.~A., {Amato}, S.~M., {et~al.} 2014, in Society of
  Photo-Optical Instrumentation Engineers (SPIE) Conference Series, Vol. 9147,
  Ground-based and Airborne Instrumentation for Astronomy V, 91473Y

\bibitem[{{Sunyaev} \& {Zeldovich}(1969)}]{Sunyaev}
{Sunyaev}, R.~A., \& {Zeldovich}, Y.~B. 1969, \nat, 223, 721

\bibitem[{{Tewes} {et~al.}(2019){Tewes}, {Kuntzer}, {Nakajima}, {Courbin},
  {Hildebrandt}, \& {Schrabback}}]{Tewes}
{Tewes}, M., {Kuntzer}, T., {Nakajima}, R., {et~al.} 2019, \aap, 621, A36

\bibitem[{{van der Wel} {et~al.}(2012){van der Wel}, {Bell}, {H{\"a}ussler},
  {McGrath}, {Chang}, {Guo}, {McIntosh}, {Rix}, {Barden}, {Cheung}, {Faber},
  {Ferguson}, {Galametz}, {Grogin}, {Hartley}, {Kartaltepe}, {Kocevski},
  {Koekemoer}, {Lotz}, {Mozena}, {Peth}, \& {Peng}}]{van_der_Wel_2012}
{van der Wel}, A., {Bell}, E.~F., {H{\"a}ussler}, B., {et~al.} 2012, \apjs,
  203, 24

\bibitem[{{van der Wel} {et~al.}(2014){van der Wel}, {Franx}, {van Dokkum},
  {Skelton}, {Momcheva}, {Whitaker}, {Brammer}, {Bell}, {Rix}, {Wuyts},
  {Ferguson}, {Holden}, {Barro}, {Koekemoer}, {Chang}, {McGrath},
  {H{\"a}ussler}, {Dekel}, {Behroozi}, {Fumagalli}, {Leja}, {Lundgren},
  {Maseda}, {Nelson}, {Wake}, {Patel}, {Labb{\'e}}, {Faber}, {Grogin}, \&
  {Kocevski}}]{van_der_Wel_2014}
{van der Wel}, A., {Franx}, M., {van Dokkum}, P.~G., {et~al.} 2014, \apj, 788,
  28

\bibitem[{{Vikhlinin} {et~al.}(2009){Vikhlinin}, {Burenin}, {Ebeling},
  {Forman}, {Hornstrup}, {Jones}, {Kravtsov}, {Murray}, {Nagai}, {Quintana}, \&
  {Voevodkin}}]{vikhlinin09b}
{Vikhlinin}, A., {Burenin}, R.~A., {Ebeling}, H., {et~al.} 2009, \apj, 692,
  1033

\bibitem[{{Viola} {et~al.}(2014){Viola}, {Kitching}, \& {Joachimi}}]{Viola}
{Viola}, M., {Kitching}, T.~D., \& {Joachimi}, B. 2014, \mnras, 439, 1909

\bibitem[{{von der Linden} {et~al.}(2014){von der Linden}, {Allen},
  {Applegate}, {Kelly}, {Allen}, {Ebeling}, {Burchat}, {Burke}, {Donovan},
  {Morris}, {Blandford}, {Erben}, \& {Mantz}}]{Von_der_linden}
{von der Linden}, A., {Allen}, M.~T., {Applegate}, D.~E., {et~al.} 2014,
  \mnras, 439, 2

\bibitem[{{Wright} \& {Brainerd}(2000)}]{Brainerd}
{Wright}, C.~O., \& {Brainerd}, T.~G. 2000, \apj, 534, 34

\bibitem[{{Zhang} {et~al.}(2007){Zhang}, {Finoguenov}, {B{\"o}hringer},
  {Kneib}, {Smith}, {Czoske}, \& {Soucail}}]{Zhang}
{Zhang}, Y.~Y., {Finoguenov}, A., {B{\"o}hringer}, H., {et~al.} 2007, \aap,
  467, 437

\end{thebibliography}

\appendix

\section{VLT/HAWK-I setup} \label{sec:HAWKI}
\begin{figure*}[t]
  \centering
 \begin{minipage}{0.45\linewidth}
  
  \includegraphics[width=\linewidth]{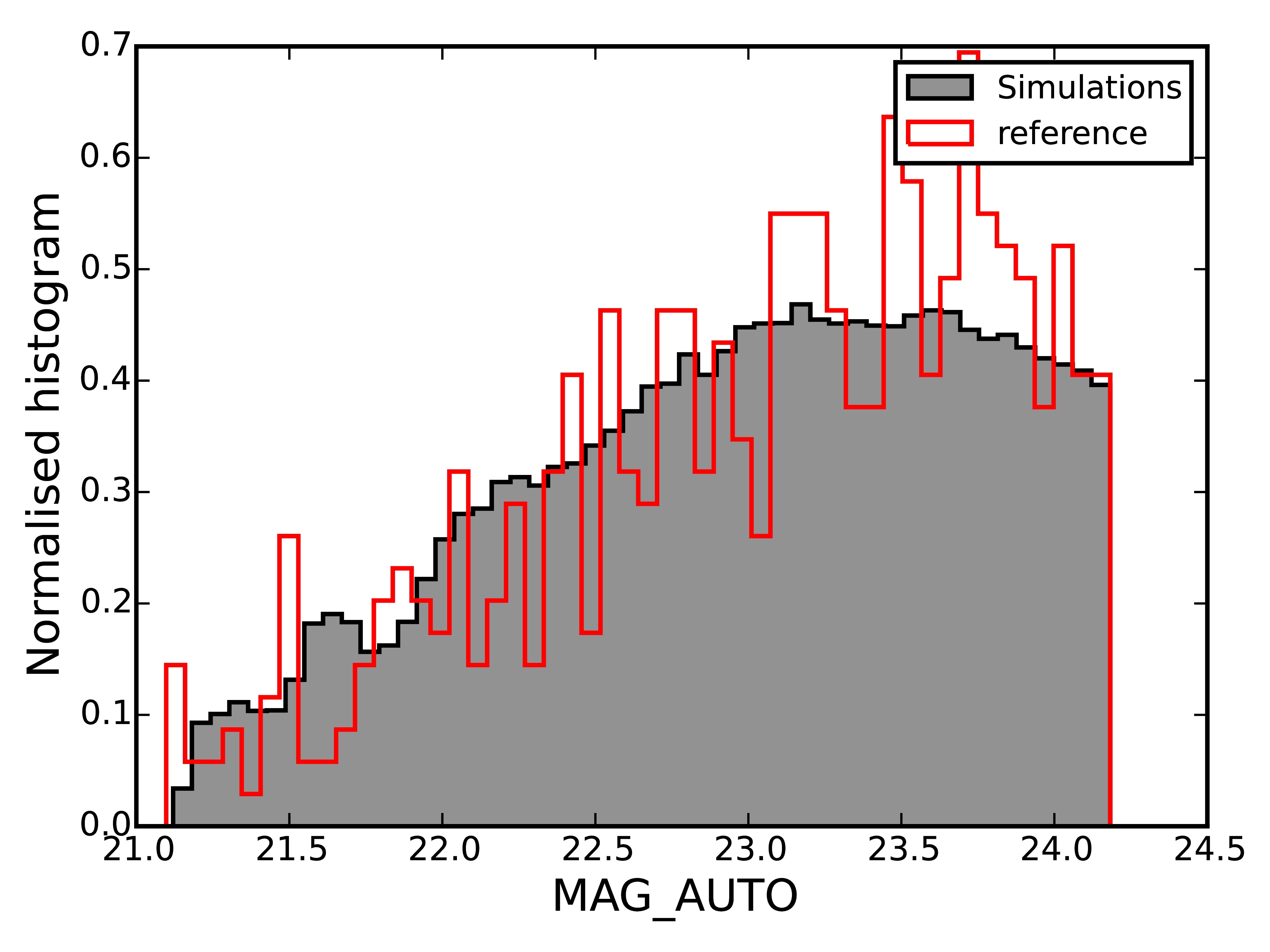}

  \end{minipage}\begin{minipage}{0.45\linewidth}
 
  \includegraphics[width=\linewidth]{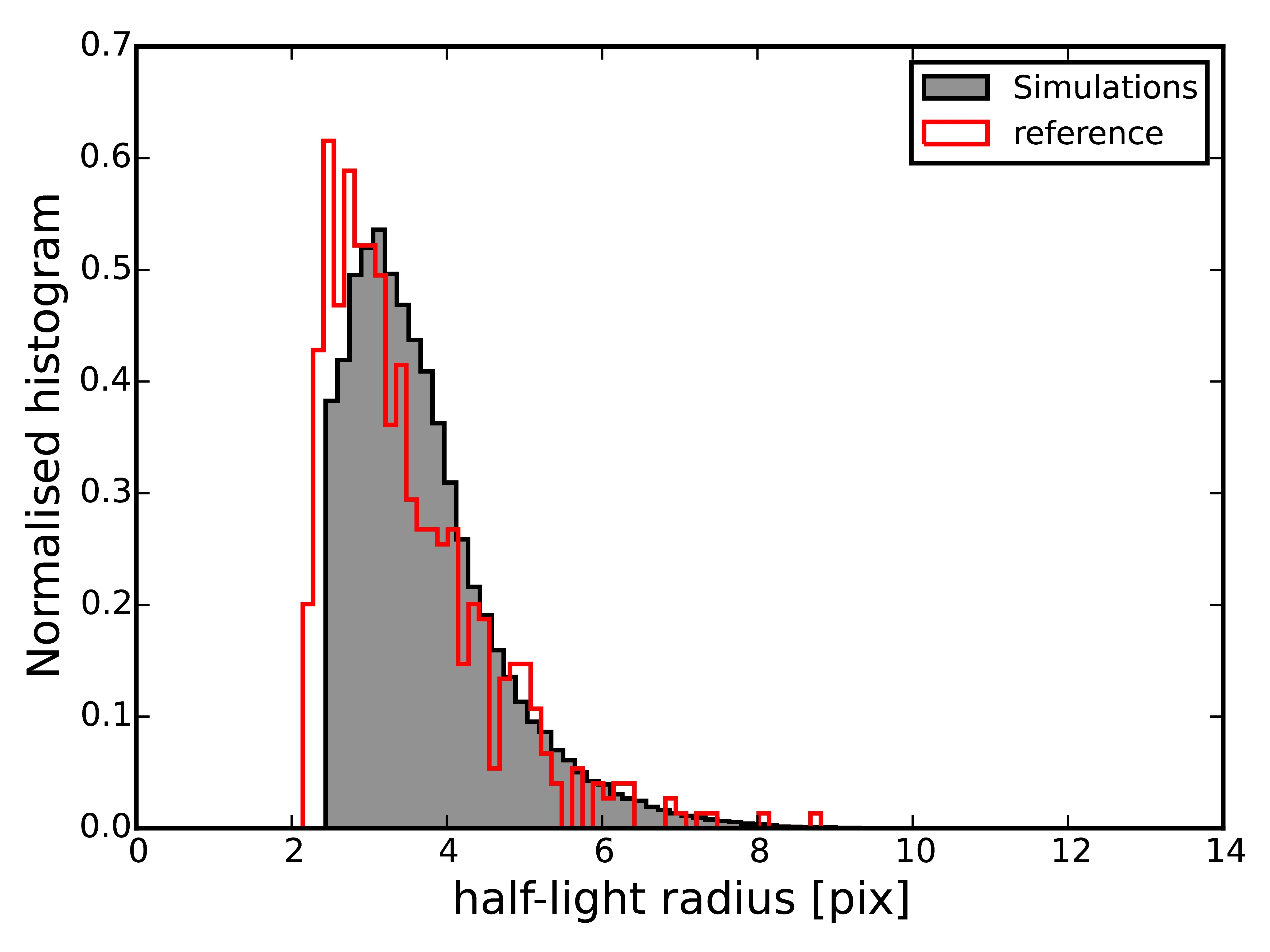}

  \end{minipage}
  \begin{minipage}{0.45\linewidth}
 
  \includegraphics[width=\linewidth]{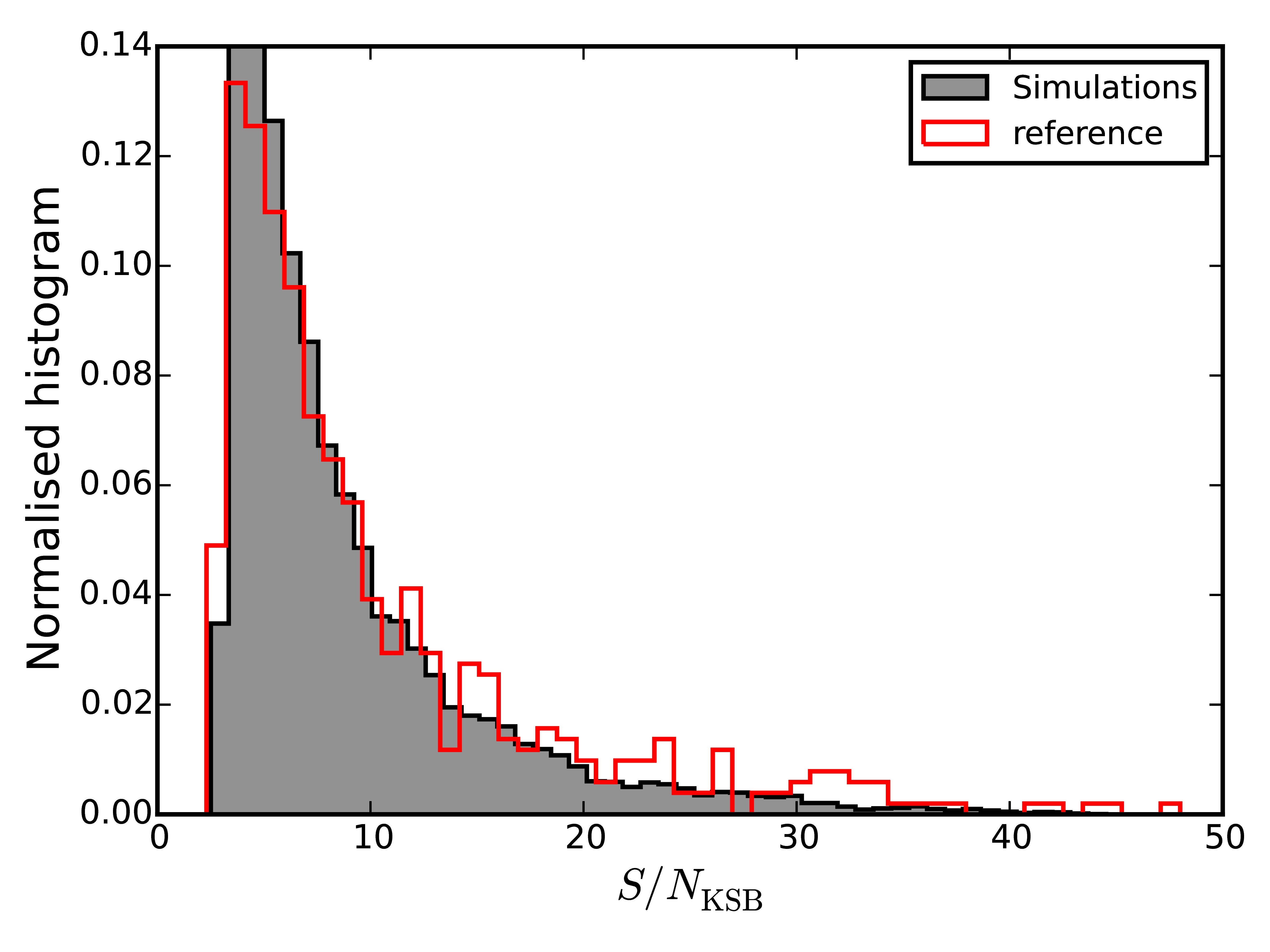}

  \end{minipage}
  \begin{minipage}{0.45\linewidth}
 
  \includegraphics[width=\linewidth]{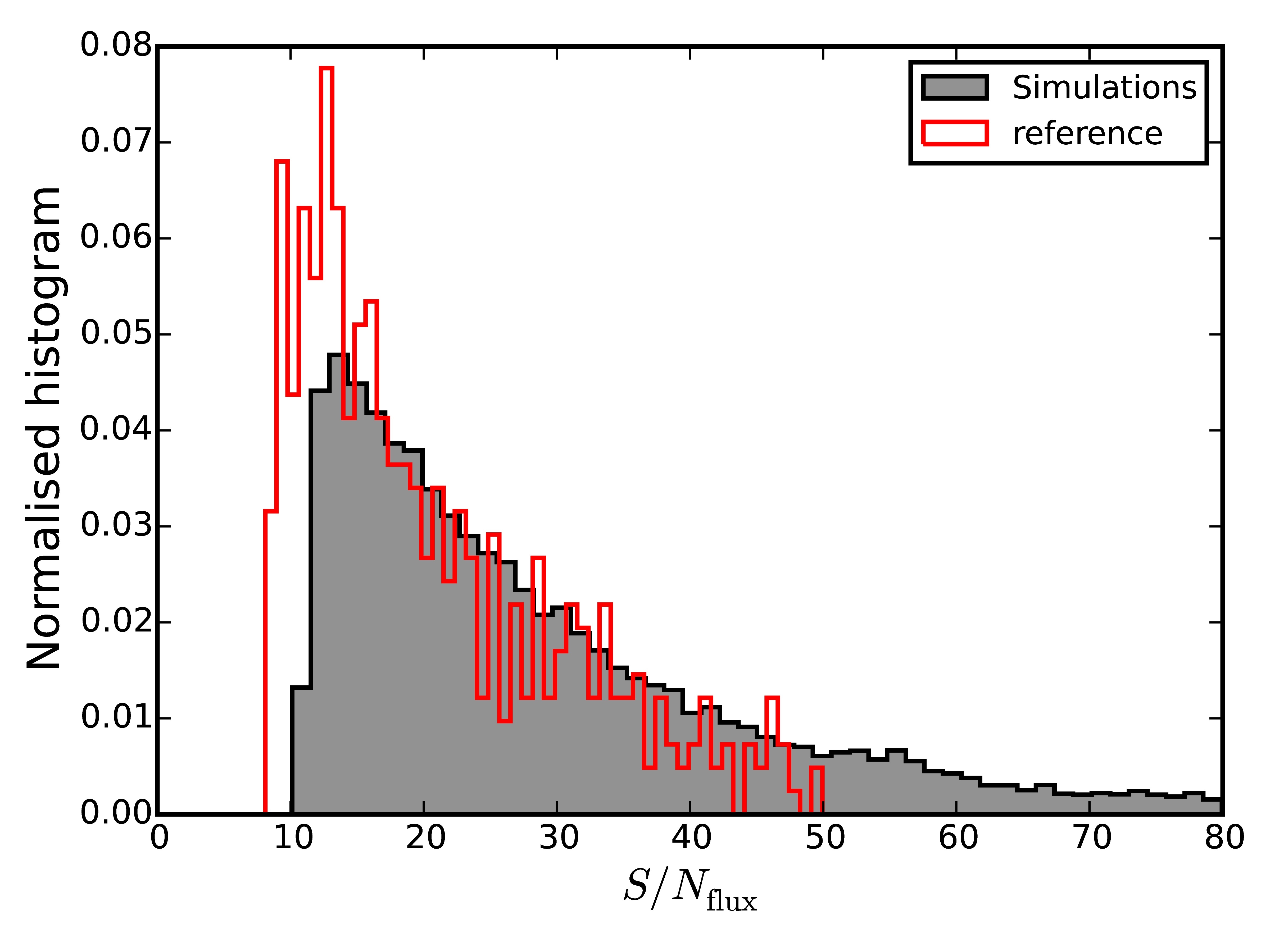}

  \end{minipage}
  \caption{Comparisons between the measured distributions in the HAWK-I-like simulations and the real image distribution for the $K_{\mathrm{s}}$ magnitudes (MAG\_AUTO), the half-light radius measured by \texttt{SExtractor}, the KSB signal-to-noise ratio $S/N_{\text{KSB}}$ and the \texttt{SExtractor} $S/N_{\text{flux}}$.} \label{fig:hawki}
 \end{figure*}

 \begin{figure}[t]

 \includegraphics[width=\linewidth]{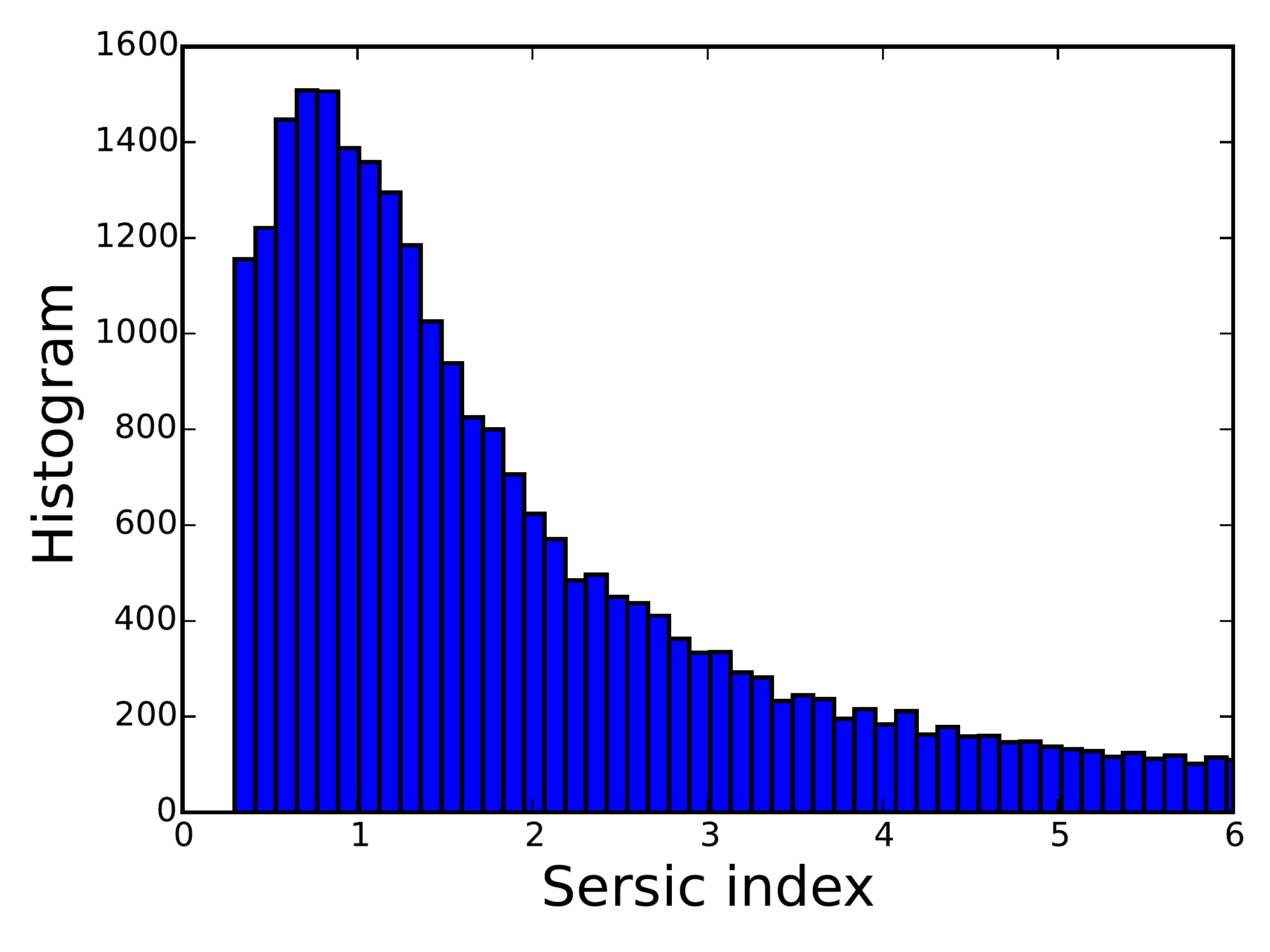}
 \caption{Distribution of S\'ersic indices in the parametric fit to real CANDELS galaxies in the F160W filter \citep{van_der_Wel_2014} used as input for the HAWK-I simulations.} \label{fig:hawki_sersic_index}
  
 \end{figure}

S18b demonstrate that good-seeing VLT/HAWK-I $K_\mathrm{s}$ images, which benefit from a sharp PSF (\mbox{$\mathrm{FWHM}\le 0\farcs4$}),  yield a similar sensitivity and efficiency for weak lensing studies of massive clusters at redshift \mbox{$0.7\lesssim z \lesssim 1$} as HST/ACS observations when combined with a background selection in $gzK_\mathrm{s}$ colour-colour space.
For their HAWK-I weak lensing analysis of the cluster RCS2{\thinspace}$J$232727.7$-$020437
they employed the same KSB+ pipeline as used in our study.
In order to provide a robust KSB+ calibration also for future HAWK-I high-$z$ cluster weak lensing studies of expanded samples, we create an additional set of simulations.
Here we mimic the HAWK-I setup and employ near-infrared-measured galaxy properties, but otherwise largely follow the procedures described in this paper for the generation and analysis of HST-like image simulations. 

To set up HAWK-I-like mock galaxies, we used the 3D-HST CANDELS light profile fits (\citealp{van_der_Wel_2012}; \citealp{van_der_Wel_2014}) as a reference, employing observations obtained in the F160W band, which is the closest available HST filter to the HAWK-I $K_\mathrm{s}$-band. The S\'ersic index distribution is shown in Fig.\thinspace\ref{fig:hawki_sersic_index}. Small differences in this distribution, which for example occur if only colour-selected galaxies are used rather than the full sample change the measured bias at the $\sim 0.5 \%$ level  only. This also indicates that the differences in the bias caused by the use of a slightly different band should be negligible. We use a similar setup as in Sect.\thinspace\ref{sec:grid}, where the galaxies were placed on a grid in order to test the behaviour of the $S/N_{\textrm{KSB}}$-dependent correction, and a more realistic scenario, using positions and magnitudes from the 3D-HST CANDELS catalogues to estimate the influence neighbours have on the bias. Following \citetalias{Schrabback_2018b} we mimick colour-selected  ($(z-K_{\mathrm{s}})>\mathrm{MIN}[g-z,2.5]$) galaxies up to $K_\mathrm{s}=24.2$. The main changes with respect to the HST-like simulations are those which are intrinsically dependent of the instrument ($0\farcs106$ pixel scale, $0\farcs4$ FWHM Moffat PSF).

In order to understand how realistic the signal-to-noise ratios (both $S/N_{\textrm{KSB}}$ and $S/N_{\textrm{flux}}$), as well as the galaxy size and $K_\mathrm{s}$ magnitude distributions are, we compared the output distributions to existing HAWK-I cluster observations analysed in \citetalias{Schrabback_2018b} with the same KSB+ implementation, as shown in Fig.\thinspace\ref{fig:hawki}, finding good agreement. 

After applying the $S/N_{\text{KSB}}$-dependent correction obtained in Eq.\thinspace(\ref{eq:sn_corr}) and the appropriate cuts, we obtained consistent results of $m_1=-0.0020 \pm 0.0039$ and $m_2=-0.0019 \pm 0.0033$ for $S/N_{\textrm{flux}}>10$ galaxies placed on a grid. 
This indicates that the $S/N_{\text{KSB}}$-dependent correction obtained for HST/ACS images is also valid for HAWK-I cluster analyses.

Likewise, very similar estimates are obtained using the CANDELS-like setup with realistic clustering: 
an estimation of the selection bias similar to Sect.\thinspace\ref{sec:selection_bias} yields a value of $m_1=-0.0135 \pm 0.0032$ and $m_2=-0.0141 \pm 0.0027$, in excellent agreement with the ACS-like analysis (compare Sect.\thinspace\ref{se:selection_bias_after_ksb}). 
Applying the calibration for both shape measurement and selection bias from Eq.\thinspace(\ref{eq:sn_corr_final}) to a HAWK-I-like version of the simulations described in  Sect.\thinspace\ref{sec:realistic}, we find residual biases of  $m_1=0.0082 \pm 0.0025$ ($0.0076 \pm 0.0030$) and $m_2=0.0088 \pm 0.0028$
($0.0069 \pm 0.0032$) 
for galaxy samples with $S/N_{\textrm{flux}}>10$ ($S/N_{\textrm{flux}}>7$).
The 3D-HST CANDELS input catalogues are only complete to $H_{160}\simeq 25$. This is approximately 1 magnitude deeper than the main weak lensing source sample, similar to the setup from Sect.\thinspace\ref{sec:realistic}.
Based on the results from Sect.\thinspace\ref{sec:faint} we therefore expect that even fainter galaxies lead to a further residual bias shift that is consistent with the entries in the second row of Table \ref{tab:final}.

\section{Simulated cluster images}
\begin{figure*}[t]
   \centering
   \includegraphics[width=\linewidth]{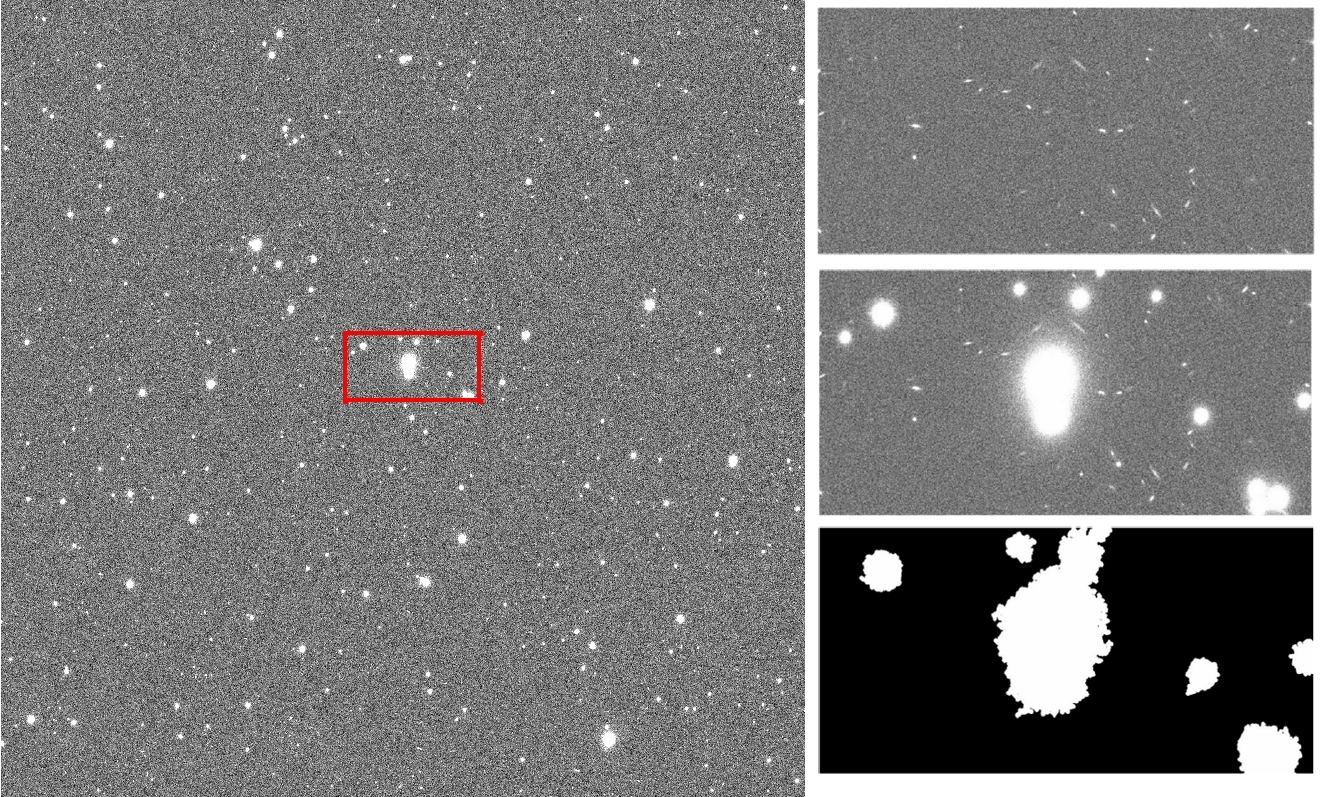}
   \caption{Example image of a simulated cluster at $z=0.28$. A cut-out of the full image, shown in red, can be seen in the right for the simulations with background galaxies only (top), with added cluster members (middle) and showing the mask used to remove bright objects (bottom). The full image and cut outs spans $300 '' \times 300''$ and $50'' \times 25''$, respectively. } \label{fig:028_cluster}
 \end{figure*}

As a comparison, we show the same image as for Fig.\thinspace\ref{fig:examples}, but for a simulated cluster at $z=0.28$ in Fig.\thinspace\ref{fig:028_cluster}, featuring much brighter cluster galaxies.

\end{document}